\documentclass[aps,twocolumn,nofootinbib,showpacs,showkeys,final,balancelastpage,superscriptaddress]{revtex4}
\usepackage{graphicx}
\usepackage{amssymb}
\usepackage{amsmath}
\usepackage{amsbsy}
\usepackage{amsfonts}
\usepackage{comment}
\usepackage{mathrsfs}
\usepackage{dsfont}
\usepackage{nicefrac}
\usepackage{epsf,color,colordvi,pifont}
\usepackage[notref,notcite]{showkeys}
\usepackage{amsfonts}

\DeclareFontFamily{OT1}{pzc}{}
\DeclareFontShape{OT1}{pzc}{m}{it}{<-> s * [1.10] pzcmi7t}{}
\DeclareMathAlphabet{\mathpzc}{OT1}{pzc}{m}{it}

\begin{document}

\title{QED with external field: Hamiltonian treatment for anisotropic medium%
\\
formed by the Lorentz-noninvariant vacuum}
\author{Selym \surname{Villalba-Ch\'avez}\footnote{%
Present address: Institut f\"{u}r Theoretische Physik I, Heinrich Heine
Universit\"{a}t D\"{u}sseldorf. Universit\"{a}tsstrasse. 1/Geb\"{a}ude
25.32.01 40225 D\"{u}sseldorf, Germany.}}
\affiliation{Max-Planck-Institut f\"ur Kernphysik, Saupfercheckweg 1 D-69117 Heidelberg,
Germany.}
\email{selym@tp1.uni-duesseldorf.de}
\author{Anatoly \surname{E. Shabad}}
\affiliation{P. N. Lebedev Physics Institute, Moscow 117924, Russia.}
\email{shabad@lpi.ru}
\date{\today }

\begin{abstract}
Nonlinear electrodynamics, QED included, is considered against the
Lorentz-noninvariant external field background, treated as an anisotropic
medium. Hamiltonian formalism is applied to electromagnetic excitations over
the background, and entities of electrodynamics of media, such as field
inductions and intensities, are made sense of  in terms of canonical variables.
Both conserved and nonconserved generators of space-time translations and
rotations are defined on the phase space, and their Hamiltonian equations of
motion and Dirac bracket relations, different from the Poincar\'e algebra,
are established. Nonsymmetric, but--in return--gauge-invariant,
energy-momentum  tensor suggests a canonical momentum density other
than the Poynting vector. A photon magnetic moment  is found to
govern the evolution of the photon angular momentum. It is determined by the
antisymmetric part of the  energy-momentum tensor.
\end{abstract}

\pacs{%
{11.30.Cp,}{}
{11.30.Qc,}{}
{12.20.-Ds,}{}
{11.10.Jj,}{}
{13.40.Em,}{}
{14.70.Bh.}{}%
}
\keywords{Lorentz symmetry breaking, Vacuum polarization, Photon Magnetic
Moment}
\maketitle


\section{Introduction}


While  relativistic invariance as a symmetry under the Lorentz group [$%
SO(3,1)$] is usually an obligatory requirement imposed on a theory,
some classes of theories in which it is, at least, weakly broken are also
of interest, especially when looking beyond the Standard Model. Quantum
electrodynamics (QED) in an external classical electromagnetic field $(%
\mathscr{F}_{\mu\nu})$ is a clear example of a Lorentz-violating theory that
may share principal features with other theories of that series. This is a 
motivation for studying it on a very general basis.

The self-coupling of electromagnetic fields by means of  the creation and
annihilation of virtual charged fermions makes QED a nonlinear
electrodynamics. Like any other nonlinear electrodynamics--for instance,
Born-Infeld electrodynamics--QED proposes an interaction between a strong
classical external field and electromagnetic fields that live against its
background even when these are small. The linearized approximation  
based on the smallness of these perturbations (``photons") will be dealt with
in this paper. The most important object responsible for the interaction of
photons with the background in the linearized theory is the vacuum
polarization  tensor, $\Pi_{\mu\nu}(x,x^\prime)$,  calculated in the external
field. Through this object the gauge sector of QED is, in the first
instance, provided with a dependence on the $\mathscr{F}_{\mu\nu}$ structure
and therefore on the reference frame. Consequently, the photon vacuum seems
to behave like an (in general, moving) anisotropic material, in which  
light propagation is strongly modified. Perhaps the most remarkable
property associated with this issue concerns the existence of photon degrees
of freedom which are not in correspondence to the standard observable--
helicity values--of the respective irreducible $SO(3,1)-$%
representations. Instead, the photon propagation modes turn out to be
closely associated with birefringent states \cite{batalin}, and their speeds
of propagation differ from the speed of light in an empty space-time. Some
interesting features that occur in the linearized approximation, when the
background is a constant and homogeneous magnetic field, have been
predicted. These are cyclotron resonance in the vacuum \cite{shabadrc}
leading to  photon capture \cite{shabadpc},  anisotropization, the
short-ranging and the dimensional reduction of the potential \cite{shabad5}
produced by a pointlike static charge in a supercritical magnetic field $%
\vert\pmb{B}\vert\gg B_c,$ $B_c=\mathrm{m^2c^3/e\hbar=4.42\cdot 10^{13} G},$
where $\mathrm{m}$ and $\mathrm{e}$ are the electron mass and charge,
respectively,\footnote{%
From now on, rationalized Heaviside-Lorentz units, $\epsilon_0=\hbar=c=1$ are
used.} and  also the production of a magnetic field by a static charge (the
magnetoelectric effect) \cite{PRD2010} that takes place in the external
field, where electric and magnetic fields coexist in parallel. Beyond the
linearized approximation, the important effect of photon splitting and
merging in a magnetic field \cite{adler} has attracted much attention. Other
consequences of the self-interaction of small electromagnetic fields are the
magnetoelectric effect in QED with an  external magnetic field \cite{gitshab}
and in the nonlinear electrodynamics generated by the $U_*$(1)
noncommutative theory \cite{stern}.

Despite the achievements reached in this area, a formal treatment of 
Lorentz Symmetry Breaking (LSB) in nonlinear electrodynamics has not yet
been fully developed. A clear understanding of this theme has  paramount
importance in theoretical physics, since the QED vacuum in an external field
constitutes an ideal laboratory for studying the unconventional properties
of other Lorentz-violating theories encompassed as possible extensions of
the minimal Standard Model of the fundamental interactions. Among the
candidates appear the Lorentz-violating electrodynamics \cite{KosteleckyI,Kostelecky:2007zz,Kostelecky} and the noncommutative field theories \cite{carroll}
with their phenomenologies closely related to those described in the
preceding paragraph. Moreover, since the classification of particles is
intimately related with the realization of the space-time symmetry, the
employing of this analysis could lead to new insights on plausible phenomena
in which other spin representations like axions (spin $0$) \cite{Gies:2007ua}
and gravitons (spin $2$) \cite{bastianelli1} are coupled to $\mathscr{F}$.

Of course, an experimental confirmation of all these processes strongly
depends on the external field strengths, whose current laboratory values are
much lower than the critical field. As a consequence, all pre\-dicted phenomena
remain elusive and far from being de\-tectable. Even so, their studies still
find a high mo\-tivation in light of upcoming laser facilities \cite{ELI,hiper}
that will achieve the unprecedented level of $\vert\pmb{E}\vert\sim 0.01-0.1E_{c}$
with $E_{c}=\mathrm{m}^{2}/\mathrm{e}=1.3\cdot 10^{16}\ \mathrm{V}/\mathrm{cm}.$ In
addition, some evidence points out the possible existence of ultrahigh
magnetic $\vert\pmb{B}\vert\gg B_{c}$ and electric $\vert\pmb{E}\vert\gg E_{c}$ fields in
the surfaces of stellar objects identified as neutron stars \cite{Manchester}
and strange stars \cite{AFO86}, respectively. In such scenarios a pronounced
LSB is expected and most of the quantum processes described above could have  
significant astro\-physical and cosmological interest.

Inspired by the importance associated with  LSB, we make an attempt to
fill some gaps in this topic. Our main purpose is to analyze how the vacuum
polarization effects modify the Lorentz and Poincar\'e generators. The
results presented in this work are based on the Poincar\'e invariance of the
photon effective action as a functional of the background field. This
implies that those Lorentz transformations which leave the external field
invariant, together with the space-time translations that do not affect the
external field either, as long as it is time and space independent,
provide the residual symmetry subgroup of the anisotropic vacuum, while the
full Poincar\'e group remains the group of broken symmetry. To define the
space-time translation and rotation generators, although only a part of them
is conserved, we appeal to the context of the Noether theorem and then
introduce them into the framework of the Hamiltonian formalism, which
requires us to impose constraints associated with the gauge invariance of the
theory, and thus the Dirac brackets. The general aspects related to the
constrained Hamiltonian dynamics were developed by Dirac \cite{Dirac:1958sq}
and have been applied to several problems in Quantum Field Theory \cite{books}, including the
analysis of Poincar\'e invariance in Yang-Mills theories quantized in
noncovariant gauges \cite{Burnel:2008zz} (for the Coulomb gauge, see 
Ref.~\cite{Besting:1989nq}). We do not know whether the Hamiltonian formalism was
ever applied to electrodynamics of an anisotropic medium or whether the
characteristic entities of the latter are known in terms of canonical
variables, but in  any event  the present exploitation of this formalism results in
some interesting features. Among them is the distinction between the
generating function of infinitesimal canonical transformations of spatial
translations, that most naturally turns out to be parallel to the wave
vector in each eigenmode, and the Poynting vector that points in the  direction of 
the energy propagation and the group velocity. The Dirac commutation
relations between the space-time generators leave intact the $SO(3)-$%
algebra of the angular momentum, although this symmetry group is broken, but
the rest of the Poincar\'e algebra commutation relations that include
transformations of the time are distorted. The Hamiltonian equation of
motion for the nonconserving angular momentum indicates that its
time evolution is determined by the interaction of the external field with a
certain magnetic moment (that also contributes to the Hamiltonian)
emerging in the special frame in which a pure magnetic field is present.
Such a quantity is interpreted as the magnetic moment of the photon; i.e., the
magnetic moment of the anisotropic medium that is not only polarized, but
also magnetized by the photon field. The possible connection of the photon
magnetic moment with the same notion introduced previously by Per\'ez Rojas
and one of the present authors (Villalba-Chavez) \cite{selhugo} (see also
Refs.~\cite{hugoelsel,Chavez:2009ia}) will be discussed in a due place
below. It looks at attribute of the photon interaction with an
anisotropic medium.

We organize the manuscript as follows. In Sec. \ref{sect2} A, we recall some
basic features of the photon propagation in an external field based on the
relativistic covariant formalism introduced in Ref.~\cite{batalin} that
involves diagonalization of the polarization tensor and analysis of its
eigenvalues and eigenvectors. In Sec. \ref{sect2B}, we consider the
contribution of the polarization operator to the effective nonlocal action
(the generating functional of irreducible vertices \cite{Weinberg:1995mt})
to give it the form of the action of the equivalent linear anisotropic
medium with time and space dispersion. The tensor decompositions and
principal values of dielectric, $\varepsilon _{ij}$, and magnetic, $\mu
_{ij} $ permittivities--related to the special class of Lorentz frames 
where there is only a magnetic, or only an electric,  external field--are 
presented in terms of the polarization operator eigenvalues in an
approximation-independent way. We point out  the uniaxial character of the vacuum
in these frames. This statement holds true also for the most general
combination of constant electric and magnetic fields in the Lorentz frames,
where these fields are parallel. In Sec. \ref{sect2C}, for the same
general external field, we obtain the covariant decompositions of the global
coordinate transformations that leave the external field intact and thus
make the invariance subgroup of our problem.

In Sec. \ref{sect3} and thenceforth, we confine ourselves to the local
approximation of the effective action, when it does not include field
derivatives as the functional arguments. This limitation will make the
subsequent development of the Hamiltonian formalism straightforward. The
local approximation corresponds to the small four-momentum (infrared) limit of
the polarization operator and to the frequency- and momentum-independent tensors
$\varepsilon _{ij}$ and $\mu _{ij}$. The latter are expressed in terms of
derivatives of the effective Lagrangian over the field invariants for the
cases reducible to a single field in a special frame (magneticlike and
electriclike cases). In Sec. \ref{contructionenmotensor}, based on
the Noether theorem, we define a nonsymmetrical, but gauge-invariant,
energy-momentum tensor of small perturbations of the vacuum that satisfies
the continuity equation with respect to only one of its tensor indices. The
antisymmetric part of this tensor will become, in what follows, responsible
for the nonconservation of those Lorentz and spatial rotation generators of
canonical transformations constructed using the energy-momentum tensor,
which do not leave the external field invariant. We stress the difference
between the momentum flux vector (canonical momentum), whose direction is
parallel to the wave vector according to Appendix \ref{mpvsew}, and the
energy flux Poynting vector, whose direction coincides with that of the
group velocity and the center-of-mass velocity of eigenmodes, as
demonstrated in Appendix \ref{pvgvsew}. The constrained Hamiltonian
formalism serving the dynamics of small perturbations over the background
field is presented in Sec. \ref{sect3B} following the procedure well
elaborated in gauge theories. It is the electric induction of the
perturbation, and not the field strength, that comes out as a variable,
canonically conjugated to its three-vector potential. With the Coulomb gauge
condition extended to the problem of the anisotropic vacuum under
consideration, the Dirac brackets are defined as performing the
infinitesimal canonical transformations in the phase space. In Sec.  
\ref{spingenerallrofjdffk}, referring again to the Noether transformations,
we define the conserved and nonconserved components of the angular momentum
and the Lorentz boost, express them in terms of the canonical variables,  and
find their Dirac brackets with the fields and inductions. In Sec. \ref{sect4}, 
the Hamiltonian equations of motion for these quantities are given (Sec. IV A), 
as well as the set of Dirac commutators for the generators of
space-time translations and rotations, defined above, that substitute for
the standard relations of the the Poincar\'e algebra in the present case  the vacuum  invariant not  under all 
Lorentz and space rotations (Sec. IV B and Appendices \ref{EMA}, \ref{MLA}). In Sec.  \ref{sect4C}, we dwell on
the algebra of the space-time invariance subgroup and define its conserved
Casimir invariants.

In Sec.  \ V, we deal with the magneticlike external field. The magnetic
moment $\pmb{\mathpzc{M}}$ of the photon propagating over the magnetized
vacuum is analyzed. It appears as an entity that governs the evolution of
the photon angular momentum $\pmb{\mathscr{J}}$ in the magnetic field $%
\pmb{B}$ following the equation of motion $d\pmb{\mathscr{J}}/dx^0=2%
\pmb{\mathpzc{M}}\times\pmb{B},$ and contributes into the photon energy as $-%
\pmb{\mathpzc{M}}\cdot \pmb{B}$. A further step in our understanding of this
quantity is given by showing its connection with the optical tensors of our
problem. In the large$-\vert\pmb{B}\vert$  region, the photon magnetic moment
treated following the one-loop approximation of quantum electrodynamics  
depends quadratically on the photon electric field alone. Its
appearance may be understood as another manifestation of the magnetoelectric
effect \cite{PRD2010,gitshab} in QED, known also in noncommutative
electrodynamics \cite{stern}.

In Sec. VI,  we write down the coefficient tensor customarily used to serve
the gauge sector in the general Lorenz-violation approach to a
$U(1)-$invariant theory, as it follows from the general covariant decomposition
of the polarization tensor in a magnetic field found in Ref.~\cite{batalin}. We
establish that the coefficient tensor is not double traceless, contrary to
what  is assumed in the above approach. The double trace is physically
meaningful as being connected with the magnetic and electric permeability of
the magnetized vacuum. We express the condition of the absence of
birefringence in terms of field derivatives of the effective Lagrangian.
Finally, we estimate the values of the magnetic field likely to  produce  the
Lorenz violation that would be equivalent to the Lorentz violations
intrinsic in the vacuum and detectable using experimental devices of 
present-day sensitivity. These magnetic fields are too large to make a
realistic cosmic background.

We present our concluding summary in Sec.  VII, while the essential steps of
many calculations have been deferred to the  appendixes.


\section{Lorentz symmetry breaking: General aspects \label{sect2}}



\subsection{The photon effective action \label{sect2A}}


In the presence of an external field $\mathscr{A}_{\mu }(x)=-\frac{1}{2}%
\mathscr{F}_{\mu \nu }x^{\nu }$ with a constant field strength $\mathscr{F}%
_{\mu \nu }=\partial _{\mu }\mathscr{A}_{\nu }-\partial _{\nu }\mathscr{A}%
_{\mu },$ the action which describes small-amplitude electromagnetic waves $%
a_{\mu }(x)$ over a constant background field reads
\begin{equation}
S=-\frac{1}{4}\int f^{\mu \nu }f_{\mu \nu }d^{4}x+\Gamma ,  \label{S}
\end{equation}
where $f_{\mu \nu }=\partial _{\mu }a_{\nu }-\partial _{\nu }a_{\mu },$ and $%
\Gamma $ is called  the effective action, connected as $\Gamma =\int \mathfrak{L}%
d^{4}x$ with the effective Lagrangian $\mathfrak{L.}$ Effective action may
be expanded in powers of the small field potential as
\begin{equation}
\Gamma =\frac{1}{2}\int d^{4}xd^{4}x^{\prime }a^{\mu }(x)\Pi _{\mu \nu
}(x,x^{\prime }|\mathscr{A})a^{\nu }(x^{\prime })+\ldots
\label{effelagrangian}
\end{equation}%
where $+\ldots $ stands for higher-order terms in $a_{\mu },$ whereas $\Pi
_{\mu \nu }(x,x^{\prime }|\mathscr{A})$ is the second-rank polarization
tensor related as
\begin{eqnarray}
&&\mathscr{D}_{\mu \nu }^{-1}(x,x^{\prime }|\mathscr{A})=\left[ \square \eta
_{\mu \nu }-\partial _{\mu }\partial _{\nu }\right] \delta ^{(4)}(x^{\prime
}-x)  \notag  \label{propagatorinversoQED} \\
&&\qquad \qquad \qquad \quad +\Pi _{\mu \nu }(x,x^{\prime }|\mathscr{A})
\end{eqnarray}%
with the inverse photon Green function $\mathscr{D}_{\mu \nu
}^{-1}(x,x^{\prime }|\mathscr{A}).$  Hereafter,  the metric tensor $\eta ^{\mu
\nu }$ has the signature $+++-$ with $\eta ^{11}=\eta ^{22}=\eta ^{33}=-\eta
^{00}=1;$ the electric field is given by $E^{i}=\mathscr{F}^{0i},$ whereas
the magnetic field is defined by $B^{i}=1/2\epsilon ^{ijk}\mathscr{F}_{jk}.$
The tensor $\tilde{\mathscr{F}}^{\mu\nu}=1/2\epsilon^{\mu\nu\rho\sigma}%
\mathscr{F}_{\rho\sigma}$ (with $\epsilon^{\mu\nu\rho\sigma}$ being the
fully antisymmetric unit tensor,  $\epsilon^{1230}=1$) represents the dual
of $\mathscr{F}_{\mu\nu}.$ The field invariants are $\mathfrak{F}=1/4%
\mathscr{F}_{\mu\nu}\mathscr{F}^{\mu\nu}=(\pmb{B}^2-\pmb{E}^2)/2$ and $%
\mathfrak{G}=1/4\mathscr{F}^{\mu\nu}\tilde{\mathscr{F}}_{\mu\nu}=-%
\pmb{E\cdot B}=0$. For the special case of an external field mainly dealt with
in the present work,  we shall choose $\mathfrak{G}=0$. Then the corresponding
external field will be referred to as magneticlike (if $\mathfrak{F}>0$) or
electriclike (if $\mathfrak{F}<0$), because in either case  a Lorentz frame
exists  wherein the field is purely magnetic or electric, respectively.

According to Eq.~(\ref{effelagrangian}),  a photon can interact with the
external field through the vacuum polarization tensor $\Pi_{\mu\nu}(
x,x^\prime\vert \mathscr{A})$. In this context, the QED Schwinger-Dyson
equation for the photon field $a_\mu(x)$ is given by
\begin{eqnarray}
\int d^4x^\prime\mathscr{D}_{\mu\nu}^{-1}(x,x^\prime\vert \mathscr{A})
a^\nu(x^\prime)=0.  \label{sdpmBF}
\end{eqnarray}
The external field strength is independent of the space-time coordinates;
therefore, the polarization tensor, as a gauge-invariant quantity, should
correspond to a spatially homogeneous optical medium  whose properties do
not change with  time. This is provided by the translational invariance
of $\Pi_{\mu\nu}$: it depends only on the coordinate difference $%
\Pi_{\mu\nu}(x,x^\prime\vert \mathscr{A})=\Pi_{\mu\nu}(x-x^\prime\vert %
\mathscr{A})$ \cite{shabadpolarization}. In this case, a Fourier transform
converts Eq.~(\ref{sdpmBF}) into a linear homogeneous algebraic equation
given by
\begin{eqnarray}
\left[k^2\eta_{\mu\nu}-k_\mu k_\nu-\Pi_{\mu\nu}(k\vert \mathscr{A})\right]%
a^\nu(k)=0  \label{tsdp}
\end{eqnarray}
with
\begin{eqnarray}
\Pi_{\mu\nu}(k\vert \mathscr{A})=\int \Pi_{\mu\nu}(x-x^\prime\vert %
\mathscr{A})\mathrm{e}^{-i k (x-x^\prime)} d^4(x-x^\prime).  \label{hermit}
\end{eqnarray}

To understand what follows, it is necessary to recall some basic results
developed in Refs.~\cite{batalin,shabadpolarization}. In the presence
of a constant magneticlike or electriclike field, the four eigenvectors of the
polarization operator $\flat_\mu^{(\lambda)}$ are known in a final,
approximation-independent form. In addition to the photon momentum
four-vector $\flat^{(4)}_\mu=k_\mu$ (its zeroth component $k^0$ being  the
frequency $\omega$), the three other mutually orthogonal four-transverse
eigenvectors $\flat_\mu^{(\lambda)}$ are
\begin{eqnarray}
\begin{array}{c}
\displaystyle \flat^{(1)}_\mu= k^2 \mathscr{F}^2_{\mu
\lambda}k^\lambda-k_\mu (k\mathscr{F}^2 k), \\
\displaystyle \flat^{(2)}_\mu=\frac{\tilde{\mathscr{F}}_{\mu
\lambda}k^\lambda}{(k\tilde{\mathscr{F}}^2k)^{1/2}}, \qquad \flat^{(3)}_\mu=%
\frac{\mathscr{F}_{\mu \lambda}k^\lambda}{(-k\mathscr{F}^2k)^{1/2}},%
\end{array}
\label{eigvec}
\end{eqnarray}
$k^\mu \flat^{(\lambda)}_{\mu}=0$ for $\lambda=1,2,3$. (The eigenvectors
relating to the most general case, $\mathfrak{F}\neq0,~\mathfrak{G}\neq0,$ are
written in Refs.~\cite{batalin,shabadpolarization} and \cite%
{footnote,PRD2011}). We remark that $\flat^{(\lambda)}_\mu$ fulfills both
the orthogonality condition,  $\flat_{\mu}^{(\lambda
)}\flat^{\mu(\lambda^\prime)}=\delta^{\lambda\lambda^\prime}\left(
\flat^{(\lambda)}\right)^2,$ and the completeness relation, 
\begin{eqnarray}
\eta_{\mu\nu}-\frac{k_\mu k_\nu}{k^2}=\sum_{\lambda=1}^3\frac{%
\flat_\mu^{(\lambda)}\flat_{\nu}^{(\lambda)}}{\left(\flat^{(\lambda)}%
\right)^2}.  \label{completenessrelation}
\end{eqnarray}
Note that from $\flat^{(\lambda)},$ one obtains the fundamental scalars
\begin{eqnarray}
\begin{array}{c}
k^2=z_1+z_2, \\
\displaystyle z_1=\frac{k\tilde{\mathscr{F}}^{2}k}{2\mathfrak{F}}\ \ \mathrm{%
and} \ \ z_2=-\frac{k\mathscr{F}^2k}{2\mathfrak{F}}.%
\end{array}
\label{casimirs}
\end{eqnarray}%
The last two scalars acquire simple forms in a special reference frame 
where the external field is purely magnetic (if $\mathfrak{F}>0$) or purely
electric (when the opposite inequality holds). The same equations hold in
reference frames that are moving parallel to the external field. For the
magnetic background $\mathfrak{F}>0$, one finds  that $z_2=k_\perp^2$ and $z_1=
k_\parallel^2-\omega^2.$ On the contrary, if the electric field is
considered, $z_2=k_\parallel^2-\omega^2,$  whereas $z_1=k_\perp^2.$ The
previous relations involve the vectors ${\pmb{k}}_{\perp}$ and $\pmb {k}%
_{\parallel},$ which denote the components of $\pmb{k}$ perpendicular to  and
along the external field,  respectively. Henceforth, boldface letters will
designate the spatial part of our four-vectors.

Besides the creation of the fundamental scalars,  i.e., Eq.~(\ref%
{casimirs}), the vectorial basis $\flat ^{(\lambda )}$ is suitable to
express the vacuum polarization tensor in a diagonal form
\begin{equation}
\Pi _{\mu \nu }=\sum_{\lambda =0}^{4}\varkappa _{\lambda }(z_{1},z_{2},%
\mathfrak{F})\frac{\flat _{\mu }^{(\lambda )}\flat _{\nu }^{(\lambda )}}{%
\left( \flat ^{(\lambda )}\right) ^{2}}.  \label{gstrpi}
\end{equation}%
Here $\varkappa _{\lambda }$ denotes the eigenvalues of the vacuum
polarization tensor
\begin{equation*}
\Pi_\mu^{~\tau}~\flat_{\tau}^{(a)}=\varkappa_{a}(k)~\flat_{\mu}^{(a)},\quad
a=1,2,3,4,
\end{equation*}
which define the energy spectrum of the electromagnetic waves and poles of
the photon propagator. Owing to the transversality property $(k^{\mu }\Pi
_{\mu \nu }=0),$ the eigenvalue corresponding to the fourth eigenvector
vanishes identically $[\varkappa ^{(4)}=0].$ Substituting Eq.~(\ref{gstrpi})
into Eq.~(\ref{tsdp}) and using the orthogonality condition, we find its
solutions in the form of a superposition of eigenmodes given by $a_{\mu
}(k)=\sum_{\lambda =1}^{3}F_{\lambda }\delta (k^{2}-\varkappa _{\lambda
})\flat _{\mu }^{(\lambda )},$ where $F_{\lambda }$ are arbitrary functions
of $k.$ According to the latter,  three nontrivial dispersion relations arise:
\begin{equation}  \label{dispequat}
k^{2}=\varkappa _{\lambda }\left( z_{2},z_{1},\mathfrak{F}\right),\ \lambda
=1,2,3,
\end{equation}
whose solutions can be written as
\begin{equation}  \label{dispequatbelow}
\omega _{\lambda }^{2}=k_{\parallel }^{2}+f_{\lambda }(k_{\perp }^{2},%
\mathfrak{F}).
\end{equation}%
The term $f_{\lambda }(k_{\perp }^{2},\mathfrak{F})$ arises as a sort of
dynamical mass. Due to the gauge invariance condition $\varkappa _{\lambda
}(0,0,\mathfrak{F})=0$ there always exist two (out of three) solutions with $%
f_{\lambda }(0,\mathfrak{F})=0$ that correspond to photons  whose rest
energy is zero,  and the number of polarization degrees of freedom is two.
Massive branches of all the three polarizations, $f_{\lambda }(0,\mathfrak{F}%
)\neq 0,~\lambda =1,2,3,$ may also exist. For more details,  we refer the
reader to Refs.~\cite{PRD2011, PRD2010}.%

Moreover, by considering $a_{\mu }\sim \flat _{\mu }^{(\lambda )}$ as the
electromagnetic four-vector describing the eigenmodes, we obtain the
corresponding electric and magnetic fields of each mode in the special frame,
provided that $\mathfrak{F}>0:$
\begin{equation}
\pmb{e}^{(\lambda )}\simeq i\left( \omega ^{(\lambda )}\pmb{\flat}^{(\lambda
)}-\pmb{k}\flat ^{0(\lambda )}\right) \ \ \mathrm{and}\ \ \pmb{b}^{(\lambda
)}\simeq i\pmb{k}\times \pmb{\flat}^{(\lambda )}.  \label{electricpromag}
\end{equation}%
Up to a nonessential proportionality factor, they are explicitly given by
\begin{equation*}
\begin{array}{c}
\pmb{e}^{(1)}\simeq -i\pmb{n}_{\perp }\omega ,\ \ \pmb{b}^{(1)}\simeq -i%
\pmb{k}_{\parallel }\times \pmb{n}_{\perp }, \\
\pmb{e}_{\perp }^{(2)}\simeq -i\pmb{k}_{\perp }k_{\parallel }/(k_{\parallel
}^{2}-\omega ^{2})^{1/2},\ \ \pmb{e}_{\parallel }^{(2)}\simeq -i\pmb{n}%
_{\parallel }(k_{\parallel }^{2}-\omega ^{2})^{1/2}, \\
\pmb{b}^{(2)}\simeq i\omega (\pmb{k}_{\perp }\times \pmb{n}_{\parallel
})/(k_{\parallel }^{2}-\omega ^{2})^{1/2}, \\
\pmb{e}^{(3)}\simeq i\omega (\pmb{n}_{\perp }\times \pmb{n}_{\parallel }),
\\
\pmb{b}_{\parallel }^{(3)}\simeq -i\pmb{n}_{\parallel }k_{\perp },\ \ \pmb{b}%
_{\perp }^{(3)}\simeq i\pmb{n}_{\perp }k_{\parallel }.%
\end{array}%
\end{equation*}%
Here, $\pmb{n}_{\parallel }=\pmb{k}_{\parallel }/|\pmb{k}_{\parallel }|$ and
$\pmb{n}_{\perp }=\pmb{k}_{\perp }/|\pmb{k}_{\perp }|$ are the unit vectors
associated with the parallel and perpendicular directions with respect to the
magnetic field $\pmb{B}$ in the special frame in which the external electric
field $\pmb{E}$ vanishes identically.


\subsection{The vacuum as an anisotropic medium within a  linear optics approximation \label{sect2B}}


In classical electrodynamics, the Maxwell action of a linear continuous
medium with  the dielectric tensor $\varepsilon _{ij}(\pmb{k},\omega )$ and
the magnetic permeability tensor $\mu _{ij}(\pmb{k},\omega )$ is given by the
expression  (quadratic in the fields)
\begin{equation}
\begin{array}{c}
\displaystyle\mathcal{S}=\int \displaystyle L=\frac{1}{2}\pmb{d}(-\pmb{k}%
,-\omega )\cdot \pmb{e}(\pmb{k},\omega )-\frac{1}{2}\pmb{h}(-\pmb{k},-\omega
)\cdot \pmb{b}(\pmb{k},\omega ).%
\end{array}
\label{accionanisotroipca}
\end{equation}

Here $\pmb{e},\pmb{b}$ are connected by means of the relations
\begin{equation}  \label{eq15}
\begin{array}{c}
\pmb{d}(\pmb{k},\omega )=\tensor{\pmb{\varepsilon}}(\pmb{k},\omega )\cdot %
\pmb{e}(\pmb{k},\omega ), \\
\pmb{h}(\pmb{k},\omega )=\tensor{\pmb{\mu}}^{-1}(\pmb{k},\omega )\cdot %
\pmb{b}(\pmb{k},\omega ).%
\end{array}%
\end{equation}%
where the double-sided arrow denotes a tensorial quantity. In this context, $%
\tensor{\pmb{\varepsilon}}\cdot \pmb{e}=\varepsilon _{ij}e_{j}$ and $%
\tensor{\pmb{\mu}}^{-1}\cdot \pmb{b}=\mu _{ij}^{-1}b_{j}.$

The optical properties of an  anisotropic medium depend primarily on the
symmetry of its tensors $\varepsilon _{ij}$ and $\mu _{ij}.$ In an uniaxial
medium, one of the principal axes of $\varepsilon _{ij}$ and $\mu _{ij}$
forms   the ``optical axis''. In what follows, we denote the principal
values of $\varepsilon _{ij}$ and $\mu _{ij}$ relating to this axis as $%
\varepsilon _{\parallel}$ and  $\mu _{\parallel },$ respectively, and the values  
relating to the plane  perpendicular to the optical axis as $\varepsilon
_{\perp }$ and $\mu _{\perp },$ respectively. With this in mind,  the  Maxwell
Lagrangian [Eq.~(\ref{accionanisotroipca})] acquires the following form:
\begin{equation}
L=\frac{1}{2}\left\{ \varepsilon _{\perp }|\pmb{e}_{\perp }|^{2}-\frac{1}{%
\mu _{\perp }}|\pmb{b}_{\perp }|^{2}+\varepsilon _{\parallel }|\pmb{e}%
_{\parallel }|^{2}-\frac{1}{\mu _{\parallel }}|\pmb{b}_{\parallel
}|^{2}\right\}  \label{actionsimple}
\end{equation}%
where we have decomposed $\pmb{e}=(\pmb{e}_{\perp },\pmb{e}_{\parallel })$
and $\pmb{b}=(\pmb{b}_{\perp },\pmb{b}_{\parallel }).$ Here the symbols  $%
\perp $ and $\parallel $ refer to the optical axis as well.

Let us consider now the quadratic part of the effective action corresponding
to the dynamical gauge field sector of QED in an external field.
Substituting Eq.~(\ref{gstrpi}) into Eq.~(\ref{effelagrangian}) and making
use of Eq.~(\ref{completenessrelation}), we find for the integrand of $S$  [Eq.~(\ref{S})] in momentum space $\displaystyle\mathscr{L}=-\frac{1}{4}f^{\mu \nu
}f_{\mu \nu }+\mathfrak{L},$
\begin{equation}  \label{generaleffectivelagragianstrongfield}
\begin{array}{c}
\displaystyle\mathscr{L}=-\frac{1}{4}\mathcal{O}^{\mu \nu }f_{\mu \nu }.%
\end{array}%
\end{equation}%
Here the second-rank antisymmetric tensor $\mathcal{O}^{\mu \nu }$ reads
\begin{eqnarray}
\mathcal{O}^{\mu \nu } &=&\left( 1-\frac{\varkappa _{1}}{k^{2}}\right)
f^{\mu \nu }-\frac{1}{2}\frac{\varkappa _{1}-\varkappa _{2}}{k\tilde{%
\mathscr{F}}^{2}k}\left( f^{\lambda \sigma }\tilde{\mathscr{F}}_{\lambda
\sigma }\right) \tilde{\mathscr{F}}^{\mu \nu }  \notag \\
&&\quad -\frac{1}{2}\frac{\varkappa _{1}-\varkappa _{3}}{k\mathscr{F}^{2}k}%
\left( f^{\lambda \sigma }\mathscr{F}_{\lambda \sigma }\right) \mathscr{F}%
^{\mu \nu }.  \label{nuevotensorinductions}
\end{eqnarray}%
The Lagrangian $\mathscr{L}$ was written in Ref.~\cite{PRD2011}; its
small-momentum form [corresponding to Eq.~(\ref{IReffelagrangianmass}) in
the next Subsection]  is present in an earlier paper \cite{dipiazza}.

The expression above is valid for  both   magneticlike  and electriclike  cases  ($%
\mathfrak{F}\lessgtr0,$ $\mathfrak{G}=0$) and defines the corresponding
induction vectors according to the following rule:
\begin{eqnarray}  \label{vectorindcutiondefinition}
\begin{array}{c}
\displaystyle d^i=\mathcal{O}^{0i}=\partial\mathscr{L}/\partial{e^i} \\
\displaystyle h^i=\frac{1}{2}\epsilon^{ijk}\mathcal{O}_{jk}=-\partial%
\mathscr{L}/\partial{b^i}%
\end{array}%
\end{eqnarray}
where $\pmb{e}(x)=\pmb{\nabla}a_0(x)-\partial_0\pmb{a}(x)$ and $\pmb{b}=%
\pmb{\nabla}\times\pmb{a}$ are the averaged (classical) electric and
magnetic fields associated with the electromagnetic wave. With these
definitions, the Maxwell equations have the recognizable form
\begin{eqnarray}  \label{19}
\begin{array}{c}
\displaystyle \pmb{\nabla}\cdot\pmb{d}=0,\quad \pmb{\nabla}\cdot\pmb{b}=0;
\\
\\
\displaystyle \pmb{\nabla}\times\pmb{e}=-\frac{\partial \pmb{b}}{\partial x^0%
}, \quad \pmb{\nabla}\times\pmb{h}=\frac{\partial \pmb{d}}{\partial x^0}. \\
\end{array}%
\end{eqnarray}
The above expressions allow us to obtain the most general structure of the
dielectric and magnetic permeability tensor. To derive it, we first note
that our effective Lagrangian $\mathscr{L}$ in Eq.~(\ref%
{generaleffectivelagragianstrongfield}) acquires the structure of Eq.~(\ref%
{accionanisotroipca}) as long as it is expressed in terms of the induction
vectors $\pmb{d},$ $\pmb{h}$ and the electric $\pmb{e}$ and magnetic $\pmb{b}
$ fields of the small electromagnetic waves. Likewise, the optical tensors
can be defined as in Eq.~(\ref{eq15}). In fact, by considering the following
relations, valid in the special frames
\begin{eqnarray}
\begin{array}{c}
-\frac{1}{4}f^{\mu\nu}\tilde{\mathscr{F}}_{\mu\nu}=\frac{1}{2}\pmb{e}\cdot %
\pmb{B},\ \ \frac{1}{4}f^{\mu\nu}\mathscr{F}_{\mu\nu}=\frac{1}{2}\pmb{b}%
\cdot \pmb{B},\ \ \mathfrak{F}>0 \\
\\
-\frac{1}{4}f^{\mu\nu}\tilde{\mathscr{F}}_{\mu\nu}=\frac{1}{2}\pmb{b}\cdot %
\pmb{E},\ \ \frac{1}{4}f^{\mu\nu}\mathscr{F}_{\mu\nu}=-\frac{1}{2}\pmb{e}%
\cdot \pmb{E},\ \ \mathfrak{F}<0%
\end{array}
\label{relationforefecyivelagraganiaf}
\end{eqnarray}
one can express them for the magnetic external field
\begin{eqnarray}
\begin{array}{c}
\displaystyle \varepsilon_{ij}(\pmb{k},\omega)=\left(1-\frac{\varkappa_1}{k^2}%
\right)\delta_{ij}+\frac{\varkappa_1-\varkappa_2} {k_\parallel^2-\omega^2}%
\frac{B_iB_j}{B^2} \\
\\
\displaystyle \mu_{ij}^{-1}(\pmb{k},\omega)=\left(1-\frac{\varkappa_1}{k^2}%
\right)\delta_{ij}+\frac{\varkappa_1-\varkappa_3} {k_\perp^2}\frac{B_iB_j}{%
B^2}.%
\end{array}
\label{anisotropicmaxwelllagragiasadsn}
\end{eqnarray}
It is notable that the components of the three-momentum vector $\pmb k$ do not
take part in forming these tensors;  only components of $\pmb B$ do. This
feature is not typical of crystal optics with spatial dispersion and can be
attributed to the explicit exploitation of the gauge invariance laid in Eq.~(%
\ref{nuevotensorinductions}). The eigenvalues of matrices in  Eq.~(\ref%
{anisotropicmaxwelllagragiasadsn})  [the principal values of the electric and
(inverse) magnetic permittivities] are 
\begin{eqnarray}
\begin{array}{c}
\displaystyle \varepsilon_{\perp}=\mu_\perp^{-1}=1-\frac{\varkappa_1}{k^2},\
\ \ \ \varepsilon_{\parallel}=1-\frac{\varkappa_1}{k^2}+\frac{%
\varkappa_1-\varkappa_2}{k_\parallel^2-\omega^2}, \\
\\
\displaystyle \mu_{\parallel}^{-1}=1-\frac{\varkappa_1}{k^2}+\frac{%
\varkappa_1-\varkappa_3}{k_\perp^2}.%
\end{array}
\label{magind}
\end{eqnarray}
The values $\varepsilon_{\parallel}$ and $\mu_{\parallel}^{-1}$ correspond
to the eigenvector directed along the external magnetic field $\pmb{B}$,
which therefore makes it  the direction of the principal optical axis. The
values $\varepsilon_{\perp}=\mu_\perp^{-1}$ correspond to the eigenvectors
directed transverse to the external magnetic field. The principal values of 
Eq.~(\ref{magind}) are rotational scalars and depend upon direction: their
arguments are the frequency $\omega$ and the scalar product $\pmb{B\cdot k},$
combined into $z_1=\omega^2-k_\|^2,~z_2=k_\perp^2$. We also find a similar
result for an electriclike background $(\mathfrak{F}<0).$ In this case,  the
optical axis is determined by $\pmb{E}$ and
\begin{eqnarray}
\begin{array}{c}
\displaystyle \varepsilon_{ij}(\pmb{k},\omega)=\left(1-\frac{\varkappa_1}{k^2}%
\right)\delta_{ij}+\frac{\varkappa_1-\varkappa_3} {k_\parallel^2-\omega^2}%
\frac{E_iE_j}{E^2} \\
\\
\displaystyle \mu_{ij}^{-1}(\pmb{k},\omega)=\left(1-\frac{\varkappa_1}{k^2}%
\right)\delta_{ij}+\frac{\varkappa_1-\varkappa_2} {k_\perp^2}\frac{E_iE_j}{%
E^2}.%
\end{array}
\label{anisotropicmaxwelllagragiasadsnelectric}
\end{eqnarray}
From these tensors, we obtain that $\varepsilon_\perp$ and $\mu_\perp^{-1}$
have the same structure as in Eq.~(\ref{magind}),  whereas
\begin{eqnarray}
\varepsilon_{\parallel}=\varepsilon_\perp+\frac{\varkappa_1-\varkappa_3}{%
k_\parallel^2-\omega^2},\qquad \mu_{\parallel}^{-1}=\mu_\perp^{-1}+\frac{%
\varkappa_1-\varkappa_2}{k_\perp^2}.  \label{elecind}
\end{eqnarray}
The results given in Eqs.~(\ref{anisotropicmaxwelllagragiasadsn})-(\ref%
{elecind}) point out that in the presence of an external magnetic or
electric field,  the vacuum behaves like an uniaxial anisotropic material.
Note that the procedure shown in this section is independent of any
approximation made in the calculation of the vacuum polarization tensor.
However, it is only valid in a class of special frames, in which the
external field is purely magnetic  (when $\mathfrak{G}=0,\mathfrak{F}>0)$ or
purely electric  (when $\mathfrak{G}=0,\mathfrak{F}<0)$.  In a general Lorentz
frame, an electric (magnetic) component is added to the primarily purely
magnetic (electric) field, as produced by the Lorentz boost. Hence, the
statement that the vacuum is uniaxial is no longer true in that frame,
because the second axis is specialized by the direction of the added
component--or, in other words, by the direction of the motion of the
reference frame with respect to the special frame. Therefore, the vacuum is
  a biaxial medium that can be rendered  uniaxial by an appropriate Lorentz
transformation. (In the case of a material anisotropic medium, which is
uniaxial in its rest frame, we can also state that it becomes biaxial  if it
moves with respect to an observer, the direction of motion specializing
additional direction in the frame of the observer.)

The same statements are readily extended to the case of a  general external
field with both the field invariants different from zero: $\mathfrak{G}%
\neq0,\ \mathfrak{F}\neq0.$ In this general case, (a class of) special frames
exist, where the external electric and magnetic fields are mutually
parallel, their common direction specializing the principal optical axis in
such frames. The point is that the diagonal representation for the
polarization operator [Eq.~(\ref{gstrpi})] remains valid,  the only
reservation being that now the eigenvectors $\flat^{(\lambda)}_\nu$ in it are not
just the vectors of  Eq.~(\ref{eigvec}), but linear combinations of them \cite%
{footnote}. A representation  analogous to Eq.~(\ref{nuevotensorinductions})
can be written in that case, and the principal values substituting for Eq.~(%
\ref{magind}) again depend on the same combinations of momenta $%
\omega^2-k_\|^2$ and $k_\perp^2,$ where now the designations $\|$ and $\perp$
mark the directions parallel and orthogonal to the common direction of the
external fields.


\subsection{The vacuum symmetry subgroup $ISO_A(3,1)$: Covariant
decomposition of transformations \label{sect2C}}


Actually,  the anisotropic character of the medium, equivalent to that of the vacuum
with an external field, 
arises due to the Lorentz and rotational symmetry breakdown [which is not
manifest in the Maxwell Lagrangian of  Eq.~(\ref{accionanisotroipca}), because
it relates only to the rest frame of the medium and does not reflect its
spatial symmetry]. The Lagrangian [Eq.~(\ref%
{generaleffectivelagragianstrongfield})] is not Lorentz and
rotational invariant, because it contains an external tensor responsible for
the external field. So, to keep it invariant, one should transform the
external field together with the photon field. Correspondingly, the explicit
forms of the scalars $z_2$ and $z_1,$ as well as  $\varkappa_i,$ when these are
expressed through the photon momentum components, depend on the reference
frame. However, the Lagrangian [Eq.~(\ref{generaleffectivelagragianstrongfield})]
turns out to be invariant under those space-time transformations which leave
the external field intact. Thus, bearing in mind the translational
invariance of our problem, the proper inhomogeneous orthochronous Lorentz
transformations relating to the symmetry group of an anisotropic homogeneous
vacuum occupied by an external space- and time-independent classical field
must fulfill the conditions
\begin{eqnarray}
\begin{array}{c}
x^{\mu}\to\Lambda^\mu_{\ \ \nu} x^\nu+\epsilon^\mu,\ \
\eta_{\lambda\rho}=\Lambda^\mu_{\ \ \lambda}\Lambda^\nu_{\ \
\sigma}\eta_{\mu\nu}, \\
\\
\mathscr{F}_{\mu\nu}= \Lambda^\rho_{\ \ \mu}\Lambda^{\sigma}_{\ \ \nu}%
\mathscr{F}_{\rho\sigma},\ \ \mathrm{{det}\Lambda=1,\ \ \Lambda_0^0>0.}%
\end{array}
\label{c1}
\end{eqnarray}
The set of pairs $\{\epsilon,\Lambda\}$ satisfying Eq.~(\ref{c1}) form a
subgroup of the Poincar\'e group $[ISO(3,1)]$ which will be
referred to as the ``Amputated Poincar\'e Group,'' $ISO_{{A}}(3,1)$.
Also, $\Lambda\in SO_A(3,1),$ where $SO_A(3,1)$ is called
the ``Amputated Lorentz Group.''  Due to Eq.~(\ref{c1}),  the infinitesimal
Lorentz transformation associated with our problem can be written as \cite%
{Bacry}
\begin{eqnarray}  \label{infintesimaltransforma}
\Lambda^{\mu}_{\ \ \nu}=\delta^{\mu}_{\ \ \nu}+\omega_{\ \ \nu}^{\mu}\ \ \ \
\mathrm{with} \ \ \ \ \omega_{\ \ \nu}^\mu=\vartheta~^\prime\mathscr{F}%
^{\mu}_{\ \ \nu}+\xi~^\prime\tilde{\mathscr{F}}^{\mu}_{\ \ \nu}  \notag \\
\end{eqnarray}
where $\vartheta^\prime$ and $\xi^\prime$ are real infinitesimal parameters.
Note that the description above is not restricted to a magneticlike  or
electriclike field. On the contrary, it holds for any other constant and
homogeneous external field configuration with the second field invariant,  $%
\mathfrak{G}\neq0,$  and it  emerges in any other nonlinear electrodynamics 
different from QED.

We exploit Eq.~(\ref{infintesimaltransforma}) to obtain explicitly the
structure of those Lorentz transformations that are associated with our
problem,  i.e., $\Lambda\in SO_{A}(3,1).$ Before doing this,  we
redefine the group parameter in Eq.~(\ref{infintesimaltransforma}), $%
\vartheta^\prime=\vartheta/(\mathcal{N}_{+}+\mathcal{N}_{-})$ and $%
\xi^\prime=\xi/(\mathcal{N}_{+}+ \mathcal{N}_{-}),$ where $\mathcal{N}_{\pm}=%
\left[(\mathfrak{F}^2+\mathfrak{G}^2)^{1/2}\pm\mathfrak{F}\right]^{1/2}$ are
eigenvalues of the external field tensor $\mathscr{F}.$ In addition, we
express the finite Lorentz transformation as an exponential of a matrix
argument
\begin{eqnarray}
\Lambda=\exp\left(\frac{\vartheta\mathscr{F}}{\mathcal{N}_{+}+\mathcal{N}_{-}%
}+\frac{\xi\tilde{\mathscr{F}}} {\mathcal{N}_{+}+\mathcal{N}_{-}}\right),
\label{FLTG2}
\end{eqnarray}
defined by its series expansion.

Substantial simplifications can be achieved by the introduction of the
matrix basis
\begin{eqnarray}
\left(\mathcal{Z}_{\pm}\right)_{\ \ \nu}^{ \mu}&=&\frac{\mathcal{N}_{\pm}%
\tilde{\mathscr{F}}_{\ \ \nu}^{ \mu}\mp\mathcal{N}_{\mp}\mathscr{F}_{\ \
\nu}^{\mu}}{\mathcal{N}_{+}^2+\mathcal{N}_{-}^2}, \\
\left(\mathcal{Z}_{\pm}^2\right)_{\ \ \nu}^{ \mu}&=&\frac{\mathscr{F}^{\mu
\lambda}\mathscr{F}_{\lambda\nu}\pm\mathcal{N}_{\pm}^2\delta_{\ \ \nu}^{\mu}%
}{\mathcal{N}_{+}^2+\mathcal{N}_{-}^2}
\end{eqnarray}
whose elements fulfill the following properties:
\begin{eqnarray}
\begin{array}{c}
\mathcal{Z}_{+}\mathcal{Z}_{-}=0, \ \ \mathcal{Z}_{+}^{2n}=\mathcal{Z}%
_+^{2}, \ \ \mathcal{Z}_{+}^{2n+1}=\mathcal{Z}_{+}, \\
\\
\mathcal{Z}_{-}^{2n}=(-1)^{n-1}\mathcal{Z}_{-}^2, \ \ \mathcal{Z}%
_{-}^{2n+1}=(-1)^{n}\mathcal{Z}_{-}.%
\end{array}
\label{matrixproperty12}
\end{eqnarray}
With these details in mind,  a finite Lorentz transformation belonging to $%
SO_{A}(3,1)$ decomposes according to
\begin{eqnarray}
\Lambda=\mathcal{Z}_-\sin\varphi+\mathcal{Z}_+\sinh\zeta-\mathcal{Z}%
_-^{2}\cos\varphi+\mathcal{Z}_+^{2}\cosh\zeta,  \label{GLT}
\end{eqnarray}
where the the arguments of the trigonometric and hyperbolic functions are
\begin{eqnarray}
\zeta=\frac{\mathcal{N}_{+}\xi}{\mathcal{N}_{+}+\mathcal{N}_{-}}-\frac{%
\mathcal{N}_{-}\vartheta}{\mathcal{N}_{+}+\mathcal{N}_{-}}\  \\
\varphi=\frac{\mathcal{N}_{-}\xi}{\mathcal{N}_{+}+\mathcal{N}_{-}}+\frac{%
\mathcal{N}_{+}\vartheta}{\mathcal{N}_{+}+\mathcal{N}_{-}}.
\end{eqnarray}
In contrast to the cases  considered in Subsections A and B, the covariant
decomposition of the Lorentz transformation [Eq.~(\ref{GLT})]  deduced here
relates to the most general case of a constant and homogeneous external
field with both its invariants $\mathfrak{G},~\mathfrak{F}$ different from
zero. (Therefore it remains valid in the crossed field system $\mathfrak{G}=%
\mathfrak{F}=0$ and $\vert \pmb{E}\vert=\vert\pmb{B}\vert).$ For the
magnetized vacuum $(\mathfrak{G}=0,$ $\mathfrak{F}>0),$ the variables in Eq.~(\ref%
{GLT}) become $\zeta = \xi$ and $\varphi = \vartheta,$ whereas in an
electric background $(\mathfrak{G}=0$, $\mathfrak{F}<0)$ they turn out to be $%
\zeta =-\vartheta$ and $\varphi = \xi.$

Now, the explicit structure of $\Lambda$ in terms of the external electric $%
\pmb{E}$ and magnetic $\pmb{B}$ fields is rather complicated in a general
Lorentz frame. However, it becomes simpler if one considers the external
field configurations inherent to special frames, where the vectors $\pmb{E}$
and $\pmb{B}$ are parallel, and directed, say, along the $z$ axis. In these 
special frames, the transformations of  Eq.~(\ref{GLT}) have a simple sense of
rotation about the axis $z$ by the angle $\varphi,$ and the Lorentz boost
along this axis is parametrized by the angle $\zeta$.

Considering the general case $\mathfrak{G}\neq0,$ $\mathfrak{F}\neq0$ in
the special frame where $\pmb{B}=(0,0,B_z),$ $\pmb{E}=(0,0,E_z),$ we found
that
\begin{eqnarray}
\Lambda^{\mu}_{\ \ \nu}=\mathscr{R}^{\mu}_{\ \ \lambda}(\varphi)\mathscr{B}%
_{\ \ \nu}^{\lambda}(\zeta)=\mathds{D}^{SO(2)}\oplus \mathds{D}^{%
SO(1,1)}.  \label{gdecompo}
\end{eqnarray}
The expression above involves the matrices
\begin{eqnarray}
\mathscr{R}(\varphi)=\mathds{D}^{SO(2)}\oplus \mathds{1}\ \ \text{%
and} \ \ \mathscr{B}(\zeta)=\mathds{1} \oplus\mathds{D}^{SO(1,1)},
\end{eqnarray}
where the $\oplus$ denotes the usual direct sum of matrices, and $\mathds{1}$
denotes a $2\times2$ identity matrix. Here both $\mathds{D}^{SO(2)}$
and $\mathds{D}^{SO(1,1)}$ are two-dimensional representations of $%
SO(2)$ and $SO(1,1),$ respectively. Explicitly, they read
\begin{eqnarray}
\mathds{D}^{\mathrm{SO}(2)}=\left[%
\begin{array}{cc}
\cos\varphi & \sin\varphi \\
-\sin\varphi & \cos\varphi%
\end{array}%
\right],\ \ \mathds{D}^{\mathrm{SO}(1,1)}=\left[%
\begin{array}{cc}
\cosh\zeta & \sinh\zeta \\
\sinh\zeta & \cosh\zeta%
\end{array}%
\right]  \notag
\end{eqnarray}
Both matrices conform two independent Abelian-invariant subgroups of $%
SO_A(3,1),$ which emphasizes its nonsemisimple structure.
Moreover, the group parameter space is the product manifold $S^{(1)} \times
\Re^{(1)},$ where $S^{(1)}$ is the circle and $\Re^{(1)}$ the real line.
Therefore, the topology of $SO_{A}(3,1)$ is the surface of a
circular cylinder, a manifold which is neither compact nor simply connected. 
(Indeed,  it is infinitely connected.)


\section{Generators of space-time transformations \label{sect3}}


The constrained Hamiltonian formulation of the electromagnetic field in an
anisotropic material remains poorly developed in the literature; we are not
aware of previous Hamiltonian treatment for these optical media. In this
section we shall see how this method can be implemented in QED with an
external constant electromagnetic field. To pursue this analysis we first
note that the effective Lagrangian $\mathscr{L}$ in Eq.~(\ref%
{generaleffectivelagragianstrongfield}) is not a local function, since it
depends on momenta in a very cumbersome way. The canonical formalism in its
standard form is not applicable to nonlocal situations. For this reason we
will restrict ourselves to the infrared approximation, $k_\mu\to 0$, in
which case the effective Lagrangian $\mathscr{L}$ becomes a local function
on the photon field $a_\mu(x).$ This approximation corresponds to
anisotropic media with no spatial or frequency dispersion. In our case, it
becomes actual in the region of very strong external fields, where the
external field dominates over the photon momentum \cite{shabad4}.

Consider the static $(\omega=0)$ case. When $k_\perp=0,$ we have
from Eq.~(\ref{magind})
\begin{equation}
\varepsilon_\|(k_\parallel,0)=1-\left.\frac{\varkappa_2} {k_\|^2}%
\right|_{\omega,k_\perp=0}.  \notag
\end{equation}
When $\omega=0$, $k_\parallel=0,$  we have that
\begin{equation}  \label{wrongsign}
\mu_\|^{-1}(k_\perp,0) = 1-\frac{\varkappa_3}{k_\perp^2}.
\end{equation} Owing to the degeneracy property \cite{PRD2010}, $\varkappa_1=%
\varkappa_2 $ at $\omega^2 -k^2_\parallel=0,$ we also that
\begin{equation}
\varepsilon_\perp(k_\perp,0)=\mu_\perp^{-1}(k_\perp,0)=1-\left.\frac{%
\varkappa_1} {k_\perp^2}\right|_{\omega,k_\|=0}.  \notag
\end{equation}
So, in the infrared limit $k_\mu\rightarrow 0$, the quantities in Eq.~(\ref%
{magind}) coincide with the permittivities defined for that case in 
Ref.~\cite{PRD2011}--according to the correspondences: $\varepsilon_\perp%
\Leftrightarrow \varepsilon_{\mathrm{tr}},\quad\mu_\perp\Leftrightarrow\mu^{%
\mathrm{w}}_{\mathrm{tr}},\quad \varepsilon_\|\Leftrightarrow \varepsilon_{%
\mathrm{long}},\quad \mu_\|^{-1}\Leftrightarrow\mu_{\mathrm{tr}}^{\mathrm{pl}%
}$--and responsible for screening charges and stationary currents of
special configurations.

In the infrared limit the eigenvalues of the vacuum polarization tensor can
be expressed in terms of the first and second derivatives of an effective
Lagrangian $\mathfrak{L}$ [connected with the generating functional $\Gamma$
of irreducible many-photon vertices in an external field,  as pointed out below  
Eq.~(\ref{S})] over the constant field with respect to the
corresponding external field invariants $\mathfrak{F}$ and $\mathfrak{G}$
(see details in Ref.~\cite{PRD2011}):
\begin{equation}  \label{poleig1}
\begin{array}{c}
\displaystyle\varkappa _{1}=k^{2}\mathfrak{L}_{\mathfrak{F}}, \\
\displaystyle\varkappa _{2}=\varkappa _{1}-2\mathfrak{F}\mathfrak{L}_{%
\mathfrak{G}\mathfrak{G}}z_{1},\ \ \varkappa _{3}=\varkappa _{1}+2\mathfrak{F%
}\mathfrak{L}_{\mathfrak{F}\mathfrak{F}}z_{2},%
\end{array}%
\end{equation}%
where $\mathfrak{L}_{\mathfrak{F}}=\partial \mathfrak{L}/\partial \mathfrak{F%
},$ $\mathfrak{L}_{\mathfrak{G}\mathfrak{G}}=\partial ^{2}\mathfrak{L}%
/\partial \mathfrak{G}^{2}$ and $\mathfrak{L}_{\mathfrak{F}\mathfrak{F}%
}=\partial ^{2}\mathfrak{L}/\partial \mathfrak{F}^{2}$ with $\mathfrak{G}$
set equal to zero after differentiation. It follows from Eqs.~(\ref{poleig1}%
),  (\ref{anisotropicmaxwelllagragiasadsn}),  and  (\ref%
{anisotropicmaxwelllagragiasadsnelectric}) that
\begin{equation}  \label{37}
\begin{array}{cc}
\left.
\begin{array}{c}
\varepsilon _{ij}=\left( 1-\mathfrak{L}_{\mathfrak{F}}\right) \delta _{ij}+%
\mathfrak{L}_{\mathfrak{G}\mathfrak{G}}B_{i}B_{j} \\
\mu _{ij}^{-1}=\left( 1-\mathfrak{L}_{\mathfrak{F}}\right) \delta _{ij}-%
\mathfrak{L}_{\mathfrak{F}\mathfrak{F}}B_{i}B_{j}%
\end{array}%
\right\} & \mathfrak{F}>0 \\
&  \\
\left.
\begin{array}{c}
\varepsilon _{ij}=\left( 1-\mathfrak{L}_{\mathfrak{F}}\right) \delta _{ij}+%
\mathfrak{L}_{\mathfrak{F}\mathfrak{F}}E_{i}E_{j} \\
\mu _{ij}^{-1}=\left( 1-\mathfrak{L}_{\mathfrak{F}}\right) \delta _{ij}-%
\mathfrak{L}_{\mathfrak{G}\mathfrak{G}}E_{i}E_{j}%
\end{array}%
\right\} & \mathfrak{F}<0.%
\end{array}%
\end{equation}

For the magneticlike case $\mathfrak{F}>0$ this is equivalent to
\begin{equation}  \label{convexityproperties}
\begin{array}{c}
\varepsilon _{\perp }=\mu _{\perp }^{-1}=1-\mathfrak{L}_{\mathfrak{F}} \\
\varepsilon _{\parallel }=1-\mathfrak{L}_{\mathfrak{F}}+2\mathfrak{F}%
\mathfrak{L}_{\mathfrak{G}\mathfrak{G}}, \\
\mu _{\parallel }^{-1}=1-\mathfrak{L}_{\mathfrak{F}}-2\mathfrak{F}\mathfrak{L%
}_{\mathfrak{F}\mathfrak{F}},%
\end{array}%
\end{equation}%
The following relations were established in Ref.~\cite{PRD2011} on the basis of
causality and unitarity principles, valid  for $\mathfrak{F}\lessgtr0$%
\begin{equation}  \label{convexityproperties2}
\begin{array}{c}
1-\mathfrak{L}_{\mathfrak{F}}\geq 0 \\
1-\mathfrak{L}_{\mathfrak{F}}+2\mathfrak{F}\mathfrak{L}_{\mathfrak{G}%
\mathfrak{G}}\geq 0,\ \ \mathfrak{L}_{\mathfrak{G}\mathfrak{G}}\geq 0 \\
1-\mathfrak{L}_{\mathfrak{F}}-2\mathfrak{F}\mathfrak{L}_{\mathfrak{F}%
\mathfrak{F}}\geq 0,\ \ \mathfrak{L}_{\mathfrak{F}\mathfrak{F}}\geq 0%
\end{array}%
\end{equation}%
which guarantees the consistency of the theory.


\subsection{The energy-momentum tensor \label{contructionenmotensor}}


The symmetry reduction by the external magnetic field does not alter the
translational group embeded in  $ISO(3,1)$.  Therefore, for a photon, the
spacetime configuration with an external classical field is translation
invariant. To find in the local approximation the associated Noether
current of the electromagnetic radiation, let us first insert Eq.~(\ref%
{poleig1}) into Eq.~(\ref{nuevotensorinductions}) to find
\begin{eqnarray}
\mathcal{O}^{\mu\nu}&=&\left(1-\mathfrak{L}_\mathfrak{F}\right)f^{\mu\nu}-%
\frac{1}{2}\mathfrak{L}_{\mathfrak{G} \mathfrak{G}}\left(f^{\rho\lambda }%
\tilde{\mathscr{F}}_{\rho\lambda}\right)\tilde{\mathscr{F}}^{\mu\nu}  \notag
\\
&&-\frac{1}{2}\mathfrak{L}_{\mathfrak{F}\mathfrak{F}}\left(f^{\rho\lambda}%
\mathscr{F}_{\rho\lambda}\right) \mathscr{F}^{\mu\nu}.
\label{varianteinductiontensor}
\end{eqnarray}
The substitution of this tensor into $\mathscr{L}$ [Eq.~(\ref%
{generaleffectivelagragianstrongfield})]  defines the Lagrangian \cite%
{dipiazza,PRD2011} of the small-amplitude, low-frequency, long-wave
electromagnetic field $a_\mu(x)$:
\begin{eqnarray}  \label{explaeff}
&&\mathscr{L}=-\frac{1}{4}f^{\mu\nu}f_{\mu\nu}+\mathfrak{L}  \notag \\
&&\quad=-\frac{1}{4}(1-\mathfrak{L}_\mathfrak{F})f^{\mu\nu}f_{\mu\nu}+\frac{1%
}{8}\mathfrak{L}_{\mathfrak{G}\mathfrak{G}}\left(f^{\mu\nu}\tilde{\mathscr{F}%
}_{\mu\nu}\right)^2  \notag \\
&&\qquad\qquad+\frac{1}{8}\mathfrak{L}_{\mathfrak{F}\mathfrak{F}%
}\left(f^{\mu\nu}\mathscr{F}_{\mu\nu}\right)^2
\end{eqnarray}%
The corresponding conserved stress-energy tensor is obtained from it by 
following the Noether theorem in the standard way:
\begin{eqnarray}
T^{\mu\nu}=\eta^{\mu\nu}\mathscr{L}-\frac{\partial \mathscr{L}}{%
\partial\left(\partial_\mu a_\lambda\right)}\partial^{\nu}a_{\lambda}.
\label{canonicalenergytensor}
\end{eqnarray}
Here
\begin{equation}  \label{O}
\frac{\partial\mathscr{L}}{\partial\left(\partial_\mu a_\nu\right)}=-\frac{%
\partial\mathscr{L}}{\partial\left(\partial_\nu a_\mu\right)}=-\mathcal{O}%
^{\mu\nu}
\end{equation}
in accordance with Eqs.~(\ref{generaleffectivelagragianstrongfield})-(\ref%
{vectorindcutiondefinition}). The antisymmetricity of this tensor is owing
to the gauge invariance manifesting in that the Lagrangian contains only the
field tensor.

Substituting Eq.~(\ref{explaeff}) into Eq.~(\ref{canonicalenergytensor}) we
obtain
\begin{eqnarray}  \label{Temunu}
T^{\mu\nu}&=&\left(1-\mathfrak{L}_\mathfrak{F}\right)f^{\mu\lambda}%
\partial^{\nu}a_{\lambda}- \frac{1}{2}\mathfrak{L}_{\mathfrak{G}\mathfrak{G}%
}\left(f^{\varrho\sigma} \tilde{\mathscr{F}}_{\varrho\sigma}\right)\tilde{%
\mathscr{F}}^{\mu\lambda}\partial^{\nu}a_{\lambda}  \notag \\
&-&\frac{1}{2}\mathfrak{L}_{\mathfrak{F}\mathfrak{F}}\left(f^{\varrho\sigma}%
\mathscr{F}_{\varrho\sigma}\right) \mathscr{F}^{\mu\lambda}\partial^{\nu}a_{%
\lambda}+\eta^{\mu\nu} \{-\frac{1}{4}\left(1-\mathfrak{L}_ \mathfrak{F}%
\right)  \notag \\
\displaystyle &\times&f_{\varrho\sigma}f^{\varrho\sigma}+\frac{1}{8}%
\mathfrak{L}_ {\mathfrak{G}\mathfrak{G}}\left(f^{\sigma\varrho}\tilde{%
\mathscr{F}}_{\sigma\varrho}\right)^2+ \frac{1}{8}\mathfrak{L}_{\mathfrak{F}%
\mathfrak{F}}\left(f^{\sigma\varrho}\mathscr{F}_{\sigma\varrho} \right)^2\}.
\notag \\
\end{eqnarray}
Let us define the tensor%
\begin{eqnarray}
\Theta^{\mu\nu}=\mathscr{L}\eta^{\mu\nu}+\mathcal{O}^{\mu\lambda}f^\nu_{\ \
\lambda}-j^\mu a^{\nu}.  \label{IReffelagrangianmass}
\end{eqnarray}
related to Eq.~(\ref{canonicalenergytensor}) as%
\begin{equation}  \label{connection}
T^{\mu\nu}=\Theta^{\mu\nu} -\partial_\lambda K^{\lambda\mu\nu}
\end{equation}%
where $K^{\lambda\mu\nu}= -\mathcal{O}^{\mu\lambda}a^{\nu} $ is
antisymmetric in its first two indices, while the electric current
\begin{equation}  \label{current}
j^\nu=\partial_\mu\frac{\partial \mathscr{L}}{\partial(\partial_\mu a_\nu)}%
=-\partial_\mu\mathcal{O}^{\mu\nu}
\end{equation}
disappears on equations of motion $j_\mu\cong0$. It conserves, $%
\partial^\mu j_\mu=0$ due to antisymmetricity of the tensor $\mathcal{O}%
^{\mu\lambda}$ [Eq.~(\ref{O})]. When taken on the equations of motion, the tensors
$T^{\mu\nu}$ and $\Theta^{\mu\nu}$ coincide up to a full derivative. However, 
the latter depends only on the field strengths of the small electromagnetic
field $f_{\alpha\beta},$ and not on its four-vector potential $a_\mu$, as  
was the case with the tensor of Eqs.~(\ref{canonicalenergytensor}) and  (\ref{Temunu}).
The achievement of this gauge-invariance property of the tensor $%
\Theta^{\mu\nu}$ [Eq.~(\ref{IReffelagrangianmass})] was  the motivation \cite%
{PRD2011} for taking it as the stress-energy tensor (on equations of
motion). It remains, however, not symmetric. Substituting Eq.~(\ref%
{varianteinductiontensor}) into Eq.~(\ref{IReffelagrangianmass}) results in
\begin{eqnarray}  \label{gaugesdlsdiif121}
\Theta^{\mu\nu}&=&\left(1-\mathfrak{L}_\mathfrak{F}\right)f^{\mu\lambda}f_{\
\ \lambda}^{ \nu}-\frac{1}{2}\mathfrak{L}_{\mathfrak{G}\mathfrak{G}%
}\left(f^{\varrho\sigma} \tilde{\mathscr{F}}_{\varrho\sigma}\right)\tilde{%
\mathscr{F}}^{\mu\lambda}f_{\ \ \lambda}^{ \nu}  \notag \\
&-&\frac{1}{2}\mathfrak{L}_{\mathfrak{F}\mathfrak{F}}\left(f^{\varrho\sigma}%
\mathscr{F}_{\varrho\sigma}\right)\mathscr{F}^{\mu\lambda}f_{\ \ \lambda}^{
\nu}+\eta^{\mu\nu}\left\{-\frac{1}{4}\left(1-\mathfrak{L}_\mathfrak{F}%
\right)\right.  \notag \\
&\times&\left. f_{\varrho\sigma}f^{\varrho\sigma}+\frac{1}{8} \mathfrak{L}_{%
\mathfrak{G}\mathfrak{G}}\left(f^{\sigma\varrho}\tilde{\mathscr{F}}%
_{\sigma\varrho}\right)^2+\frac{1}{8} \mathfrak{L}_{\mathfrak{F}\mathfrak{F}%
}\left(f^{\sigma\varrho}\mathscr{F}_{\sigma\varrho}\right)^2\right\}  \notag
\\
&-&j^\mu a^\nu
\end{eqnarray}

In an empty Minkowski space ($\mathscr{F}=0$), the possibility  of
symmetrization of the canonical stress-energy tensor by adding a full
derivative is provided by the conservation of $SO(3,1)$ generators. 
(For details,  we refer the reader to Ref.~\cite{Weinberg:1995mt} and references
therein.)  In our case,  there is a lack of isotropy due to the external field
which implies that only a subset of the Lorentz generators are conserved
(see the Secs. \ref{spingenerallrofjdffk} and \ref{consequenceLSB}). This
fact, therefore, prevents us from obtaining an equivalently symmetrized version 
of Eq.~(\ref{canonicalenergytensor}), and some dramatic differences arise in
comparison with the case of an empty vacuum. The relation obeyed by Eq.~(\ref%
{gaugesdlsdiif121}) to substitute for the symmetricity property is $\check{f}%
\check{\Theta}=\check{\Theta}^T\check{f}$, where $\check{\Theta}$ and $%
\check{f}$ are the matrices $\Theta_{\mu\nu}$ and $f_{\mu\nu}$, while $%
\check{\Theta}^T$ is the transposed matrix. When there is no Lorentz
breaking, the matrix $\check{\Theta}$ commutes with the matrix $\check{f}$,
because in this case the stress-energy tensor is built only of the field
tensor and  the unit metric tensor, its explicit dependence on the
coordinate vector not being  admitted. In this case,  the symmetricity of $%
\Theta$ is in agreement with the above relation.

The stress-energy tensor $\Theta^{\mu\nu}$, as well as $T^{\mu\nu},$
satisfies the continuity equation with respect to the first index on
equations of motion,
\begin{eqnarray}
\partial_\mu\Theta^{\mu\nu}=0,  \label{conservationequataion}
\end{eqnarray}
with their difference satisfying the same property $\partial_\mu\partial_\lambda
K^{\lambda\mu\nu}=0.$ Its components, explicitly, are
\begin{equation}  \label{explicit1}
\Theta^{00}=\frac{1}{2}\pmb{d}\cdot\pmb{e}+\frac{1}{2}\pmb{h}\cdot\pmb{b}+a_0%
\pmb{\nabla}\cdot\pmb{d},
\end{equation}
\begin{equation}  \label{explicit2}
\Theta^{0i} = [\pmb{d}\times\pmb{b}]^i-a^i\pmb{\nabla}\cdot\pmb{d},
\end{equation}
\begin{equation}  \label{poytinvectordensisty}
\Theta^{i0} = [\pmb{e}\times \pmb{h}]^i-a_0\left[\partial_0 d^i-\left(%
\pmb{\nabla}\times\pmb{h}\right)^i\right],
\end{equation}
\begin{eqnarray}  \label{explicit4}
\Theta^{ij} = &-&d^ie^j-b^ih^j+\frac{1}{2}\delta^{ij}(\pmb{e}\cdot\pmb{d}+%
\pmb{b}\cdot\pmb{h})  \notag \\
&+&a^j\left[\partial_0 d^i-\left(\pmb{\nabla}\times\pmb{h}\right)^i\right],
\end{eqnarray}
where the electric and magnetic induction vectors are defined in Eq.~(\ref%
{vectorindcutiondefinition}). Explicitly,
\begin{eqnarray}  \label{momentamagnetized}
\begin{array}{cc}
\pmb{d}=\left(1-\mathfrak{L}_\mathfrak{F}\right)\pmb{e}+\mathfrak{L}_{%
\mathfrak{G}\mathfrak{G}}\left(\pmb{e}\cdot\pmb{B}\right) \pmb{B} & \mathrm{%
for}\ \ \mathfrak{F}>0, \\
&  \\
\pmb{d}=\left(1-\mathfrak{L}_\mathfrak{F}\right)\pmb{e}+\mathfrak{L}_{%
\mathfrak{F}\mathfrak{F}}\left(\pmb{e}\cdot\pmb{E}\right) \pmb{E} & \mathrm{%
for}\ \ \mathfrak{F}<0,%
\end{array}%
\end{eqnarray}
and 
\begin{eqnarray}  \label{magneinduc}
\begin{array}{cc}
\pmb{h}=\left(1-\mathfrak{L}_\mathfrak{F}\right)\pmb{b}-\mathfrak{L}_{%
\mathfrak{F}\mathfrak{F}}\left(\pmb{b}\cdot\pmb{B}\right) \pmb{B} & \mathrm{%
for}\ \ \mathfrak{F}>0, \\
&  \\
\pmb{h}=\left(1-\mathfrak{L}_\mathfrak{F}\right)\pmb{b}-\mathfrak{L}_{%
\mathfrak{G}\mathfrak{G}}\left(\pmb{b}\cdot\pmb{E}\right) \pmb{E} & \mathrm{%
for}\ \ \mathfrak{F}<0.%
\end{array}%
\end{eqnarray}

Let us consider the case in which the equations of motion  are fulfilled, i.e.,
when the current $j^\mu=0$ vanishes identically. By integrating the
continuity equation [Eq.~(\ref{conservationequataion})] with $\nu=0$ over a final
spatial volume $\mathrm{V}$ and defining the energy in this volume as
\begin{eqnarray}
\begin{array}{c}
\displaystyle \mathscr{P}_{\mathrm{V}}^0=\int_{\mathrm{V}} d^3x
\Theta^{00}(x)=\int_{\mathrm{V}} d^3x\left(\frac{1}{2}\pmb{d}\cdot\pmb{e}+%
\frac{1}{2}\pmb{h}\cdot\pmb{b}\right),%
\end{array}
\label{transalationalchargesiniciales}
\end{eqnarray}
we get
\begin{equation}
\frac{\partial\mathscr{P}_{\mathrm{V}}^0}{\partial x^0}=\oint_{\mathrm{S}%
}\Theta_{i0}d\sigma_i=\oint_{\mathrm{S}}(\pmb{e}\times \pmb{h})_id\sigma_i,
\end{equation}
where the integral in the right-hand side is run over the surface $\mathrm{S}
$ surrounding the volume $\mathrm{V}$. Therefore, the Poynting vector $%
\Theta^{i0}=(\pmb{e}\times\pmb{h})^i$ accounts for the energy per unit of time,
per unit area, transported by the small electromagnetic waves. It is
parallel to the group velocity of eigenmodes and to their center-of-mass
velocity, as described in Appendix  \ref{pvgvsew}. In the infinite-volume
limit, V$=\infty$,  and under the assumption that the fields $\pmb{e}, \pmb{h}$
fall off at the spatial infinity, we find that the energy inside the infinite
volume $\mathscr{P}^0=\mathscr{P}_\infty^0$ does not depend on time:
\begin{equation}
\frac{\partial\mathscr{P}^0}{\partial x^0}=0
\end{equation}%
By integrating the continuity equation [Eq.~(\ref{conservationequataion})] with $%
\nu=j$ over a final spatial volume $\mathrm{V}$,  we find that the momentum in
this volume,
\begin{equation}  \label{transalationalchargesiniciales2}
\mathscr{P}_{\mathrm{V}}^j=\int_{\mathrm{V}} d^3x\Theta^{0j}(x)=\int_{%
\mathrm{V}} d^3x\left(\pmb{d}\times\pmb{b}\right)^j,
\end{equation}%
satisfies the equation
\begin{equation}
\frac{\partial\mathscr{P}_{\mathrm{V}}^j}{\partial x_0}=\oint_{\mathrm{S}%
}\Theta^{ij}d\sigma_i,
\end{equation}
which  indicates that the total momentum contained inside the infinite volume $%
\pmb{\mathscr{P}}=\pmb{\mathscr{P}}_{\infty}$ conserves,
\begin{equation}
\frac{\partial\pmb{\mathscr{P}}}{\partial x^0}=0,
\end{equation}
under the same assumption that the fields decrease at spatial infinity.
Note that the momentum density $\sim\pmb{d}\times\pmb{b}$ and the
Poynting vector $\sim \pmb{e}\times\pmb{h}$ describe different quantities,
which  does not take place in an empty space-time. Observe that, on the
equation of motion, the structure of Eqs.~(\ref{explicit1})-(\ref{explicit4})
does not differ from the case of light propagation in an anisotropic
material \cite{iwo}. \footnote{%
Remember  that we restrict ourselves to the low-frequency, low-momentum limit
in the present subsection and in the rest of the article.}

In the general case in which the current $j^\mu$ does not vanish, the
spatial integral of Eqs.~(\ref{explicit1})-(\ref{explicit2}) defines the
translation generators. For further convenience,  we express the latter in
terms of the momentum $\pi^{i}=\partial\mathscr{L}/\partial\left(\partial_0
a_i\right)$, canonically conjugated to the field $a_i$ taken as a canonical
coordinate. It coincides with the electric induction, defined in Eq.~(\ref%
{vectorindcutiondefinition}), $\pmb{d}=-\boldsymbol{\pi}.$ We invert Eq.~(%
\ref{eq15}) so that the electric field of the wave can be expressed as $%
e_i=\varepsilon_{ij}^{-1}d_j$,  where $\varepsilon_{ij}^{-1}$ must be
understood as the inverse of the tensors given in Eq.~(\ref{37})
\begin{equation}
\begin{array}{cc}
\displaystyle \varepsilon_{ij}^{-1}=\frac{1}{1-\mathfrak{L}_\mathfrak{F}}%
\left[\delta_{ij}-\frac{\mathfrak{L}_{\mathfrak{G}\mathfrak{G}}}{1-\mathfrak{%
L}_\mathfrak{F}+2\mathfrak{F}\mathfrak{L}_{\mathfrak{G}\mathfrak{G}}}B_iB_j%
\right] & \mathfrak{F}>0, \\
&  \\
\displaystyle \varepsilon_{ij}^{-1}=\frac{1}{1-\mathfrak{L}_\mathfrak{F}}%
\left[\delta_{ij}-\frac{\mathfrak{L}_{\mathfrak{F}\mathfrak{F}}}{1-\mathfrak{%
L}_\mathfrak{F}-2\mathfrak{F}\mathfrak{L}_{\mathfrak{F}\mathfrak{F}}}E_iE_j%
\right] & \mathfrak{F}<0.%
\end{array}
\label{invertdieleccosn}
\end{equation}
With these details in mind,  the translation generators turn out to be
\begin{eqnarray}
\begin{array}{c}
\displaystyle \pmb{\mathscr{P}}=-\int d^3x\left\{\pmb{\pi}\times\pmb{b}-%
\pmb{a}\left(\pmb{\nabla}\cdot\pmb{\pi}\right)\right\}, \\
\\
\displaystyle \mathscr{P}^0=\int d^3x\left\{\frac{1}{2}\pi_i%
\varepsilon_{ij}^{-1}\pi_j+\frac{1}{2}b_i\mu_{ij}^{-1}b_j- a_0 \left(%
\pmb{\nabla}\cdot\pmb{\pi}\right)\right\}.%
\end{array}
\label{transalationalcharges}
\end{eqnarray}%
Some comments are in order. First of all, we point out that these generators
and their  respective translational chargess  Eqs.~(\ref%
{transalationalchargesiniciales2}) and (\ref{transalationalchargesiniciales}),  
differ from each other in the terms  which are proportional to  Gauss's 
law,  $\pmb{\nabla}\cdot\pmb{\pi}=0.$ Such terms are intrinsically associated
with the constrained Hamiltonian formalism \cite{Dirac:1958sq} (see the next
subsection). Correspondingly, we shall show that the expression involved in
Eq.~(\ref{transalationalcharges}) canonically realizes the space-time
translations, at least when acting on a phase space defined by constraints
associated with the gauge symmetry.


\subsection{Gauge fixing and Dirac brackets \label{sect3B}}


The local approximation [Eq.~(\ref{explaeff})] of our effective Lagrangian $%
\mathscr{L}$ does not depend on the velocity $\partial a_{0}/\partial x^0,$
so that the related momentum vanishes identically. Obviously, this leads us 
to introduce
\begin{equation}
\varphi_1\equiv\pi^0\approx0
\end{equation}
as a ``primary constraint.''  Note that the symbol $\approx$ must be
understood as ``weak equality'';  i.e.,  the constraints cannot be assumed to equal 
zero until the Poisson bracket between two arbitrary functionals $%
\mathscr{Q}$ and $\mathcal{Q}$ of the field variables $(a_\mu,\pi_\mu)$ is
calculated: 
\begin{eqnarray}
\left\{\mathscr{Q},\mathcal{Q}\right\}=\int d^3x\left[\frac{\delta\mathscr{Q}%
}{\delta a_\mu(\pmb{x})}\frac{\delta\mathcal{Q}}{\delta \pi^\mu(\pmb{x})}-%
\frac{\delta\mathscr{Q}}{\delta \pi_\mu(\pmb{x})}\frac{\delta\mathcal{Q}}{%
\delta a^\mu(\pmb{x})}\right].  \label{generalizedpoissonbracket}
\end{eqnarray}

According to the Dirac algorithm, $\varphi_1$ must be implemented within the
canonical Hamiltonian
\begin{eqnarray}
\mathscr{H}_C\equiv\mathscr{P}^0=\int d^3x\left(\boldsymbol{\pi}\frac{%
\partial\pmb{a}}{\partial x^0}-\mathscr{L}\right),
\label{generalhamitlonianform}
\end{eqnarray}%
by means of a Lagrangian multiplier $\mathscr{C}$,  so  that the ``total''
Hamiltonian turns out to be
\begin{eqnarray}
\mathscr{H}=\mathscr{P}^0+\int d^3x\ \mathscr{C}(x)\pi^0(x).
\label{juegodefields}
\end{eqnarray}
In this context, the equation  of motion for $\mathscr{Q}$ reads
\begin{eqnarray}
\frac{d\mathscr{Q}}{dx^0}=\frac{\partial \mathscr{Q}}{\partial x^0}+\left\{%
\mathscr{Q},\mathscr{H}\right\}  \label{hijodeputa}
\end{eqnarray}
and, in particular, for $\mathscr{Q}=\pi_0$,  one has
\begin{eqnarray}
\frac{d{\pi}^0}{dx^0}=\left\{\pi^0,\mathscr{H}\right\}=\pmb{\nabla}\cdot%
\pmb{\pi}.
\end{eqnarray}
The constraint $\varphi_1$ should hold at all time. In consequence, Gauss's law
\begin{eqnarray}
\varphi_2\equiv\pmb{\nabla}\cdot\pmb{\pi}\approx0,  \label{gausslawdiel}
\end{eqnarray}%
which is one of the field equations,  $j^0=\nabla\cdot\boldsymbol{d}=0$,
arises as a ``secondary constraint.''  However, the latter is already
present in $\mathscr{H}_C=\mathscr{P}^0$ in the form $a_0\pmb{\nabla}\cdot%
\pmb{\pi}$ [see Eq.~(\ref{transalationalcharges})]. Therefore, $a_0$ can be
considered as a Lagrange multiplier and thus an arbitrary function of $x.$
Its equation of motion implies that
\begin{eqnarray}
\frac{da_0}{dx^0}=\left\{a_0,\mathscr{H}\right\}=\mathscr{C}(x).
\label{caorelation}
\end{eqnarray}

The string of constraints stops here because  Gauss's law commutes with
the Hamiltonian. Moreover,  our primary and secondary constraint are ``
first class''  with a  vanishing Poisson bracket:
\begin{eqnarray}
\left\{\pi_0(x),\pmb{\nabla}\cdot\pmb{\pi}(x^\prime)\right\}=0.
\end{eqnarray}
The remaining algebra of the constraints and the Hamiltonian is given by
\begin{eqnarray}
\begin{array}{c}
\left\{\pi_0(x),\pi_0(x^\prime)\right\}=0, \ \ \left\{\pmb{\nabla}\cdot%
\pmb{\pi}(x),\pmb{\nabla}\cdot\pmb{\pi}(x^\prime)\right\}=0, \\
\\
\left\{\pi_0, \mathscr{H}_{C}\right\}=-\pmb{\nabla}\cdot\pmb{\pi},\ \ \left\{%
\pmb{\nabla}\cdot\pmb{\pi}, \mathscr{H}_{C}\right\}=0.%
\end{array}%
\end{eqnarray}
Observe that the Lagrangian multipliers $\mathscr{C}$ and $a_0$ transfer an
arbitrariness to the Hamiltonian [Eq.~(\ref{juegodefields})]. As a consequence,  
we are forced to deal with a phase space plagued by  nonphysical degrees of
freedom. This problem is closely associated with the gauge invariance
property and is formally removed by imposing two gauge-fixing conditions.
Because of this fact,  the two existing multipliers are rendered to precise
dependences on the physical fields,  and can  eventually   be removed from the
theory.

A suitable set of gauge conditions can be found by solving  Gauss's  law
[Eq.~(\ref{gausslawdiel})]  with respect to  $a_0$:
\begin{equation}  \label{gausssolution}
a_0=\frac{1}{\nabla\varepsilon\nabla} \partial_0\left(\nabla_i%
\varepsilon_{ij}a_{j}\right)
\end{equation}%
where $\nabla\varepsilon\nabla=\nabla_i\varepsilon_{ij}\nabla_j.$ Guided by
this result,  we are led to choose a generalized version of the Coulomb gauge
as the third constraint of the theory
\begin{equation}  \label{gaugechoose}
\varphi_3\equiv \nabla_i\varepsilon_{ij}a_{j}\approx0.
\end{equation}
The consistency consequence of this gauge condition can be found by
Poisson-commuting the Hamiltonian [Eq.~(\ref{juegodefields})] with $\varphi_3$.
However,  it can be read off directly from Eq.~(\ref{gausssolution}) and
promotes the last constraint, 
\begin{eqnarray}  \label{phi4}
\varphi_4\equiv a_0\approx0.\quad
\end{eqnarray}
Since this is found to be stationary as well, Eq.~(\ref{caorelation})
provides a vanishing value of the Lagrangian multiplier $\mathscr{C}$,  and
there are no  further constraints.

The accessibility of these gauge conditions can be checked by using the gauge
function
\begin{equation}
\varLambda=-\frac{1}{\nabla\varepsilon\nabla} \nabla_i\varepsilon_{ij}a_{j}.
\end{equation}
In fact, for any value of $\pmb{a},\ a^0$,  the gauge-transformed fields
\begin{eqnarray}  \label{gaugetramnsformation1}
&&a_i^\prime(x)=a_i(x)+\nabla_i\varLambda,\ \ a_0^\prime(x)=a_0(x)+\partial_0%
\varLambda 
\end{eqnarray}
obey the gauge conditions of Eqs.~(\ref{gaugechoose}) and  (\ref{phi4}). Explicitly,%
\footnote{%
It is worth observing at this point that the electric field associated with
the small-amplitude waves $\pmb{e}=\pmb{\nabla}a_0-\partial_0\pmb{a}$ is a
gauge-invariant quantity. As a consequence,  the canonical momenta $%
\pi_i=-\varepsilon_{ij}e_j$ are invariant as well. So, under the gauge
transformation of Eq.~(\ref{gaugetramnsformation1}), $\pmb{\pi}^\prime=\pmb{\pi}.$}
\begin{eqnarray}
\begin{array}{c}
\displaystyle \nabla_i\varepsilon_{ij} a_j^\prime=\nabla_i\varepsilon_{ij}
a_j+\nabla_i\varepsilon_{ij}\nabla_j\varLambda=0, \\
\\
\displaystyle \pmb{\nabla}\cdot\pmb{\pi}^\prime=-\nabla_i\varepsilon_{ij}%
\nabla_j a_0^\prime=\nabla_i\varepsilon_{ij}\nabla_j
a_0+\nabla_i\varepsilon_{ij}\nabla_j\partial_0\varLambda=0.%
\end{array}
\label{80}
\end{eqnarray}
Note, in addition, that the constraints $\{\varphi_i\}$ defined above
restrict the original phase space of the theory to a four-dimensional
hypersurface, 
\begin{eqnarray}
\pmb{\varOmega}\equiv\left\{\left(a_\mu,\pi_\mu\right)\vert \ \
\varphi_i\approx0, \ \ i=1\ldots4\right\}  \label{hyphyssurf}
\end{eqnarray}
in which the time evolution of two physical degrees of freedom takes place.

Certainly the set $\{\varphi_i\}$ is second class with a characteristic
matrix $C_{\alpha\beta}(\pmb{x},\pmb{x}^\prime)\equiv\left\{\varphi_\alpha(%
\pmb{x}),\varphi_\beta(\pmb{x}^\prime)\right\}$ given by
\begin{eqnarray}
C_{\alpha\beta}=\left(
\begin{array}{cccc}
0 & 0 & 0 & 1 \\
0 & 0 & \nabla^{\pmb{x}}\varepsilon\nabla^{\pmb{x}} & 0 \\
0 & -\nabla^{\pmb{x}}\varepsilon\nabla^{\pmb{x}} & 0 & 0 \\
-1 & 0 & 0 & 0%
\end{array}
\right)\delta^{(3)}(\pmb{x}-\pmb{x}^\prime).
\end{eqnarray}
Since $C_{\alpha\beta}(\pmb{x},\pmb{x}^\prime)$ is regular by construction,
we can also write down its inverse
\begin{eqnarray}  \label{inversematricsdd}
C_{\alpha\beta}^{-1}=\left(
\begin{array}{cccc}
0 & 0 & 0 & -1 \\
0 & 0 & -\frac{1}{\nabla^{\pmb{x}}\varepsilon\nabla^{\pmb{x}}} & 0 \\
0 & \frac{1}{\nabla^{\pmb{x}}\varepsilon \nabla^{\pmb{x}}} & 0 & 0 \\
1 & 0 & 0 & 0%
\end{array}
\right)\delta^{(3)}(\pmb{x}-\pmb{x}^\prime).
\end{eqnarray}
With the help of the latter we introduce the Dirac bracket:
\begin{eqnarray}  \label{diracbracket}
&&\left\{\mathscr{Q}(\pmb{x}),\mathcal{Q}(\pmb{x}^\prime)\right\}_*= \left\{%
\mathscr{Q}(\pmb{x}),\mathcal{Q}(\pmb{x}^\prime)\right\}-\int d^3y\left\{%
\mathscr{Q}(\pmb{x}),\varphi_\alpha(\pmb{y})\right\}  \notag \\
&&\qquad \int d^3zC_{\alpha\beta}^{-1}(\pmb{y},\pmb{z})\left\{\varphi_\beta(%
\pmb{z}),\mathcal{Q}(\pmb{x}^\prime)\right\}
\end{eqnarray}
In this context, the fundamental bracket of the theory can be calculated
straightforwardly and reads
\begin{equation}
\left\{a_i(\pmb{x}),\pi_j(\pmb{x}^\prime)\right\}_*=\pmb{\mathpzc{t}}_{ij}(%
\pmb{x},\pmb{x}^\prime)  \label{fundDirabrac}
\end{equation}
where
\begin{equation}  \label{projtranscv}
\pmb{\mathpzc{t}}_{ij}(\pmb{x},\pmb{x}^\prime)\equiv\left[%
\delta_{ij}-\nabla_i^{\pmb{x}}\frac{1}{\nabla\varepsilon\nabla}%
\varepsilon_{jk}\nabla_k^{\pmb{x}}\right]\delta^{(3)}(\pmb{x}-\pmb{x}^\prime)
\end{equation}
is a projector-valued distribution which fulfills the relation
\begin{equation}  \label{tpro}
\int d^3 y \pmb{\mathpzc{t}}_{il}(\pmb{x},\pmb{y})\pmb{\mathpzc{t}}_{lj}(%
\pmb{y},\pmb{x}^\prime)=\pmb{\mathpzc{t}}_{ij}(\pmb{x},\pmb{x}^\prime).
\end{equation}
This, however, is not symmetric $\pmb{\mathpzc{t}}_{ij}(\pmb{x},\pmb{x}%
^\prime)\neq\pmb{\mathpzc{t}}_{ji}(\pmb{x},\pmb{x}^\prime)$\footnote{%
For a vanishing external field, this reduces to the symmetric transversal
projector associated with the Coulomb gauge $\pmb{\mathpzc{t}}_{ij}^{\mathrm{%
Coul}}(\pmb{x},\pmb{x}^\prime)=\left[\delta_{ij}-\nabla_i\nabla_j/%
\pmb{\nabla}^2\right]\delta^{(3)}(\pmb{x}-\pmb{x}^\prime).$} and, depending
on the field to be projected, one must contract one index or  the  other. For
instance, let us decompose the canonical field into two mutually orthogonal
pieces: $\pmb{a}(\pmb{x},t)=\pmb{a}^{\tau}(%
\pmb{x},t)+\pmb{a}^\ell(\pmb{x},t).$ Here the transversal component is
obtained by contracting the field $\pmb{a}(\pmb{x,t})$ with the index of $%
\pmb{\mathpzc{t}}_{ij}(\pmb{x},\pmb{x}^\prime)$,  which is provided by the
optical tensor $\varepsilon_{lj}.$ Thus,
\begin{eqnarray}
a^\tau_i(\pmb{x},t)&&=\int d^3x^\prime \pmb{\mathpzc{t}}_{ij}(\pmb{x},\pmb{x}%
^\prime)a_j(\pmb{x}^\prime,t)  \notag \\
&=& \left[\delta_{ij}-\nabla_i\frac{1}{\nabla\varepsilon\nabla}%
\nabla_k\varepsilon_{kj}\right]a_j(\pmb{x},t)
\end{eqnarray}
Accordingly,  the longitudinal components of $\pmb{a}$ turn out to be $%
a_i^\ell(\pmb{x},t)=\int d^3x^\prime\mathpzc{l}_{ij}(\pmb{x},\pmb{x}%
^\prime)a_j(\pmb{x}^\prime,t)$, where
\begin{equation}
\mathpzc{l}_{ij}(\pmb{x},\pmb{x}^\prime)\equiv \nabla_i^{\pmb{x}}\frac{1}{%
\nabla\varepsilon\nabla}\nabla_k^{\pmb{x}}\varepsilon_{kj}\delta^{(3)}(%
\pmb{x}-\pmb{x}^\prime)
\end{equation}
is the nonsymmetric longitudinal projector. Likewise,  we decompose $\pmb{\pi}=%
\pmb{\pi}^\tau+\pmb{\pi}^\ell.$ However, to extract its transversal and
longitudinal elements, the canonical momentum must be  contracted with the
index provided by the gradient operator:
\begin{eqnarray}
\pi^\tau_j(\pmb{x},t)&=&\int d^3 x^\prime\pmb{\mathpzc{t}}_{ij}(\pmb{x},%
\pmb{x}^\prime)\pi_i(\pmb{x}^\prime,t)  \notag \\
&=& \left[\delta_{ij}-\nabla_i^{\pmb{x}}\frac{1}{\nabla\varepsilon\nabla}%
\nabla_k^{\pmb{x}}\varepsilon_{kj}\right]\pi_i(\pmb{x},t).
\label{pipotrasversal}
\end{eqnarray}
Due to the gauge-fixing condition [Eq.~(\ref{gaugechoose})] and  Gauss's law [Eq.~(%
\ref{gausslawdiel})],   the longitudinal components ${a}^\ell_i=\int d^3
x^\prime \mathpzc{l}_{ij}(\pmb{x}, \pmb{x}^\prime)a_j(\pmb{x}^\prime,t)$ 
and $\pi^\ell_{j}(\pmb{x},t)=\int d^3 x^\prime \mathpzc{l}_{ij}(\pmb{x}, %
\pmb{x}^\prime)\pi_i(\pmb{x}^\prime,t)$ vanish identically. Keeping these
details in mind,  the fundamental Dirac bracket of our system [Eq.~(\ref%
{fundDirabrac})]  acquires the following structure:
\begin{equation}
\left\{a_i^\tau(\pmb{x}),\pi_j^\tau(\pmb{x}^\prime)\right\}_*=%
\pmb{\mathpzc{t}}_{ij}(\pmb{x},\pmb{x}^\prime).  \label{fundDirabractr}
\end{equation}%
On the other hand,  the Dirac bracket of $a_0$ or $\pi_0$ with an arbitrary
functional $\mathscr{Q}$ vanishes identically by construction:
\begin{equation}
\left\{\mathscr{Q}(\pmb{x}),a_0(\pmb{x}^\prime)\right\}_*=0,\quad \left\{%
\mathscr{Q}(\pmb{x}),\pi_0(\pmb{x}^\prime)\right\}_*=0.
\label{elliminationao}
\end{equation}
Because of this,  the sector $(a_0,\pi_0)$ can be formally eliminated
from the phase space,  and the theory is fully described in terms of $\left(%
\pmb{a}^\tau,\pmb{\pi}^\tau\right).$ Moreover,with $\mathscr{Q}$ being a
generic functional one has
\begin{equation}
\{\mathscr{Q}(\pmb{x}),\overbrace{\pmb{\nabla}\cdot\pmb{\pi}(\pmb{x}^\prime)}%
^{\varphi_2}\}_*=\{\mathscr{Q}(\pmb{x}),\overbrace{\nabla_i%
\varepsilon_{ij}a_j(\pmb{x}^\prime)}^{\varphi_3}\}_*=0.
\end{equation}%
Thus, both the second and third constraints [Eqs.~(\ref{gausslawdiel}) and (%
\ref{gaugechoose})]  are no longer ``weak'' equalities,  and instead they can be
used as ``strong'' equations. The latter terminology is conceptually
equivalent to replacing the symbol $\approx$ by $=$ in the set of
constraints $\{\varphi_i\}$. So, once the Dirac brackets [Eq.~(\ref{diracbracket}%
)]  are constructed, they can be set to zero everywhere.

Making use of Eq.~(\ref{diracbracket}) and Eqs.~(\ref{tpro})-(\ref%
{pipotrasversal}),  we find
\begin{eqnarray}
\begin{array}{c}
\displaystyle \left\{\mathscr{P}^0,a_i(x)\right\}_*=\partial^0 a_i(x),\ \
\left\{\mathscr{P}^0,\pi_i(x)\right\}_*=\partial^0\pi_i(x), \\
\\
\left\{\pmb{\mathscr{P}},a_i(x)\right\}_*=\pmb{\nabla}a_i(x),\ \ %
\displaystyle \left\{\pmb{\mathscr{P}},\pi_i(x)\right\}_*=\pmb{\nabla}%
\pi_i(x)%
\end{array}
\notag
\end{eqnarray}
which are the well-known time and spatial transformation properties of
fields. It follows that for any polynomial functional $\mathscr{Q}$ of $%
\pmb{a}$ and $\pmb{\pi}$ that does not depend explicitly on $x,$ one has
\begin{eqnarray}
\left\{\mathscr{P}^\mu,\mathscr{Q}(x)\right\}_*=\partial^\mu\mathscr{Q}(x).
\label{trucogenial}
\end{eqnarray}

To conclude this subsection,  we derive the modified Maxwell equations.
Whatever the nature of the electromagnetic background ($\mathfrak{F}>0$
or $\mathfrak{F}<0$), the Hamiltonian equation of motion for $\pmb{\pi}$
becomes Ampere's law: 
\begin{eqnarray}  \label{maxwell1}
\frac{d\pmb{\pi}}{dx^0}=\left\{\pmb{\pi},\mathscr{P}^0\right\}_*=\pmb{\nabla}%
\times\frac{\delta\mathscr{P}^0} {\delta{\pmb{b}}}=-\pmb{\nabla}\times\pmb{h}%
.  \label{eq.1far}
\end{eqnarray}
Together with the constraint $\varphi_2=0$ (Gauss's law), they make the second
pair of Maxwell equations. The Hamiltonian equation for $\pmb{b}$, 
\begin{eqnarray}  \label{maxwell2}
\frac{d\pmb{b}}{dx^0}=\left\{\pmb{b},\mathscr{P}^0\right\}_*=\pmb{\nabla}
\times\frac{\delta\mathscr{P}^0}{\delta{\pmb{\pi}}}=-\pmb{\nabla}\times%
\pmb{e},  \label{eq.2far}
\end{eqnarray}%
where  $\pmb{e}=-\partial_0\pmb{a},$ becomes Faraday's equation. It is
fulfilled as an identity. Together with another identity, Gauss's law for
magnetism, $\pmb{\nabla}\cdot\pmb{b}=0$,  which is not a Hamilton equation of
motion, they make the first pair of Maxwell equations.

We stress that the procedure developed in this subsection is also applicable
to any other linear approximation of  electrodynamics in which the
optical tensors  depend on neither  the space-time coordinates nor the
derivates with respect to the latter. Observe that it is even suitable to
describe the situation in which there exists a certain biaxiality associated
with the crossed fields configuration; i.e., where the external field
invariants vanish identically, $\mathfrak{F}=\mathfrak{G}=0.$


\subsection{Generators of rotations and Lorentz transformations \label%
{spingenerallrofjdffk}}


In this subsection we first obtain the conserved generators associated with
the Amputated Lorentz Group $SO_{A}(3,1)$. The Noether theorem
for infinitesimal transformations from the $SO_A(3,1)$ group over
the field $a^\mu(x)$ that leave the action $\Gamma$ in Eq.~(\ref{effelagrangian}) invariant reads
\begin{eqnarray}  \label{noethercurrentlorrent}
\partial_{\mu}\left\{\frac{\partial \mathscr{L}}{\partial
(\partial_{\mu}a^{\nu})}\delta a^\nu+ T^{\mu}_{\ \ \nu}\delta x^\nu\right\}=0,
\end{eqnarray}
where $\mathscr{L}$ is the quadratic Lagrangian of  Eq.~(\ref%
{generaleffectivelagragianstrongfield}), $T^{\mu}_{\ \ \nu}$ is the
canonical stress-energy tensor of Eq.~(\ref{canonicalenergytensor}),  and the
transformation laws are
\begin{eqnarray}  \label{vectorialphotongenerallaw}
\begin{array}{c}
\displaystyle \delta a^\rho(x)=\frac{i}{2}\omega^{\alpha\beta}\left(%
\mathfrak{J}_{\alpha\beta}\right)^{\rho}_{\ \ \sigma} a^{\sigma}( x), \\
\\
\displaystyle \delta x^\rho =\frac{i}{2}\omega^{\alpha\beta}\left(\mathfrak{J%
}_{\alpha\beta}\right)^{\rho}_{\ \ \sigma} x^{\sigma}%
\end{array}%
\end{eqnarray}
with $\omega^{\alpha\beta}$ determined by Eq.~(\ref{infintesimaltransforma})
and with the vectorial representation of the Lie algebra generator of $SO(3,1)$
\begin{eqnarray}
\left(\mathfrak{J}_{\alpha\beta}\right)^{\rho}_{\ \
\sigma}=i\left(\delta^\rho_{\ \ \beta}\eta_{\alpha\sigma} -\delta^\rho_{\ \
\alpha}\eta_{\beta\sigma}\right).  \label{vectrepresentationlorents}
\end{eqnarray}
With these details in mind,  the Noether conservation equation [Eq.~(\ref%
{noethercurrentlorrent})]  may be written as
\begin{eqnarray}
\partial_\mu\left[\frac{1}{2}\vartheta\mathscr{F}_{\alpha\beta}J^{\mu\alpha
\beta}+\frac{1}{2}\xi\tilde{\mathscr{F}}_{\alpha\beta}J^{\mu\alpha \beta}%
\right]=0.  \label{current2122}
\end{eqnarray}
Correspondingly, the conserved currents associated with the $SO_A(3,1)$ symmetry are given by
\begin{eqnarray}  \label{ggconservation1}
J^{\mu}=\frac{1}{2}\mathscr{F}_{\alpha\beta}J^{\mu\alpha \beta}\quad \mathrm{%
and}\quad \tilde{J}^{\mu}=\frac{1}{2}\tilde{\mathscr{F}}_{\alpha\beta}J^{\mu%
\alpha \beta}.
\end{eqnarray}%
Their respective continuity equations read
\begin{eqnarray}
\partial_\mu J^\mu&=&\frac{1}{2}\mathscr{F}_{\alpha\beta}\partial_\mu
J^{\mu\alpha \beta}=0,  \label{conJ1} \\
\partial_\mu \tilde{J}^\mu&=&\frac{1}{2}\tilde{\mathscr{F}}%
_{\alpha\beta}\partial_\mu J^{\mu\alpha \beta}=0.  \label{conJ2}
\end{eqnarray}
To provide the fulfillment of Eqs.~(\ref{current2122}) and  (\ref{conJ1}), and  (\ref%
{conJ2}),  it is sufficient to define the in generally  nonconserved current $%
J^{\mu\alpha\beta}$ in the form
\begin{eqnarray}  \label{decompositionJMAB}
J^{\mu\alpha\beta}=-\left[T^{\mu\nu}x^{\sigma}+\frac{\partial\mathscr{L}}{%
\partial(\partial_\mu a_\nu)}a^\sigma\right]i\left(\mathfrak{J}%
^{\alpha\beta}\right)_{\nu \sigma}, 
\end{eqnarray}
which  imitates the elements associated with $SO(3,1)$ invariance. For further convenience,  we express $J^{\mu\alpha\beta}$ in terms
of $\Theta^{\mu\nu}$ [Eq.~(\ref{gaugesdlsdiif121})]. To this end, we substitute Eq.~(\ref{connection}) and make use of the identity
\begin{eqnarray}
\frac{\partial\mathscr{L}}{\partial(\partial_\mu a^\nu)}a^\sigma&=&
\partial_\lambda\left[x_\nu\frac{\partial\mathscr{L}}{\partial(\partial_\mu
a_\lambda)}a^\sigma\right]-x_\nu\partial_\lambda\left[\frac{\partial%
\mathscr{L}}{\partial (\partial_\mu a_\lambda)} a^\sigma\right].  \notag
\end{eqnarray}%
Consequently, 
\begin{eqnarray}  \label{gudalupe1}
J^{\mu \alpha\beta}=-i\Theta^{\mu\nu}x^{\sigma}\left(\mathfrak{J}%
^{\alpha\beta}\right)_{\nu
\sigma}=x^\alpha\Theta^{\mu\beta}-x^\beta\Theta^{\mu\alpha}.
\end{eqnarray}%
Observe that the nonconservation of $J^{\mu\alpha\beta}$ is intrinsically
related to the nonsymmetric feature of the energy momentum tensor [Eq.~(\ref%
{gaugesdlsdiif121})]. In fact,
\begin{equation}
\partial_\mu J^{\mu\alpha\beta}=\Theta^{\alpha\beta}-\Theta^{\beta\alpha}.
\label{muyimportante}
\end{equation}
Since the continuity equations [Eqs.~(\ref{conJ1}) and (\ref{conJ2})] involve the
projection of Eq.~(\ref{muyimportante}) onto the external field tensors, we
end up with the following identities:
\begin{eqnarray}
\mathscr{F}_{\alpha\beta}\Theta^{\alpha\beta}=0\quad \mathrm{and}\quad
\tilde{\mathscr{F}}_{\alpha\beta}\Theta^{\alpha\beta}=0.
\label{ggconservationimplication}
\end{eqnarray}

Now, the respective spatial integrals of the time components of the currents
contained in Eq.~(\ref{ggconservation1}) provide the conserved charges
\begin{eqnarray}
\mathscr{G}=\frac{1}{2}\mathscr{F}_{\mu\nu}\mathscr{J}^{\mu\nu} \quad
\mathrm{and}\quad \tilde{\mathscr{G}}=\frac{1}{2}\tilde{\mathscr{F}}_{\mu\nu}%
\mathscr{J}^{\mu\nu}  \label{chargeconserved}
\end{eqnarray}%
with
\begin{equation}
\frac{\partial \mathscr{G}}{\partial x^0}=0\quad\ \mathrm{and} \quad \ \frac{%
\partial \tilde{\mathscr{G}}}{\partial x^0}=0.  \label{vacuumcharge}
\end{equation}
These scalars involve a second-rank tensor whose structure extends the known
representation of the Lorentz generator
\begin{eqnarray}
\mathscr{J}^{\mu\nu}&\equiv&\int d^3x J^{0\mu\nu}(\pmb{x},t)  \notag \\
&=&\int d^3x\left(x^\mu\Theta^{0\nu}-x^\nu\Theta^{0\mu}\right)
\label{lorentzgeneratorscur}
\end{eqnarray}%
to the case of the violated Lorentz invariance under consideration.
Correspondingly, we can define both  the photon angular momentum $%
\pmb{\mathscr{J}}=(\mathscr{J}^{23},\mathscr{J}^{31},\mathscr{J}^{12})$ and
the photon boost generator $\pmb{\mathscr{K}}=(\mathscr{J}^{10},\mathscr{J}%
^{20},\mathscr{J}^{30})$.

Considering the prescription above we can express the conserved charges of $%
SO_A(3,1)$ in the most general case of a constant and homogeneous
external field as
\begin{eqnarray}
\begin{array}{c}
\mathscr{G}=\pmb{B}\cdot\pmb{\mathscr{J}}-\pmb{E}\cdot\pmb{\mathscr{K}}, \\
\\
\tilde{\mathscr{G}}=\pmb{B}\cdot\pmb{\mathscr{K}}+\pmb{E}\cdot%
\pmb{\mathscr{J}}.%
\end{array}
\label{cucucucucucucssuuuuu}
\end{eqnarray}
Thus, as soon as the effects of the vacuum polarization tensor are
considered the number of Lorentz generators is reduced from 6 to 2.
Therefore the vacuum symmetry group $ISO_A(3,1)$ has dimension $2.$

The explicit structure of $\pmb{\mathscr{J}}$ and $\pmb{\mathscr{K}}$ in
terms of the canonical fields $(\pmb{a},\pmb{\pi})$ follows from Eqs.~(\ref%
{lorentzgeneratorscur}) and   (\ref{IReffelagrangianmass}),  and reads
\begin{eqnarray}  \label{Lorentzgenerators1}
&&\pmb{\mathscr{J}}=-\int d^3x\left\{ \pmb{x}\times\left[\pmb{\pi}\times%
\pmb{b}\right]-(\pmb{x}\times \pmb{a})(\pmb{\nabla}\cdot\pmb{\pi})\right\},
\\
&&\pmb{\mathscr{K}}=-x^0\pmb{\mathscr{P}}+\int d^3x\left\{\pmb{x}\left(\frac{%
1}{2} \pi_i\varepsilon_{ij}^{-1}\pi_j+\frac{1}{2}b_i
\mu_{ij}^{-1}b_j\right.\right.  \notag \\
&&\qquad\qquad-\left.\left. \frac{}{} a_0\pmb{\nabla}\cdot\pmb{\pi}%
\right)\right\},  \label{Lorentzgenerators2}
\end{eqnarray}%
where $\pmb{b}=\pmb{\nabla}\times\pmb{a}.$ Note how the secondary constraint
of our problem is implemented in both generators through terms proportional
to $\sim \pmb{\nabla}\cdot\pmb{\pi}$. In the phase subspace defined in Eq.~(\ref%
{hyphyssurf}),  such terms vanish identically,  and the resulting expression of $%
\pmb{\mathscr{J}}$ coincides with the angular momentum of light in an
optical medium \cite{iwo}.

Now, using the definition of the Dirac bracket [Eq.~(\ref{diracbracket})] and
equipped with the ``Lorentz-like'' generators $\mathscr{J}_i$ and $%
\mathscr{K}_i$ in Eqs.~(\ref{Lorentzgenerators1}) and (\ref{Lorentzgenerators2}), 
we are able to express
\begin{eqnarray}  \label{59}
&&\{\mathscr{J}_i,a_j(\pmb{x})\}_{*}=(\pmb{x}\times\pmb{\nabla})_ia_j(\pmb{x}%
)+\epsilon_{ijl} a_l(\pmb{x})  \notag \\
&&\qquad-\epsilon_{ikm}\frac{\nabla_j\varepsilon_{lk}\nabla_m}{%
\nabla\varepsilon\nabla} a_l(\pmb{x})- \epsilon_{ikl}\frac{%
\nabla_j\varepsilon_{km}\nabla_m}{\nabla\varepsilon\nabla} a_l(\pmb{x}), \\
&& \{\mathscr{J}_i,\pi_j(\pmb{x})\}_{*}=(\pmb{x}\times\pmb{\nabla})_i\pi_j(%
\pmb{x})+\epsilon_{ijl} \pi_l(\pmb{x}),  \label{rotopt2} \\
&& \{\mathscr{K}_i,a_j(\pmb{x})\}_{*}= x^i \partial^0 a_j(\pmb{x}%
)-x^0\nabla_i a_j(\pmb{x})-\frac{\nabla_j\varepsilon_ {kl}\nabla_l}{%
\nabla\varepsilon\nabla}  \notag  \label{61} \\
&&\qquad\times \left[x_i \partial^0 a_k(\pmb{x})\right], \\
&& \{\mathscr{K}_i,\pi_j(\pmb{x})\}_*=-x^0
\nabla_{i}\pi_j-\epsilon_{ijk}\mu_{km}^{-1}\left(\pmb{\nabla}\times\pmb{a}%
\right)_m  \notag  \label{62} \\
&&\qquad+ x_i\epsilon_{jkl}\mu_{lm}^{-1}\pmb{\nabla}_k\left(\pmb{\nabla}%
\times\pmb{a}\right)_m
\end{eqnarray}
These expressions deserve some comments. The first pair of brackets realizes
the infinitesimal rotation on the canonical variables $\pmb{a}$ and $%
\pmb{\pi},$ respectively. Note that, in contrast to the momentum $\pmb{\pi},$
the field $\pmb{a}$ transforms as a vector up to terms associated with the
the gauge-fixing condition [Eq.~(\ref{gaugechoose})]. A similar statement applies to
the boost transformation properties contained in Eqs.~(\ref{61}) and (\ref{62}).
We find it convenient to emphasize that all these brackets guarantee that the
infinitesimal canonical transformations induced by the generating functions $%
\mathscr{J}_i,~\mathscr{K}_i$ do not lead out from the constrained phase
subspace of Eq.~(\ref{hyphyssurf}). Indeed,
\begin{align}  \label{63}
\nabla_m\varepsilon_{mj}\{\mathscr{J}_i,a_j(\pmb{x})\}_{*}=0,\ \nabla_j\{%
\mathscr{J}_i,\pi_j(\pmb{x})\}_{*}=0, \\
\nabla_m\varepsilon_{mj}\{\mathscr{K}_i,a_j(\pmb{x})\}_{*}=0,\ \nabla_j\{%
\mathscr{K}_i,\pi_j(\pmb{x})\}_{*}=0.  \label{64}
\end{align}
While the first  column realizes $\varphi_3=0$ [Eq.~(\ref{gaugechoose})],  the
second one verifies $\varphi_2=0$  [Eq.~(\ref{gausslawdiel})]. It is worth
observing at this point that in the limit in which the external field
vanishes, $\varepsilon_{ij}=\mu_{ij}^{-1}=\delta_{ij}$,  the gauge condition is
reduced to the standard Coulomb gauge $(\pmb{\nabla}\cdot\pmb{a}=0).$ Due to
this fact,  the rotation transformation property of the gauge field [Eq.~(\ref{59})]
becomes similar to the one associated with the canonical momentum [Eq.~(\ref%
{rotopt2})].

The transformation properties of the gauge field $\pmb{a}(x)$ in Eqs.~(\ref{59})
and (\ref{61}) are not very helpful as they stand. For computing more
cumbersome brackets involving gauge-invariant quantities, it is more
convenient to represent the $\boldsymbol{a}$-containing part of Eqs.~(\ref%
{59})-(\ref{62}) in terms of the magnetic field $\boldsymbol{h}$ and magnetic
induction $\boldsymbol{b}$, respectively. To this end we apply $\boldsymbol{%
\nabla}\times$ to Eqs.~(\ref{59}) and (\ref{61}). Then
\begin{eqnarray}  \label{agua0}
&&\{\mathscr{J}_i,b_j(\pmb{x})\}_* =(\pmb{x}\times\pmb{\nabla})_ib_j(\pmb{x}%
)+\epsilon_{ijk}b_k(\pmb{x}), \\
&&\{\mathscr{J}_i,\pi_j(\pmb{x})\}_{*}=(\pmb{x}\times\pmb{\nabla})_i\pi_j(%
\pmb{x})+\epsilon_{ijk} \pi_k(\pmb{x}), \\
&&\{\mathscr{K}_i,b_j(\pmb{x})\}_{*}=-\epsilon_{ijk}e_k+x_i(\pmb{\nabla}%
\times\pmb{e})_j-x^0\nabla_i b_j,  \label{agua1} \\
&&\{\mathscr{K}_i,\pi_j(\pmb{x})\}_*=-\epsilon_{ijk}h_k+x_i(\pmb{\nabla}%
\times\pmb{h})_j-x^0\nabla_i\pi_j.  \label{agua2}
\end{eqnarray}%
These brackets constitute the starting point for determining the effects
induced by the vacuum polarization within Lorentz algebra. Note the
remarkable symmetry under the interchange $\pmb{\pi}\leftrightarrows \pmb{b}$ of the
first pair of these equations. The second pair, however, is invariant under
the simultaneous replacement $\pmb{\pi}\leftrightarrows\pmb{b}$ and $\pmb{e}%
\leftrightarrows\pmb{h}$.


\section{Dirac commutators and equations of motion for generators \label%
{sect4}}


The goal of this section is to determine the equations of motion of the
generators of the Lorentz rotations and to establish the Dirac commutation
relations distorting the Lie algebra of the Poincar\'e group. By contracting
these relations with the field tensor and with its dual,  the closed algebra
of the vacuum symmetry subgroup is obtained. As with  other algebraic
relations, the vacuum invariance holds in the physical subspace of the phase
space specialized by constraints, where the evolution of the physical
degrees of freedom takes place.


\subsection{Equations of motion for the angular momentum and for the boost
generator\label{consequenceLSB}}


The vacuum polarization tensor provides an effective coupling between photons
and the external field $\mathscr{F}$. In order to explore the role of this
quantity within LSB, we first consider the equation of motion associated
with the total angular momentum of the electromagnetic waves:
\begin{equation}
\frac{d\pmb{\mathscr{J}}}{d x^0}=\frac{\partial\pmb{\mathscr{J}}}{\partial
x^0}+\left\{\pmb{\mathscr{J}},\mathscr{P}^0\right\}_*.  \label{newton-law}
\end{equation}
The first term on the right-hand side vanishes identically, since $%
\pmb{\mathscr{J}}$ does not depend explicitly  on time $(x^0=t)$, whereas the
Dirac bracket provides the total torque exerted over the photon field. In
our context (see details in Appendix \ref{EMA1}), this can be  expressed as
\begin{eqnarray}  \label{gpbracketbetwJH}
\left\{\pmb{\mathscr{J}},\mathscr{P}^0\right\}_*&=&\int d^3 x\left[\pmb{\pi}%
\times\pmb{e}+\pmb{h}\times\pmb{b}\right].
\end{eqnarray}
In the rotation-invariant case--say, a  vacuum or isotropic material--when
the dielectric permeability is a unit tensor, Eq.~(\ref{gpbracketbetwJH})
disappears in correspondence with the momentum conservation. In this case,  $%
\boldsymbol{\pi}\|\boldsymbol{e}$, $\boldsymbol{h}\|\boldsymbol{b}$, and
their vector products are zero.  As a consequence,  the equation of motion for
the photon angular momentum can be written in terms of the spatial part of
the energy momentum tensor [Eq.~(\ref{explicit4})]:
\begin{equation}
\frac{d\mathscr{J}^i}{d x^0}=\frac{1}{2}\epsilon^{ijk}\int d^3x
(\Theta^{jk}-\Theta^{kj}).  \label{expladfaang}
\end{equation}

On the contrary, the equation of motion associated with the photon
boost generator is given by
\begin{equation}
\frac{d\pmb{\mathscr{K}}}{d x^0}=\frac{\partial\pmb{\mathscr{K}}}{\partial
x^0}+\left\{\pmb{\mathscr{K}},\mathscr{P}^0\right\}_*.
\label{boostnewton-law1}
\end{equation}
Because of the explicit dependence of $\pmb{\mathscr{K}}$ on time [see Eq.~(%
\ref{Lorentzgenerators2})], the first term on the right-hand side
contributes to the equation of motion $\partial\pmb{\mathscr{K}}/\partial
x^0=-\pmb{\mathscr{P}}.$ The Dirac bracket involved in this expression is
calculated in Appendix  \ref{EMA2} and reads
\begin{equation}
\left\{\pmb{\mathscr{K}},\mathscr{P}^0\right\}_*=\int d^3x \ \pmb{e}\times%
\pmb{h}.  \label{boostnewton-law2}
\end{equation}%
Combining these details, Eq.~(\ref{boostnewton-law1}) acquires the following
structure
\begin{eqnarray}  \label{boostnewton-law2v}
\frac{d\mathscr{K}^i}{ d x^0}&=&\int d^3x[(\pmb{e}\times\pmb{h})^i+(\pmb{\pi}%
\times\pmb{b})^i]  \notag \\
&=&\int d^3x[\Theta^{i0}-\Theta^{0i}]
\end{eqnarray}
where $\Theta^{i0}$ and $\Theta^{0i}$ are  the Poynting vector and the density
of momentum, respectively.

Both Eq.~(\ref{expladfaang}) and Eq.~(\ref{boostnewton-law2}) can  be
embedded in a compact four-dimensional expression
\begin{equation}
\frac{d\mathscr{J}^{\mu\nu}}{d x^0}=\mathpzc{T}^{\mu\nu}
\label{torqueficovariant}
\end{equation}
in which the antisymmetric tensor $\mathpzc{T}^{\mu\nu}$ is connected to the
stress-energy tensor by means of the following expression
\begin{equation}
\mathpzc{T}^{\mu\nu}\equiv\int d^3x \tau^{\mu\nu}(x), \qquad
\tau^{\mu\nu}\equiv\Theta^{\mu\nu}-\Theta^{\nu\mu}.  \label{4dtorque}
\end{equation}
The structure of Eq.~(\ref{torqueficovariant}) with Eq.~(\ref{4dtorque}) is  
somewhat expected: the spatial integration of Eq.~(\ref{muyimportante})
reproduces the same equation for the Lorentz-like  generators up to a
surface integral $\sim\oint_\mathrm{S} d\sigma_i J^{i\alpha\beta}$,  which
vanishes identically when the rapid falloff of the canonical fields at 
spatial infinity $(\mathrm{S}\to\infty)$ is provided.

Observe that the projections of Eq.~(\ref{torqueficovariant}) onto the
external field tensors are consistent with the conservation law of $%
\mathscr{G}$ and $\tilde{\mathscr{G}}$ [Eq.~(\ref{vacuumcharge})]. In fact, due to
Eq.~(\ref{ggconservationimplication}), we obtain that
\begin{equation}  \label{torqueprojection}
\frac{1}{2}\mathscr{F}_{\mu\nu}\mathpzc{T}^{\mu\nu}=0\quad \mathrm{and}\quad
\frac{1}{2} \tilde{\mathscr{F}}_{\mu\nu}\mathpzc{T}^{\mu\nu}=0.
\end{equation}
We shall see that these identities, together with the equation of motion of $%
\mathscr{J}^{\mu\nu}$,  turn out to be very convenient to evaluate the vacuum
polarization effects on ``Poincar\'e-like'' algebra.


\subsection{Distorted Poincar\'e algebraic relations}


Let us consider Eq.~(\ref{trucogenial}) with $\mathscr{Q}$ replaced by $%
\Theta^{0\nu}$. After an integration over $\pmb{x}$ we obtain
\begin{eqnarray}  \label{abelianmoentumpsjf}
\left\{\mathscr{P}^\mu,\mathscr{P}^\nu\right\}_*=\int d^3 x\partial^\mu
\Theta^{0\nu}(\pmb{x},t).
\end{eqnarray}%
Using the divergence theorem and assuming that the canonical fields $\pmb{a}$
and $\pmb{\pi}$ vanish  sufficiently rapidly at infinity,  we find
\begin{eqnarray}
\left\{\mathscr{P}^i,\mathscr{P}^\nu\right\}_*=\int d^3 x\partial^i
\Theta^{0\nu}(\pmb{x},t)=0.  \label{poissbracktraslakdu}
\end{eqnarray}%
Therefore, 
\begin{equation}
\left\{\mathscr{P}^i,\mathscr{P}^j\right\}_*=0\quad \mathrm{and}\quad \left\{%
\mathscr{P}^i,\mathscr{P}^0\right\}_*=0.
\end{equation}%
Because of the antisymmetry of the Dirac bracket,  we have that $\left\{%
\mathscr{P}^0,\mathscr{P}^0\right\}_*$ vanishes identically as well. Having
these aspects in mind, we  can write Eq.~(\ref{abelianmoentumpsjf}) as
\begin{eqnarray}
\left\{\mathscr{P}^\mu,\mathscr{P}^\nu\right\}_*=0.
\label{standardtrasnlationcomuiur}
\end{eqnarray}

We also apply Eq.~(\ref{trucogenial}) to the following case:
\begin{eqnarray}
&&\left\{\mathscr{P}^\mu,J^{0\lambda\sigma}\right\}_*=x^\lambda\left\{%
\mathscr{P}^\mu,\Theta^{0\sigma}\right\}_*-x^\sigma\left\{\mathscr{P}%
^\mu,\Theta^{0\lambda}\right\}_*  \notag \\
&&\qquad\qquad\qquad=x^\lambda\partial^\mu\Theta^{0\sigma}-x^\sigma\partial^%
\mu\Theta^{0\lambda}  \label{boostmomentum111}
\end{eqnarray}%
where the integrand in Eq.~(\ref{lorentzgeneratorscur}) has been considered.
Thanks to the  identity $x^\alpha\partial^\beta\Theta^{0\tau}=%
\partial^\beta(x^\alpha\Theta^{0\tau})-\eta^{\beta\alpha}\Theta^{0\tau}$,  we
are able to express Eq.~(\ref{boostmomentum111}) as
\begin{eqnarray}
\left\{\mathscr{P}^\mu,J^{0\lambda\sigma}\right\}_*&=&\eta^{\mu\sigma}%
\Theta^{0\lambda}-\eta^{\mu\lambda}\Theta^{0\sigma}-\partial^\mu
J^{0\lambda\sigma}.  \label{rotmomentum}
\end{eqnarray}%
Integrating the expressions above over $\pmb{x}$,  we find
\begin{eqnarray}
\left\{\mathscr{P}^\mu,\mathscr{J}^{\lambda\sigma}\right\}_*&=&\eta^{\mu%
\sigma}\mathscr{P}^\lambda-\eta^{\mu\lambda}\mathscr{P}^\sigma  \notag \\
&&\qquad\qquad-\int d^3x \partial^\mu J^{0\lambda\sigma}.
\label{preliminary1PBSdf0}
\end{eqnarray}
Because of the rapid vanishing of the field at infinity, we obtain that, for $%
\mu=i,$ $\int d^3x\partial^iJ^{0\lambda\sigma}$ vanishes identically.
However, if $\mu=0,$ the last integral can be written as $\partial^0 \int
d^3xJ^{0\lambda\sigma}=\partial^0\mathscr{J}^{\lambda\sigma}.$ With these
details in mind, and using Eq.~(\ref{torqueficovariant}), we end up with
\begin{eqnarray}
\left\{\mathscr{P}^\mu,\mathscr{J}^{\lambda\sigma}\right\}_*=\eta^{\mu\sigma}%
\mathscr{P} ^\lambda-\eta^{\mu\lambda}\mathscr{P}^\sigma-\eta^{\mu0}%
\mathpzc{T}^{\lambda\sigma},  \label{preliminary1PBSdf}
\end{eqnarray}
where $\mathpzc{T}^{\lambda\sigma}$ is specified in Eq.~(\ref{4dtorque}).
The bracket above reproduces the commutators associated with $ISO(3,1)$ 
Lie algebra up to a term manifesting LSB.

Our analysis on the Poincar\'e-like algebra is to be completed by
deriving the Lorentz-like  algebra,  which includes $\left\{\mathscr{K}^i,%
\mathscr{J}^j\right\}_*$, $\left\{\mathscr{J}^i,\mathscr{J}^j\right\}_*$,  and
$\left\{\mathscr{K}^i,\mathscr{K}^j\right\}_*$. A detailed derivation of
these Dirac brackets can be found in Appendix  \ref{MLA}. In particular, we have
obtained that
\begin{eqnarray}  \label{psplitLA1}
\begin{array}{c}
\displaystyle \left\{\mathscr{J}^i,\mathscr{J}^j\right\}_*=\epsilon^{ijk}%
\mathscr{J}^k, \\
\\
\displaystyle \left\{\mathscr{J}^i,\mathscr{K}^j\right\}_*=\epsilon^{ijk}%
\mathscr{K}^k+\frac{1}{2}\epsilon^{ilm}\int d^3 x\ x^j\tau^{lm}, \\
\\
\displaystyle \left\{\mathscr{K}^i,\mathscr{K}^j\right\}_*=-\epsilon^{ijk}%
\mathscr{J}^k-\int d^3 x\left(x^i\tau^{j0}-x^j\tau^{i0}\right),%
\end{array}%
\end{eqnarray}
with $\tau^{ij}$ being the spatial part of the tensorial density $%
\tau^{\mu\nu}$ defined in Eq.~(\ref{4dtorque}). Naturally, the brackets
above can be combined in a four-dimensional expression,
\begin{eqnarray}
&&\left\{\mathscr{J}^{\mu\nu},\mathscr{J}^{\rho\sigma}\right\}_*=\eta^{\mu%
\rho}\mathscr{J}^{\nu\sigma}-\eta^{\nu\rho}\mathscr{J}^{\mu\sigma}+\eta^{%
\sigma\mu}\mathscr{J}^{\rho\nu}  \notag \\
&&\qquad\quad-\eta^{\sigma\nu}\mathscr{J}^{\rho\mu}-\int
d^3x(x^\nu\eta^{\mu0}-x^\mu\eta^{\nu0})\tau^{\rho\sigma}  \notag \\
&&\qquad\quad-\int
d^3x(x^\rho\eta^{\sigma0}-x^\sigma\eta^{\rho0})\tau^{\mu\nu}
\label{Liealgebra}
\end{eqnarray}
where the the right-hand side reproduces the standard well-known result of
the the Lorentz algebra up to terms that contain the antisymmetric part of
the energy-momentum tensor. Note that in spite of its nonconservation, the
spatial rotation generators retain the standard angular momentum algebra
according to the first line in Eq.~(\ref{psplitLA1}). This is a usual situation
in classical and quantum mechanics with  broken symmetries under
transformations that do not touch the time variable. It cannot be the same
with the whole of the Poincar\'e algebra, because the Hamiltonian itself is
one of its generators, whereas its commutators with other members of the
algebra that do not conserve cannot help being affected. That is why the
violation of the algebra in the second and third lines of Eq.~(\ref{psplitLA1})
is not unexpected. This result indicates, in addition, that the Lorentz
invariance in pure Coulomb-gauge $\pmb{\nabla}\cdot\pmb{a}=0$ Maxwell theory
is established only when the Lorentz generators are conserved. We conclude
our analysis by pointing out that the brackets in Eqs.~(\ref{preliminary1PBSdf}) and
(\ref{Liealgebra}) coincide with algebra obtained in Ref.~\cite%
{Burnel:2008zz} within the context of the free field quantization in a 
noncovariant gauge.


\subsection{Algebra of the symmetry subgroup \label{sect4C}}


In order to pursue our research, we proceed to contract Eq.~(\ref%
{preliminary1PBSdf}) with $\frac{1}{2}\mathscr{F}_{\lambda\sigma}$ and $%
\frac{1}{2}\tilde{\mathscr{F}}_{\lambda\sigma}$. As a
consequence,
\begin{eqnarray}
\left\{\mathscr{G},\mathscr{P}^\mu\right\}_*&=&\mathscr{F}^{\mu\nu}%
\mathscr{P}_\nu-\eta^{\mu 0}\frac{1}{2}\mathscr{F}_{\lambda\sigma}\mathpzc{T}%
^{\lambda\sigma}, \\
\left\{\tilde{\mathscr{G}},\mathscr{P}^\mu\right\}_*&=&\tilde{\mathscr{F}}%
^{\mu\nu}\mathscr{P}_\nu-\eta^{\mu 0}\frac{1}{2}\tilde{\mathscr{F}}%
_{\lambda\sigma}\mathpzc{T}^{\lambda\sigma},  \label{preliminary1PBSdf234}
\end{eqnarray}%
where the definitions of $\mathscr{G}$ and $\tilde{\mathscr{G}}$ have been
used. According to Eq.~(\ref{torqueprojection}),  the last terms on the
righ-hand sides of these brackets vanish identically. Therefore,
\begin{eqnarray}
\left\{\mathscr{G},\mathscr{P}^\mu\right\}_*=\mathscr{F}^{\mu\nu}\mathscr{P}%
_\nu,\quad \left\{\tilde{\mathscr{G}},\mathscr{P}^\mu \right\}_*=\tilde{%
\mathscr{F}}^{\mu\nu}\mathscr{P}_\nu.  \label{preliminary1PBSq}
\end{eqnarray}

The remaining Dirac bracket involving $\mathscr{G}$ and $\tilde{\mathscr{G}}$
can be determined by projecting Eq.~(\ref{Liealgebra}) twice onto the
external field tensors. This procedure generates the following brackets:
\begin{eqnarray}
\left\{\mathscr{G},\mathscr{G}\right\}_*&=&\mathscr{F}_{\mu\lambda}%
\mathscr{F}^{\ \ \lambda}_{\nu}\mathscr{J}^{\mu\nu},\ \left\{\tilde{%
\mathscr{G}},\tilde{\mathscr{G}}\right\}_*=\tilde{\mathscr{F}}_{\mu\lambda}%
\tilde{\mathscr{F}}^{\ \ \lambda}_{\nu}\mathscr{J}^{\mu\nu},  \notag \\
\left\{\mathscr{G}, \tilde{\mathscr{G}}\right\}_*&=&\frac{1}{2}\int d^3x%
\left[\mathscr{F}_\mu^{\ \ 0}x^\mu\tilde{\mathscr{F}}_{\rho\sigma}\tau^{\rho%
\sigma}\right.  \notag \\
&&\qquad\qquad\qquad\qquad+\left.\tilde{\mathscr{F}}_\mu^{\ \ 0}x^\mu%
\mathscr{F}_{\rho\sigma}\tau^{\rho\sigma}\right].
\end{eqnarray}
Both $\mathscr{F}_{\mu\lambda}\mathscr{F}^{\ \ \lambda}_{\nu}$ and $\tilde{%
\mathscr{F}}_{\mu\lambda}\tilde{\mathscr{F}}^{\ \ \lambda}_{\nu}=2\mathfrak{F%
}\eta_{\mu\nu}+\mathscr{F}_{\mu\lambda}\mathscr{F}^{\ \ \lambda}_{\nu}$ are
symmetric tensors. Hence,  their contractions with the antisymmetric tensor $%
\mathscr{J}^{\mu\nu}$ vanish  identically,  and $\left\{\mathscr{G},%
\mathscr{G}\right\}_*=\left\{\tilde{\mathscr{G}}, \tilde{\mathscr{G}}%
\right\}_*=0.$ The latter result is  expected because it comes out from Eq.~(%
\ref{generalizedpoissonbracket}) that $\left\{\mathscr{Q},\mathscr{Q}%
\right\}=0,$ with $\mathscr{Q}$ being a generic function of the canonical
variables. Furthermore, by considering Eq.~(\ref{torqueprojection}), we can
claim that the right-hand side of the last Dirac bracket vanishes
identically as well.

We can then summarize the Lie algebra of $ISO_A(3,1)$ as follows:
\begin{eqnarray}
\begin{array}{c}
\left\{\mathscr{P}^\mu,\mathscr{P}^\nu\right\}_*=0, \ \ \left\{\mathscr{G},%
\tilde{\mathscr{G}}\right\}_*=0, \\
\\
\left\{\tilde{\mathscr{G}},\mathscr{P}^\mu \right\}_*=\tilde{\mathscr{F}}%
^{\mu\nu}\mathscr{P}_\nu\qquad \left\{\mathscr{G},\mathscr{P}^\mu\right\}_*=%
\mathscr{F}^{\mu\nu}\mathscr{P}_\nu.%
\end{array}
\label{ISOApreliminary1PBSq}
\end{eqnarray}
with the external field tensors $\mathscr{F}$ and $\tilde{\mathscr{F}}$
playing the roles  of group structure constants. Certainly the translation
generators induce an Abelian-invariant subalgebra which defines the
nonsemisimple structure of $SO_A(3,1).$  Moreover, the Casimir
invariants of our problem are given as
\begin{eqnarray}
\begin{array}{c}
\mathscr{P}^2=\mathcal{Z}_1+\mathcal{Z}_2, \\
\\
\displaystyle \mathcal{Z}_1=\frac{\mathscr{P}\tilde{\mathscr{F}}^2\mathscr{P}%
}{2\mathfrak{F}}\ \ \mathrm{{and} \ \ \mathcal{Z}_2=-\frac{\mathscr{P}%
\mathscr{F}^{2}\mathscr{P}}{2\mathfrak{F}}.}%
\end{array}
\label{uncharged.invariants}
\end{eqnarray}
It is worth mentioning at this point that the scalars involved in Eq.~(\ref%
{casimirs}) are, therefore, maps of the invariants above: $\mathscr{P}%
^2\mapsto k^2,$ $\mathcal{Z}_1\mapsto z_1,$ $\mathcal{Z}_2\mapsto z_2.$ 
Moreover, in the special frame where the field is purely
magnetic $(\mathfrak{F}>0)$ or purely electric $(\mathfrak{F}<0)$, Eq.~(\ref%
{Liealgebra}) expands into
\begin{eqnarray}
\begin{array}{ccc}
\left\{\mathscr{P}_x,\mathscr{P}_y\right\}_*=0, &  & \left\{\mathscr{P}_z,%
\mathscr{P}_0\right\}_*=0 \\
&  &  \\
\left\{\mathscr{J}_z,\mathscr{P}_x\right\}_*=\mathscr{P}_y, &  & \left\{%
\mathscr{K}_z,\mathscr{P}_0\right\}_*=-\mathscr{P}_z, \\
&  &  \\
\left\{\mathscr{J}_z,\mathscr{P}_y\right\}_*=-\mathscr{P}_x, &  & \left\{%
\mathscr{K}_z,\mathscr{P}_z\right\}_*=-\mathscr{P}_0,%
\end{array}
\notag
\end{eqnarray}
where Eq.~(\ref{cucucucucucucssuuuuu}) has been used. Each column in this
set of commutators manifests a subalgebra: the first one corresponds to the 
two-dimensional Euclidean group $ISO(2),$ whereas the second one
corresponds to $(1+1)-$dimensional pseudo-Euclidean group $ISO(1,1)$.  
The latter groups are associated with the transverse and pseudoparallel
planes with respect to the $\pmb{B}(\pmb{E})$ direction. Therefore, the
symmetry subgroup $SO_A(3,1)$, down to which the Poincar\'e group
is broken due to the presence of an external field, reduces in the reference
frame--where that field is purely magnetic or electric--to the direct
product of $ISO(2)$ and $ISO(1,1)$. Besides, we want
to remark that as long as a photon propagates transverse to the external
field, the square of the Pauli-Lubanski operator $\mathpzc{w}%
^\mu=1/2\epsilon^{\mu\lambda\sigma\varrho}\mathscr{J}_{\lambda\sigma}%
\mathscr{P}_{\varrho}$ is no longer a Casimir invariant. This fact reflects
the underlying difference between the vacuum in an external field $%
\mathscr{F} $ and the case of an empty space-time in which the $SO(3,1)$ 
symmetry is preserved and all particles are classified according
to the spin and helicity representations encoded in the $\mathpzc{w}^2$ %
eigenvalues.


\section{Magnetic moment of small electromagnetic perturbations of the
vacuum \label{torquemagnetico}}


In this section we analyze some consequences associated with the equation of
motion for the photon angular momentum [Eq.~(\ref{newton-law}) and (\ref%
{gpbracketbetwJH})]:
\begin{eqnarray}
\begin{array}{c}
\displaystyle\frac{d\pmb{\mathscr{J}}}{d x^0}=\pmb{\mathpzc{T}}\quad\mathrm{%
with}\quad \displaystyle \pmb{\mathpzc{T}}=\int d^3 x(\pmb{\pi}\times\pmb{e}+%
\pmb{h}\times\pmb{b} ).%
\end{array}
\label{torquefibn}
\end{eqnarray}
In the special case where the external field is magneticlike $(\mathfrak{F}%
>0)$, the explicit substitution of $\pmb{\pi}$ and $\pmb{h}$  [Eqs.~(\ref%
{momentamagnetized}) and  (\ref{magneinduc})]  into $\pmb{\mathpzc{T}}$ [Eq.~(\ref%
{torquefibn})] allows us  to express the latter in a rather meaningful form:
\begin{eqnarray}
\begin{array}{c}
\displaystyle \pmb{\mathpzc{T}}=2\pmb{\mathpzc{M}}\times\pmb{B}, \\
\\
\displaystyle \pmb{\mathpzc{M}}=\frac{1}{2}\int d^3x\left\{\frac{\mathfrak{L}%
_{\mathfrak{G}\mathfrak{G}}}{\varepsilon_\perp\varepsilon_\parallel}\left(%
\pmb{\pi}\cdot\pmb{B}\right)\pmb{\pi}+\mathfrak{L}_{\mathfrak{F}\mathfrak{F}%
}\left(\pmb{b}\cdot\pmb{B}\right)\pmb{b}\right\}.%
\end{array}
\label{nuevmagemopedb}
\end{eqnarray}
where Eq.~(\ref{convexityproperties}) has been used. The expression above
mimics the torque exerted by the external field on the magnetic dipole $%
\pmb{\mathpzc{M}}$. The torque $\pmb{\mathpzc{T}}$  [Eq.~(\ref{nuevmagemopedb})]  
 vanishes when projected onto $\pmb{B},$ so Eq.~(\ref{torquefibn})
implies that the parallel component of the photon angular momentum, $%
\mathscr{J}_\parallel$,  is a constant of motion. On the contrary, the
projection (helicity) $\mathfrak{h} \sim\pmb{\mathscr{J}}\cdot%
\pmb{\mathscr{P}}$ of the angular momentum of a photon onto its canonical
momentum [Eq.~(\ref{transalationalcharges})]\footnote{%
Recall that this is the direction of the momentum flux and  the wave
vector (see Appendix \ref{mpvsew}),  not of the energy flux, whose
direction coincides with the Poynting vector and with that of the group
velocity.} is not a conserved quantity unless $\pmb{\mathscr{P}}$ turns out
to be parallel to the external field.

The magnetic moment $\pmb{\mathpzc{M}}$ is a feature of the small
electromagnetic perturbations of the vacuum (photons in the first place), when
the former interact with it through  virtual electron-positron pairs.
This interaction makes the photon behave like a magnetic dipole. Note that $%
\pmb{\mathpzc{M}}$ [Eq.~(\ref{nuevmagemopedb})] is a gauge-invariant quantity,
also orthogonal to the photon canonical momentum,  $\pmb{\mathscr{P}}\cdot%
\pmb{\mathpzc{M}}=0$. Moreover, it has two components in correspondence with
the cylindrical symmetry imposed by the external magnetic field. One of them,
\begin{eqnarray}
\pmb{\mathpzc{M}}_\perp=\frac{1}{2}\int d^3x\left\{\mathfrak{L}_{\mathfrak{G}%
\mathfrak{G}}\frac{\pmb{\pi}\cdot\pmb{B}}{\varepsilon_\perp\varepsilon_%
\parallel}\pmb{\pi}_\perp+\mathfrak{L}_{\mathfrak{F}\mathfrak{F}}\left(%
\pmb{b}\cdot\pmb{B}\right)\pmb{b}_\perp\right\}  \label{perp-magnet-photon}
\end{eqnarray}is perpendicular to $\pmb{B}$;  whereas the remaining one,
\begin{eqnarray}
\pmb{\mathpzc{M}}_\parallel=\frac{1}{2}\int d^3x\left\{\mathfrak{L}_{%
\mathfrak{G}\mathfrak{G}}\frac{\pmb{\pi}\cdot\pmb{B}}{\varepsilon_\perp%
\varepsilon_\parallel}\pmb{\pi}_\parallel+\mathfrak{L}_{\mathfrak{F}%
\mathfrak{F}}\left(\pmb{b}\cdot\pmb{B}\right)\pmb{b}_\parallel\right\}
\label{parallel-magnet-photon}
\end{eqnarray}
is parallel to the axis in which the external field lies,  and therefore 
invariant under rotation about the magnetic field direction. This, however,
does not contribute to $\pmb{\mathpzc{T}}$  [Eq.~(\ref{nuevmagemopedb})].
Hence,  it does not play any role within the equation of motion of the photon
angular momentum [Eq.~(\ref{torquefibn})]. Nevertheless, $\pmb{\mathscr{M}}%
_\parallel$ contributes to the effective Hamiltonian [Eq.~(\ref%
{transalationalcharges}),  $H \equiv\mathscr{P}^0$],  and thus  to the photon
energy. Note that with the use of Eq.~(\ref{convexityproperties}),  the latter can
be conveniently written as
\begin{equation}  \label{conven}
H=\int d^3x\left(\frac{1}{2\varepsilon_\perp}\pmb{\pi}^2+\frac{1}{2}%
\mu_\perp^{-1}\pmb{b}^2\right)-\pmb{\mathpzc{M}}\cdot\pmb{B}.
\end{equation}
In contrast to $\pmb{\mathpzc{M}}_\parallel,$ the perpendicular component of
$\pmb{\mathpzc{M}}$ neither remains invariant under a rotation around the  $%
\pmb{B}$ direction nor contributes to the photon energy since it is
projected out from the scalar product involved in Eq.~(\ref{conven}). However,  it
turns out to be a clear manifestation of  LSB, since it specifies, by
means of Eq.~(\ref{torquefibn}), that not all components of the photon
angular momentum are conserved quantities. Furthermore, it follows from Eqs.~(%
\ref{convexityproperties}) and  (\ref{parallel-magnet-photon}) that $%
\mathpzc{M}_\parallel\geq0$ behaves paramagnetically. By  contrast, it is
not possible to establish a definite magnetic behavior in $\mathpzc{M}_\perp$
because it contains terms which mix not only $\pi_\parallel$ and $\pi_\perp$,  but also
$b_\perp$ and $b_\parallel.$

Let us consider the case in which the external magnetic field is
asymptotically large, $\mathfrak{b}=\vert\pmb{B}\vert/\mathrm{B}_\mathrm{c}%
\to\infty.$ In this limit, the basic entities contained in Eq.~(\ref{conven})
[see Eq.~(\ref{convexityproperties2})]  are written in Ref.~\cite{PRD2011},  referring
to $\mathfrak{L}$ as the one-loop term of the Euler-Heisenberg Lagrangian
\cite{euler,Schwinger,ritus,VillalbaChavez:2010bp}:
\begin{eqnarray}
\begin{array}{c}
\displaystyle \mathfrak{L}_{\mathfrak{F}}\approx\frac{\alpha}{3\pi}\ln%
\mathfrak{\ b}, \\
\\
\displaystyle 2\mathfrak{F}\mathfrak{L}_{\mathfrak{G}\mathfrak{G}}\approx%
\frac{\alpha}{3\pi}\mathfrak{b},\ \ 2\mathfrak{F}\mathfrak{L}_{\mathfrak{F}%
\mathfrak{F}}\approx\frac{\alpha}{3\pi}.%
\end{array}
\label{poleiginfrared}
\end{eqnarray}
Observe that for a magnetic field with $10< \mathfrak{b}\ll 3\pi/\alpha$ one
can treat $\varepsilon_\perp=\mu_\perp^{-1}\sim1$ and therefore $\pmb{\pi}%
\sim-\pmb{e}-\mathfrak{L}_{\mathfrak{G}\mathfrak{G}}(\pmb{e}\cdot\pmb{B})%
\pmb{B}.$ The resulting effective Hamiltonian $H$ reads
\begin{equation}
H\approx\int d^3x\left(\frac{1}{2}\pmb{\pi}^2+\frac{1}{2}\pmb{b}^2\right)-%
\pmb{\mathpzc{M}}\cdot\pmb{B}.
\end{equation}%
In this approximation, the second term in Eq.~(\ref{nuevmagemopedb})
decreases as $1/\mathfrak{b}$,  and thus  contributes in $H$ as a small
constant. Hence, it can be disregarded in comparison with the term provided
by the first term  of Eq.~(\ref{nuevmagemopedb}),  which turns out to be a
linear function on the external field strength. As a consequence, the
magnetic dipole acquires the following structure: 
\begin{eqnarray}  \label{off-magnet-optical1}
\pmb{\mathpzc{M}}\approx \mathpzc{g}\frac{\mathrm{e}}{2\mathrm{m}}%
\pmb{\mathcal{S}},
\end{eqnarray}%
with $\mathpzc{g}=\alpha/3\pi$ being a sort of Land\'e factor, whereas
\begin{eqnarray}  \label{spinphtoon}
\pmb{\mathcal{S}}&=&\frac{1}{\varepsilon_\parallel\mathrm{m}}\int_\mathrm{V}
d^3x\ \pi_\parallel(\pmb{x},x^0) \pmb{\pi}(\pmb{x},x^0)  \notag \\
&\approx& \frac{1}{\mathrm{m}}\int_\mathrm{V} d^3x\ e_\parallel(\pmb{x},x^0) %
\pmb{e}(\pmb{x},x^0).
\end{eqnarray}
It is worth mentioning that $\pmb{\mathcal{S}}$ is only determined by the
electric induction vector associated with the small electromagnetic waves.
Thereby $\pmb{\mathpzc{M}}$ can be interpreted as a magnetic moment,  with $%
\pmb{\mathcal{S}}$ playing the role of  the ``spin'' of the small electromagnetic
waves. This terminology, however, is used just to establish an analogy with
the case of the electron magnetic moment. In contrast to any massive
particle, a photon lacks  a rest frame,  and thus  one cannot define  a spin
for it. Besides, $\pmb{\mathcal{S}}$ does not fulfill the standard Dirac
bracket of the angular momentum, i.e., the first bracket in Eq.~(\ref%
{psplitLA1}).

We continue our research by considering the electric field of each eigenmode
as a monochromatic plane wave,  so that
\begin{eqnarray}  \label{fieldsaveragedfff}
\pmb{e}^{(\lambda)}(\pmb{x},x^0)=\mathpzc{E}_0^{(\lambda)}\frac{\pmb{e}%
^{(\lambda)}(\pmb{k})}{\vert\pmb{e}^{(\lambda)}(\pmb{k})\vert}\cos[%
\omega_\lambda x^0-\pmb{k}\cdot\pmb{x}].
\end{eqnarray}
Here $\mathpzc{E}_0^{(\lambda)}$ and $\omega_\lambda(\pmb{k})$ are  the
amplitude and  frequency of mode $\lambda$, respectively.  Note that the shape of $%
\pmb{e}^{(\lambda)}(\pmb{k})$ can be found below Eq.~(\ref{electricpromag}).
Observe, in addition, that only mode $2$ has an electric field parallel to $%
\pmb{B}.$ As a consequence, $\pmb{\mathcal{S}}$ becomes physically relevant
for the second polarization mode. Inserting the expression above into Eq.~(%
\ref{spinphtoon}),  we find
\begin{eqnarray}
&&\pmb{\mathcal{S}}\backsimeq\int_{\mathrm{V}} d^3 x\ \frac{k_\perp %
\mathpzc{E}_0^{(2)2}}{\mathrm{m}\vert\pmb{k}\vert}\pmb{\mathpzc{s}}%
\cos^2[\omega_2 x^0-\pmb{k}\cdot\pmb{x}]
\end{eqnarray}%
where the leading term has been withheld so that $\omega^{(2)}\approx\vert%
\pmb{k}\vert$ and $\pmb{\mathpzc{s}}\equiv\pmb{e}^{(2)}/\vert\pmb{e}%
^{(2)}\vert\approx\pmb{n}\times(\pmb{n}_\parallel\times\pmb{n}_\perp)$.
Equipped with these approximations, the time average of $\pmb{\mathpzc{M}}$
reads
\begin{eqnarray}
\langle\pmb{\mathpzc{M}}\rangle=\mathpzc{g}\frac{\mathrm{e}}{2\mathrm{m}}%
\langle\pmb{\mathcal{S}}\rangle\quad \mathrm{with} \quad \langle%
\pmb{\mathcal{S}}\rangle\approx\frac{1}{2}\frac{\mathpzc{E}_0^{(2)2}\mathrm{V%
}k_\perp}{\vert\pmb{k}\vert\mathrm{m}}\pmb{\mathpzc{s}}.
\label{fdggfodslhdhdms}
\end{eqnarray}
Here,  $\mathrm{V}$ denotes the volume over which the integral contained in $%
\pmb{\mathcal{S}}$ is considered. It is worth observing at this point that $%
\pmb{\mathpzc{M}}$ is proportional to the average energy associated with the
second propagating mode: $\sim\frac{1}{2}\mathpzc{E}_0^{(2)2}.$

We should also mention at this point that the notion of the photon magnetic
moment was introduced in Ref.~\cite{selhugo} and discussed in Refs.~\cite{hugoelsel}
and \cite{Chavez:2009ia}. In those works, it was defined as contributing to
the photon energy as a function of both its momentum and the external field
strength. Moreover, it was conceptually analyzed in different energy regimes
and magnetic field contexts of the photon dispersion curve. Only in Ref.~\cite{Chavez:2009ia} 
were the asymptotic conditions investigated in the present
work  considered. The connection between the respective photon magnetic
moment and $\pmb{\mathpzc{M}}$ in Eq.~(\ref{fdggfodslhdhdms}) can only be
established under the second quantization of our problem, which requires us  to
substitute the Dirac brackets with  standard commutator relations.\footnote{%
The second quantization of the small electromagnetic field in an external
field becomes a necessary issue as far as one wishes to go beyond the purely
electromagnetic sector by including not only virtual, as here, but also free
charged particles. This task, however, is not the issue the present work,
although some needed building blocks are prepared in it.} In such a case, 
the frequencies $\omega_{2}(\pmb{k})$ and $\omega_3(\pmb{k})$ are related to
the respective photon energies, and in correspondence, the mode $2$ photon
density turns out to be $\mathpzc{N}=\frac{1}{2}\mathpzc{E}%
_0^{(2)2}/\omega_2.$ Dividing $\langle\pmb{\mathpzc{M}}\rangle$ by the total
number of mode $2$ photons contained in the volume $\mathrm{V}$, we obtain
\begin{equation}  \label{magneticmmoement}
\pmb{\mathpzc{m}}=\frac{\langle\pmb{\mathpzc{M}}\rangle}{\mathpzc{N}\mathrm{V%
}}=\mathpzc{g}\frac{\mathrm{e}}{2\mathrm{m}} \mathpzc{f}(k_\perp)%
\pmb{\mathpzc{s}},
\end{equation}%
where $\mathrm{m}$ is the electron mass and $\mathpzc{f}(k_\perp)= k_\perp/%
\mathrm{m}$ is a dimensionless form factor which guarantees the gauge
invariance of the theory; i.e., $\pmb{\mathpzc{m}}$ does not provide a photon
rest mass $\pmb{\mathpzc{m}}(k_\perp\to0)\to 0.$ Equation  (\ref{magneticmmoement}%
) coincides with the photon anomalous magnetic moment previously obtained by
one of the authors in Ref.~\cite{Chavez:2009ia}. However, in that work, $%
\pmb{\mathpzc{m}}$ was defined as the coefficient of $\pmb{B}$ when the
dispersion curve $\omega_2(\pmb{k})$ is linearly approximated in terms of
the external magnetic field.


\section{Correspondences with the General Lorentz -Violating Electrodynamics \label{lastsecVII}}


The present work is not concerned with  all thinkable Lorentz violations
associated with extensions beyond the Standard Model  like Refs.~\cite{KosteleckyI,Kostelecky:2007zz,Kostelecky}. Just the opposite; it is
developed entirely within  Standard Model  and deals specifically with such Lorentz
symmetry violations as  are stimulated by a background electromagnetic
field, mostly by a time- and space-independent magneticlike field ($
\mathfrak{G}$\ $=0,$ $\mathfrak{F}>0$) within quantum electrodynamics (QED).

Nevertheless, it makes sense to try to reduce the approaches of Refs.~\cite%
{KosteleckyI,Kostelecky:2007zz, Kostelecky} and of related studies [henceforth
referred to as General Lorentz-Violating Electrodynamics (GLVE)] to a common
denominator with the results of many works dealing in a
relativistic, covariant way with external fields, nontrivial metrics,  and/or a
medium, as  was proposed by a referee of PRD. When treated in the
framework of conventional physics, these are acting as Lorentz and
$SO(3)$ invariance-violating agents,  and therefore may supply special examples
to serve as models for verifying general constructions in GLVE. A full
analysis in this field would require a quite separate study. So as not to
deviate too far from our principal theme, we are now only  listing--for an
external magneticlike field--the CPT-even $(\kappa _{F})^{\kappa \mu
\lambda \nu }$ coefficients, whose combinations are subject to measurements
in various experiments intended for detecting Lorentz violations, as they
follow from the general covariant decomposition of the polarization tensor
in a nonlinear electrodynamics. We shall see that, contrary to the postulate
accepted in GLVE, in our context this tensor is not double traceless, but
quite the opposite; its trace is physically meaningful and associated with
the diffractive properties of the anisotropic medium formed by the
background field. Even if admitted, the case where
the double trace would vanish identically could not introduce any
modification to the distorted Poincar\'e algebra, i.e., Eqs.~(\ref%
{preliminary1PBSdf}) and (\ref{Liealgebra}). The double tracelessness
condition modifies the optical tensors of the theory, but the
Poincar\'e-like generators keep their structure as long as they are
expressed in terms of the canonical variables. Besides, the energy-momentum
tensor remains nonsymmetric, also as in Ref.~\cite{KosteleckyI}, a fact needed
to save the nontrivial equation of motion of the angular momentum of light.

We also comment on how the sensitivity achieved in experiments aimed at
detecting possible Lorentz violations in the vacuum might be confronted with
measuring equivalent magnetic fields. We made sure  that the accuracy
available would not be sufficient to detect the Lorentz violation produced
even by the hitherto strongest laboratory magnetic field in the vacuum. To
detect an effect of Lorentz violation presumably inherent in the vacuum,
which might be equivalent to the one stemming from QED with an  external
magnetic field on the order of the cosmic background ($10^{-6}$ G), the  
experimentalist would  need to achieve a  sensitivity of $10^{-44}$
in experiments that might, besides, exclude the influence of any dielectric
material involved in the experimental device. This surpasses the boldest prospects of
sensitivity under present-time considerations by at least 30 orders of
magnitude.


\subsection{The components of the CPT-even tensor}


The quadratic part of the effective Lagrangian $\mathfrak{L}$ [Eqs.~(\ref%
{generaleffectivelagragianstrongfield}), and  (\ref{nuevotensorinductions})  and (%
\ref{explaeff})] in an external magnetic or electric field can be expressed
as
\begin{equation}
\mathfrak{L}={-\frac{1}{4}}\left( \kappa _{F}\right) ^{\kappa \lambda \mu
\nu }f_{\kappa \lambda }f_{\mu \nu }.  \label{el}
\end{equation}%
In accordance with Eq.~(\ref{effelagrangian}), this is the same as
\begin{equation*}
\mathfrak{L=}\frac{1}{2}a^{\mu }\Pi _{\mu \nu }a^{\nu }.
\end{equation*}%
Thanks to the antisymmetry of the field tensor $f_{\mu \nu },$ the
coefficient tensor $(\kappa _{F})^{\kappa \lambda \mu \nu }$ is not defined
by Eq.~(\ref{el}) in a unique way: its parts, symmetric under the permutations
within the first and  second pair of indices,  are left undetermined. For
this reason,  this tensor is to be understood as antisymmetric under these
permutations:

\begin{equation}
(\kappa _{F})^{\kappa \lambda \mu \nu }=-(\kappa _{F})^{\lambda \kappa \mu
\nu }=-(\kappa _{F})^{\kappa \lambda \nu \mu }.  \label{anti}
\end{equation}%
Analogously, only the part of the tensor $(\kappa _{F})^{\kappa \lambda \mu
\nu },$ which is symmetric under the permutation of the first and second
pairs of indices, contributes to Eq.~(\ref{el}). For this reason,  this tensor
is also to be understood as symmetric under these permutations:%
\begin{equation}
(\kappa _{F})^{\kappa \lambda \mu \nu }=(\kappa _{F})^{\mu \nu \kappa
\lambda }.  \label{sym}
\end{equation}%
By using the definition $f_{\mu \nu }=i(k_{\mu }a_{\nu }-k_{\nu }a_{_{\mu
}}) $ in Eq.~(\ref{el}),  we see that the polarization tensor that has direct
physical meaning is connected with combinations of tensor $\left( \kappa
_{F}\right) ^{\kappa \lambda \mu \nu }$\ components in momentum space as 
\begin{eqnarray}
&&\Pi _{\mu \nu }(k)=\frac{1}{2}k_{\kappa }k_{\lambda }  \notag \\
&&\qquad \times \left[ (\kappa _{F})^{\kappa \nu \lambda \mu }-(\kappa
_{F})^{\kappa \mu \nu \lambda }-(\kappa _{F})^{\mu \kappa \lambda \nu
}+(\kappa _{F})^{\nu \kappa \mu \lambda }\right]  \notag \\
&&\qquad=2k_{\kappa }k_{\lambda }(\kappa _{F})^{\lambda \mu \kappa \nu }.
\end{eqnarray}%
The properties of Eqs.~(\ref{anti}) and (\ref{sym}) provide that this polarization
tensor will be symmetric, $\Pi _{\mu \nu }=\Pi _{\nu \mu },$ as it should be in the
vacuum with a background field \cite{shabadpolarization};  and transverse, $%
\Pi _{\mu \nu }k_{\nu }=0,$ as   is prescribed by the gauge invariance.
Using the relation
\begin{eqnarray*}
&&\tilde{\mathscr{F}}^{\alpha \beta }\tilde{\mathscr{F}}_{\xi }^{\ \sigma }+%
\mathscr{F}^{\alpha \beta }\mathscr{F}_{\xi }^{\ \sigma } \\
&&\qquad =2\mathfrak{F}(\eta ^{\alpha \sigma }\delta _{\ \xi }^{\beta }-\eta
^{\sigma \beta }\delta _{\ \xi }^{\alpha })+\eta ^{\alpha \sigma }\mathscr{F}%
_{\ \lambda }^{\beta }\mathscr{F}_{\ \xi }^{\lambda } \\
&&\qquad -\delta _{\ \xi }^{\alpha }\mathscr{F}^{\beta \lambda }\mathscr{F}%
_{\lambda }^{\ \sigma }+\delta _{\ \xi }^{\beta }\mathscr{F}^{\alpha \lambda
}\mathscr{F}_{\lambda }^{\ \sigma }-\eta ^{\beta \sigma }\mathscr{F}^{\alpha
\lambda }\mathscr{F}_{\lambda \xi } 
\end{eqnarray*}%
in Eq.~(\ref{explaeff}), the coefficients $(\kappa _{F})^{\kappa \lambda \mu \nu
}$ for the infrared limit in QED with a constant magnetic field may be
chosen as
\begin{eqnarray}
\left( \kappa _{F}\right) ^{\kappa \lambda \mu \nu } &=&{\frac{1}{2}}\left( -%
\mathfrak{L}_{\mathfrak{F}}+2\mathfrak{FL}_{\mathfrak{GG}}\right) \left(
\eta ^{\kappa \mu }\eta ^{\lambda \nu }-\eta ^{\kappa \nu }\eta ^{\lambda
\mu }\right)  \notag \\
&{-}&{\frac{1}{2}}\mathfrak{L}_{\mathfrak{GG}}\left( \eta ^{\lambda \mu }%
\mathscr{F}^{\kappa \sigma }\mathscr{F}_{\sigma }^{\ \nu }-\eta ^{\lambda
\nu }\mathscr{F}^{\kappa \sigma }\mathscr{F}_{\sigma }^{\ \mu }\right.
\notag \\
&+&\left. \eta ^{\kappa \nu }\mathscr{F}^{\lambda \sigma }\mathscr{F}%
_{\sigma }^{\ \mu }-\eta ^{\kappa \mu }\mathscr{F}^{\lambda \sigma }%
\mathscr{F}_{\sigma }^{\ \nu }\right)  \notag \\
&{+}&{\frac{1}{2}}\left( \mathfrak{L}_{\mathfrak{GG}}-\mathfrak{L}_{%
\mathfrak{FF}}\right) \mathscr{F}^{\kappa \lambda }\mathscr{F}^{\mu \nu }.
\label{kasubef}
\end{eqnarray}%
The (anti)symmetry properties of Eqs.~(\ref{anti}) and  (\ref{sym}) are obeyed by Eq.~(\ref%
{kasubef}). The Lorentz violation induced by the magnetic field is
characterized by $6-1=5$ components of the antisymmetric external field
tensor $\mathscr{F}_{\mu \nu }$ subjected to one condition $\mathfrak{G}=0$,
and by the three scalars $\mathfrak{L_{F},}$ $\mathfrak{L_{GG},}$ $\mathfrak{%
L_{FF}}$ determined by the dynamics of the interaction.

Beyond the infrared limit,  Eqs.~(\ref{generaleffectivelagragianstrongfield}%
) and  (\ref{nuevotensorinductions}) imply that the extension of the tensor $%
(\kappa _{F})^{\kappa \mu \lambda \nu }$ to include the infinite series of
the space-time derivatives $(\hat{{\kappa }}_{F})^{\kappa \mu \lambda \nu
}=\sum_{n}(\kappa _{F})^{\kappa \mu \lambda \nu \alpha _{1}...\alpha
_{n}}\partial _{1}...\partial _{n}$ reduces to multiplications--in the
momentum space--of its separate parts by the scalar functions $\frac{%
\varkappa _{1}}{k^{2}}$,  $\frac{\varkappa _{1}-\varkappa _{2}}{k\tilde{%
\mathscr{F}}^{2}k}$,  and $\frac{\varkappa _{1}-\varkappa _{3}}{k\mathscr{F}%
^{2}k}$ built of the polarization operator eigenvalues, or to equivalent
action of the corresponding (nonlocal) integral operators in the coordinate
space. This factorization feature is not a consequence of Eqs.~(\ref%
{accionanisotroipca}) and  (\ref{19}), since these equations  hold true already in a
more general case of nonlocality \cite{Kostelecky:2009}. On the contrary, it
follows from the structure of the  polarization operator in a magnetic field [Eq.~(\ref%
{gstrpi})].


\subsection{The double trace}


The double trace of Eq.~(\ref{kasubef}) turns out to be
\begin{equation}
\left( \kappa _{F}\right) _{\quad \mu \nu }^{\mu \nu }=-6\mathfrak{L}_{%
\mathfrak{F}}+2\mathfrak{F}\left( \mathfrak{L}_{\mathfrak{GG}}-\mathfrak{L}_{%
\mathfrak{FF}}\right) .  \label{doubletrace}
\end{equation}%
This is, in general, a nonvanishing quantity. This statement can be
verified, for instance, by considering the weak- field approximation of the
Euler-Heisenberg Lagrangian. In QED, ``weak" means small as compared to the
Schwinger characteristic value $B_{c}=\mathrm{m^{2}/e=4.42\times 10^{13}\
\mathrm{G}},$ provided $\mathfrak{F}>0$. In this asymptotic regime the field
derivatives read \cite{landau}
\begin{equation}
\begin{array}{c}
\displaystyle\mathfrak{L}_{\mathfrak{F}}={\frac{2\alpha }{45\pi }}\frac{B^{2}%
}{B_{c}^{2}}, \\
\\
\displaystyle\mathfrak{L}_{\mathfrak{FF}}={\frac{4\alpha }{45\pi }}\frac{1}{%
B_{c}^{2}},\qquad \mathfrak{L}_{\mathfrak{GG}}={\frac{7\alpha }{45\pi }}%
\frac{1}{B_{c}^{2}}.%
\end{array}
\label{lowweak}
\end{equation}%
Then the double trace of Eq.~(\ref{doubletrace}) becomes
\begin{equation*}
\left( \kappa _{F}\right) _{\quad \mu \nu }^{\mu \nu }=-\frac{\alpha }{3\pi }%
\frac{B^{2}}{B_{c}^{2}}.
\end{equation*}%
It is worth mentioning at this point that the double tracelessness condition
is customarily  taken in GLVE on the grounds that the trace may be
absorbed into the field renormalization \cite{KosteleckyI}. In QED,  the
(infinite) renormalization has been fulfilled at the stage of the one-loop
calculations that underlie the QED expressions for $\mathfrak{L}$ and for its
field derivatives, the polarization tensors. The standard renormalization
procedure of QED relates only to the zero-background field limit, while the
background field-dependent  part is fixed and obeys the condition reflecting
the correspondence principle: $\mathfrak{L}_{\mathfrak{F}}\rightarrow 0$ as $%
B\rightarrow 0$. This condition is respected by Eq.~(\ref{lowweak}) and
establishes the absence of radiative corrections to the Maxwell Lagrangian
for small and steady fields. So, the Maxwell Lagrangian remains untouched in
this limit, which is the physically necessary requirement. Therefore,  no 
renormalization of the electromagnetic field additional to the
one performed in the course of the infinite renormalization procedure is
admitted. All the terms in Eq.~(\ref{doubletrace}) are physically important,  as
they serve various components responsible for the vacuum refraction
processes [Eq.~(\ref{convexityproperties})]. We shall see in Sec. VID that
this situation is retained  even in the nonbirefringence case.


\subsection{Magnetoelectric coefficients}


Let us introduce, as is customarily, the matrix combina\-tions $%
(\kappa_{DE})^{ij}=-2\left(\kappa _{F}\right)^{0i0j},$~$(\kappa_{HB})^{ij}=%
\frac{1}{2}\epsilon^{ikl}\epsilon^{jpq}\left(\kappa _{F}\right)^{klpq},$  and $%
(\kappa_{DB})^{ij}=-(\kappa_{HE})^{ji}=\epsilon^{kpq}\left(\kappa
_{F}\right)^{0jpq}.$ In terms of these quantities,  the most general form of
the quadratic Lagrangian is \cite{Kostelecky:2007zz}
\begin{eqnarray}
&&\mathscr{L}= \frac{1}{2}e_i\left[\delta^{ij}+\left(\kappa_{DE}\right)^{ij}%
\right]e_j- \frac{1}{2}b_i\left[\delta^{ij}+\left(\kappa_{HB}\right)^{ij}%
\right]b_j  \notag \\
&&\quad+ e_i\left(\kappa_{DB}\right)^{ij}b_j.  \label{glveLangangian}
\end{eqnarray}
We wish to specialize this expression to the case where a magneticlike
background ($\mathfrak{F}>0$, $\mathfrak{G}=0$)  induces  LSB. To this end, 
we compare Eq.~(\ref{glveLangangian}) with  Eq.~(\ref{accionanisotroipca}).
As a consequence,  the following relations are established:
\begin{eqnarray}
\begin{array}{c}
\varepsilon_{ij}=\delta_{ij}+(\kappa_{DE})_{ij},\quad \mu_{ij}^{-1}
=\delta_{ij}+(\kappa_{HB})_{ij} \\
\\
(\kappa_{DE})_{ij}=-\mathfrak{L_F}\delta_{ij}+\mathfrak{L_{GG}}B^iB^j, \\
\\
(\kappa_{HB})_{ij}=-\mathfrak{L_F}\delta_{ij}-\mathfrak{L_{FF}}B^iB^j.%
\end{array}
\label{KDEHB}
\end{eqnarray}

The derivation of these relations requires the use of the corresponding
optical tensors given in Eq.~(\ref{37}) and is in agreement with Eq.~(\ref%
{kasubef}). The remaining matrices,  i.e., $(\kappa_{DB})^{ij}$ and $%
(\kappa_{HE})^{ji}$,  are responsible for a magnetoelectric effect, which  is the  magnetic  linear response to an applied 
electric field  and, reciprocally, the electric linear response to an  applied magnetic field. These matrices vanish identically in our framework. 
For the magnetoelectric effect, and hence for the matrices $(\kappa_{DB})^{ij}$ and
$(\kappa_{HE})^{ji}$, to exist,  it is necessary to admit \cite{PRD2010} the
nonvanishing of the pseudoscalar invariant of the external field, $\mathfrak{%
G}\neq0$, i.e.,  to take a general combination of an electric and a magnetic
field as the external field.


\subsection{The anisotropic nonbirefringent case}


The special option of anisotropy without birefringence is often paid
attention in GLVE; for example, in Ref.~\cite{Klinkhamer:2008ky}. To establish the
conditions for the absence of birefringence in our context,  we must equalize
the dispersion laws for the two different eigenmodes of  Eq.~(\ref{dispequatbelow}),  
$f_{2}(k_{\perp}^{2})=f_{3}(k_{\perp }^{2})$.   It follows from Eq.~(\ref{poleig1}) 
that  in the infrared limit,  one has \cite{PRD2011}
\begin{eqnarray}  \label{displaw0}
&&f_{2}(k_{\perp }^{2})=k_{\perp }^{2}\left(\frac{1-\mathfrak{L}_{\mathfrak{F%
}}}{1-\mathfrak{L}_{\mathfrak{F}}+2\mathfrak{F}\mathfrak{L}_{\mathfrak{GG}}}%
\right), \\
&&f_{3}(k_{\perp }^{2})=k_{\perp}^{2}\left(1- \frac{2\mathfrak{F}\mathfrak{%
L_{FF}}}{1-\mathfrak{L_{F}}}\right) .  \label{displaw1}
\end{eqnarray}

Hence, there is no birefringence, provided that the effective Lagrangian is
subject to the condition
\begin{equation}
2\mathfrak{FL_{GG}L_{FF}=}(1-\mathfrak{L_{F}})\left( \mathfrak{L_{GG}-L_{FF}}%
\right) .  \label{nobi}
\end{equation}%
This condition is Lorentz invariant: once there is no birefringence in a
special frame, there is none in any inertial frame. Note that the
nonbirefringence condition is not the condition of coincidence of the
eigenvalues $\varkappa _{2}$ and $\varkappa _{3}$ [Eq.~(\ref{poleig1})] (which
does not take place even on the common mass shell of the two eigenmodes \cite%
{PRD2011}), nor the coincidence of the two dielectric permeability values [Eq.~(%
\ref{convexityproperties})], contrary to what one might think. The condition  of Eq.~(\ref%
{nobi})  is not fulfilled in QED, where the Heisenberg-Euler Lagrangian \cite%
{euler,Schwinger,ritus,VillalbaChavez:2010bp} is taken for
$\mathfrak{L.}$ The only Lagrangian where the background
field tensor makes up   its single argument free of  birefringence is (as
inferred in Ref.~\cite{Jerzy}) the Born-Infeld
Lagrangian \cite{Borninfeld},\footnote{Another example of the absence of
birefringence is supplied by noncommutative electrodynamics in
an external field \cite{gitman}} wherein
\begin{equation*}
\begin{array}{c}
\displaystyle\mathfrak{L_{F}}^{\mathrm{BI}}=1-\left(1+\frac{2\mathfrak{F}}{%
\mathfrak{a}^{2}}\right) ^{-\nicefrac{1}{2}}, \\
\displaystyle\quad \mathfrak{L_{FF}}^{\mathrm{BI}}=\frac{1}{\mathfrak{a}^{2}}%
\left( 1+\frac{2\mathfrak{F}}{\mathfrak{a}^{2}}\right) ^{-\nicefrac{3}{2}},\
\mathfrak{L_{GG}^{\mathrm{BI}}}=\frac{1}{\mathfrak{a}^{2}}\left( 1+\frac{2%
\mathfrak{F}}{\mathfrak{a}^{2}}\right) ^{-\nicefrac{1}{2}}.%
\end{array}%
\end{equation*}%
Here,  $\mathfrak{a}$\ \ is the dimensional parameter inherent in that model,
and the requirement of the correspondence principle $\left. \mathfrak{L_{F}}%
\right\vert _{\mathfrak{F}=0}=0$ is obeyed. In the small external field
domain the two dispersion curves [Eqs.~(\ref{displaw0}) and (\ref{displaw1})] become
\begin{eqnarray}
\left. f_{2}\left( k_{\perp }^{2}\right) \right\vert _{\mathfrak{F}%
\rightarrow 0} &=&k_{\perp }^{2}\left(1-2\mathfrak{F}\mathfrak{L_{GG}}%
\right) , \\
\left. f_{3}\left(k_{\perp }^{2}\right) \right\vert _{\mathfrak{F}%
\rightarrow 0} &=&k_{\perp }^{2}\left( 1-2\mathfrak{F}\mathfrak{L_{FF}}%
\right) .
\end{eqnarray}%
Therefore,  in this domain,  the nonbirefringence condition [Eq.~(\ref{nobi})] 
reduces to $\mathfrak{L_{GG}=}\left. \mathfrak{L_{FF}}\right\vert
_{\mathfrak{F}=0}$ and to the disappearance of the last (Weyl-like)
term in Eq.~(\ref{kasubef}), the same as in Ref.~\cite{Klinkhamer:2008ky}.
This fact allows us to represent Eq.~(\ref{kasubef}) in the
same form as in GLVE \cite{Klinkhamer:2008ky}:
\begin{eqnarray}
\left(\kappa _{F}\right)^{\kappa \lambda \mu \nu }=\frac{1}{2}\left(
\eta^{\kappa \mu }\kappa^{\lambda \nu }-\eta^{\kappa \nu }\kappa^{\lambda
\mu}-\eta^{\lambda \mu }\kappa ^{\kappa \nu } +\eta^{\lambda \nu
}\kappa^{\kappa \mu }\right).  \notag \\
\end{eqnarray}
However, the symmetric tensor $\kappa^{\mu \nu }$ contained
in this expression  
\begin{eqnarray}\label{178}
\kappa^{\mu \nu }&\equiv&\left(\kappa_{F}\right)_{\lambda}^{\ \mu \lambda \nu }+\mathfrak{L_F}\eta^{\mu\nu}=\nonumber\\
&=&\frac{1}{2}\left(-\mathfrak{L_F}+2\mathfrak{F}\mathfrak{L}_{\mathfrak{FF}}\right)
\eta^{\mu \nu }+\mathfrak{L_{FF}}\mathscr{F}^{\mu \lambda
}\mathscr{F}_{\lambda }^{\ \nu },
\end{eqnarray}where $\left(\kappa_{F}\right)_{\lambda}^{\ \mu \lambda \nu }$ is the trace
of Eq.~(\ref{kasubef}), is not traceless. Its trace, 
$\kappa_\mu^\mu=-2\mathfrak{L_{F}}$,   disappears only in the zero
external field limit, but otherwise, in the small-field regime, it
depends quadratically on the external field strength--the same as
other terms in Eq.~(\ref{kasubef}) or in Eq.~(\ref{178})--and forms  the whole of
the isotropic part of the dielectric and magnetic permeability tensors in Eq.~(\ref%
{37}). In other words, it determines the isotropic part of the
vacuum polarization in the external magnetic field. This is a
physically meaningful quantity that cannot be expelled from
the Born-Infeld model, the same as the double trace
[Eq.~(\ref{doubletrace})] from QED.

The case under consideration is  parameterized  by two
quantities  $\mathfrak{L_F}$ and $\mathfrak{L_{FF}}$, with the
external magnetic field $\pmb{B}$ driving  LSB.    There
exists, however, an  additional parameter
$\tilde{\kappa}_{\mathrm{tr}}=\frac{2}{3}\mathfrak{F}\mathfrak{L_{FF}}-\mathfrak{L_F}$, 
whose definition is intrinsically associated with the matrices that 
characterize the theory [see Eq.~(\ref{correspondencefinal})  
in the next subsection].  We combine the latter expression
with the convexity properties of the effective Lagrangian
[Eq.~(\ref{convexityproperties2})] to establish the condition\footnote{Inequalities
[Eq.~(\ref{convexityproperties2})] are fulfilled in the Born-Infeld model
}
\begin{equation}  \label{connectionLVE}
\tilde{\kappa}_{\mathrm{tr}}\geqslant-1.
\end{equation}  Let us finally remark that the relation above  must be understood as a direct consequence of fulfilling the   fundamental
unitarity and causality principles.


\subsection{Numerical estimates}


We now estimate to what precision the coefficients $(\kappa
_{F})^{\kappa \mu \lambda \nu }$ should be measured in order that
the Lorentz violation caused by the magnetic field of a given
magnitude might be detected in the vacuum,  and confront it with
sensitivities attained in existing experiments aimed at detecting
the intrinsic Lorentz violations within GLVE, as these are listed in
Ref.~\cite{Kostelecky:2008}. To this end, we begin with
the matrix  combinations $\left(\tilde{\kappa}_{e^+}\right)^{ij}=\frac{1}{2}%
\left[(\kappa_{DE})^{ij}+(\kappa_{HB})^{ij}\right],$ $\left(\tilde{\kappa}%
_{e^-}\right)^{ij}=\frac{1}{2}\left[(\kappa_{DE})^{ij}-(\kappa_{HB})^{ij}%
\right]-\frac{1}{3}\delta^{ij}(\kappa_{DE})^{ll}$,  and $\left(\tilde{\kappa}%
_{o^\pm}\right)^{ij}=\frac{1}{2}\left[(\kappa_{DB})^{ij}\pm(\kappa_{HE})^{ij}%
\right]$,  which are frequently used in determining the parameter
space to which the experiments on birefringence are sensitive. We
find it convenient to remark that these  refer  to the
context of GLVE. The substitution of Eq.~(\ref{KDEHB}) into the
latter set of matrices allows us  to  to express
$\left(\tilde{\kappa}_{o^\pm}\right)^{ij}=0$: 
\begin{equation}
\begin{array}{c}
\displaystyle \tilde{\kappa}_{\mathrm{tr}}\equiv\frac{1}{3}%
(\kappa_{DE})^{ll}=\frac{2}{3}\mathfrak{F}\mathfrak{L_{GG}}-\mathfrak{L}_{%
\mathfrak{F}}\ <10^{-14}, 
\\
\displaystyle \left(\tilde{\kappa}_{e^+}\right)^{ij}=-\mathfrak{L_F}%
\delta^{ij}+\frac{1}{2}\left[\mathfrak{L}_{\mathfrak{GG}}-\mathfrak{L}_{%
\mathfrak{FF}}\right]B^iB^j \ <10^{-32}, 
\\
\displaystyle \left(\tilde{\kappa}_{e^-}\right)^{ij}=\tilde{\kappa}_{\mathrm{%
tr}}\delta^{ij}+\frac{1}{2} \left[\mathfrak{L}_{\mathfrak{GG}}+\mathfrak{L}_{%
\mathfrak{FF}}\right]B^iB^j\ <10^{-17}.%
\end{array}
\label{correspondencefinal}
\end{equation}
Here the inequalities indicate the experimental sensitivity related to the 
measurement  of  the corresponding coefficient combinations.

In considering a possible magnetic field-like Lorentz symmetry violation,  we
must restrict ourselves to very small magnetic fields. With QED in mind, we
should then refer to Eq.~(\ref{lowweak}) in this case. Note that ${\frac{%
2\alpha }{45\pi }=1.03\times 10^{-4}.}$ The accuracy of $10^{-32}$ would be
enough to detect a  Lorentz violation produced by the magnetic field on  the
Earth, $B_{\mathrm{earth}}=0.3-0.6\ \mathrm{G}$; the accuracy of 10$^{-14}$
would allow one to fix a Lorentz violation equivalent to the presence of a
magnetic field on the order of $10^{9}$ G. This is an unearthly large  value,
on the pulsar scale. However, an increase in experimental accuracy--say, to 10$%
^{-21}$  would give us the possibility  of detecting  the electromagnetic wave
refraction due to an external magnetic field already at  the laboratory value of
$10^{5}$ G. That would be great, because up to now no effect of
birefringence of QED has been seen in the vacuum in a direct experiment \cite%
{Cameron:1993mr,Zavattini:2007ee}. Unfortunately, as long as any
material strongly enhances the effect of the magnetic field as
compared to the vacuum, the above considerations may only relate to
experimental devices that do not exploit matter as a medium for
electromagnetic wave propagation. Apart from such devices, electron-positron pair creation 
by a single  photon,  and the photon splitting and merging are the well-recognized, efficient processes that are
efficient in pulsar magnetospheres at  magnetic field strengths
above 10$^{12}$ G. The exclusion of  one-photon pair creation at
the accuracy level of $10^{-20}$ from an ultrahigh-energy cosmic
ray event in Ref.~\cite{{Klinkhamer:2008ky}}, if viewed within  within QED,
only implies that there is no magnetic field larger than
$10^{6}$ G in the space region where that event occurred anyway, which is
not unexpected.

The general conclusion of this subsection is that the prospects of
detecting  Lorentz violations in the vacuum by perfecting the
existing experimental means might be based only   on the belief that
these violations are for some reason much larger than the ones
induced via QED by magnetic fields (presumably present in the
Galactic background).


\section{Summary and outlook}


The main effort in this work has been to perform an analysis of  Lorentz
symmetry breaking by an external field in nonlinear electrodynamics, the
gauge sector of QED included. We defined how  transformations from the
residual symmetry space-time subgroup, left after the external time- and
space-independent magnetic field had been imposed, act on coordinates and
other vector entities via the external field tensor. For small and steady
electromagnetic excitations over the magnetic field background,  we have
developed the Hamiltonian formalism to serve the linear electrodynamics of
the equivalent anisotropic medium, for which purpose a quadratic Lagrangian
of these excitations is written with the help of the polarization operator
in the external magnetic field. The electric and magnetic permeability
tensors in this linear electrodynamics are shown to be those of an
equivalent uniaxial medium in any special Lorentz frame, where the external
field is purely magnetic or purely electric. Their principal values are
expressed in the paper in terms of the field derivatives of the effective
Lagrangian. The fields and inductions are given the sense of canonical
variables, the necessary primary and secondary constraints are determined,
and the Dirac  brackets are defined on constrained physical phase space in
accordance with the $U(1)$ gauge invariance of the theory. The conserved
Noether currents, corresponding to the residual symmetry transformations, as
well as the nonconserved Noether currents corresponding to the Lorentz
transformations,  the symmetry under which is violated by the
external field, are defined on the physical phase subspace of the problem.
Among the former, and of some  importance,  is a nonsymmetric  but gauge-invariant
energy-momentum tensor, used to form the Poynting vector and the momentum
density, which are not the same quantities due to the antisymmetric part of
the  tensor. We have calculated the Dirac bracket commutation relations
between all the generators of infinitesimal space-time rotations and
translations to see that the $SO(3)$ algebra of the photon angular momentum
remains intact, despite the violation of the rotation symmetry by the
external field, whereas the Poincar\'e algebra is distorted. We derived the
evolution equation for the photon angular momentum, which is governed by the
photon magnetic moment depending on the antisymmetric part of the
energy-momentum tensor. The polarization effects entering the equation of
motion of the photon angular momentum are closely associated with the
existence of an optical torque. This is a  phenomenon inherent  to
conventional electrodynamics in anisotropic media,  which manifests the
breakdown of the rotational invariance. We argued that a small-amplitude
electromagnetic wave propagating in a strong magnetic field behaves as a
quasiparticle carrying a gauge-invariant magnetic moment orthogonal to the
wave-vector. The corresponding analysis of the equation of motion for the
angular momentum of light in a weak magnetic field was not developed here.
This limiting case seems to be very convenient for probing the nonlinear
behavior of the quantum vacuum. The latter could be achieved by transferring
the angular momentum from the small waves to a microscopic absorptive object
(e.g., tweezers). However, a detailed analysis of this issue will be given in a
forthcoming publication.

It would be also a challenge to find a closure to the distorted Poincar\'e
algebra for the present case--or, perhaps, other, simpler, cases--of the
Lorentz-symmetry violations.

In the last Sec. \ref{lastsecVII},  we confronted the structure of the polarization operator
in the magnetic field with prescriptions of a general Lorentz-violating
electrodynamics, and discussed the common features and differences.

\begin{acknowledgments}
S. Villalba-Chavez is very grateful to Reinhard Alkofer and Helios
Sanchis-Alepuz for helpful discussions. A.E. Shabad acknowledges the support
of FAPESP, Processo No. 2011/51867-9, and of RFBR under  Project
No. 11-02-00685-a. He also thanks D.M. Gitman for the  kind  hospitality extended to
him in at USP during his stay in $\mathrm{S\tilde{a}}$o Paulo, Brazil, where a
part of this work was fulfilled, and  analogously, V.V. Usov for the
hospitality at the  Weizmann Institute of Science, Rehovot, Israel.
\end{acknowledgments}

\appendix


\section{TWO IMPORTANT BRACKETS \label{EMA}}


\subsection{The case of $\left\{\pmb{\mathscr{J}},\mathscr{P}^0\right\}_*$
\label{EMA1}}

The aim of this appendix is to compute the Dirac bracket $\left\{%
\pmb{\mathscr{J}},\mathscr{P}^0\right\}_*$. In order to obtain the latter,
we divide it into three terms:
\begin{eqnarray}
\begin{array}{c}
\displaystyle \left\{\mathscr{J}^i,\mathscr{P}^0\right\}_*=\mathscr{I}_1+%
\mathscr{I}_2+\mathscr{I}_3, \\
\\
\displaystyle \mathscr{I}_1^i=\int d^3x \frac{\delta \mathscr{J}^i}{\delta
a_l}\frac{\delta \mathscr{P}^0}{\delta\pi^l}, \ \ \mathscr{I}_2^i=\int d^3x
\frac{\delta \mathscr{J}^i}{\delta\pi_l}\frac{\delta \mathscr{P}^0}{\delta
a^l}, \\
\\
\mathscr{I}_3=-\int d^3yd^3z\left\{ \mathscr{J}^i,\varphi_\alpha(\pmb{y}%
)\right\}C_{\alpha\beta}^{-1}(\pmb{y},\pmb{z})\left\{\varphi_\beta(\pmb{z}), %
\mathscr{P}^0 \right\}%
\end{array}
\label{a33}
\end{eqnarray}
An explicit derivation of the expressions above requires us  to know the
functional derivatives associated with the photon angular momentum [Eq.~(\ref%
{Lorentzgenerators1})]. For further convenience, we write the latter as
\begin{eqnarray}  \label{apenb1}
\pmb{\mathscr{J}}&=&\int d^3x\left[-(\pmb{x}\cdot\pmb{b})\pmb{\pi}+(\pmb{x}%
\cdot\pmb{\pi})\pmb{b}+(\pmb{x}\times\pmb{a})\pmb{\nabla}\cdot\pmb{\pi}%
\right].
\end{eqnarray}%
In correspondence we obtain
\begin{eqnarray}  \label{derivatives1}
\begin{array}{c}
\displaystyle \frac{\delta\mathscr{J}^j}{\delta \pi^l}=-(\pmb{x}\cdot\pmb{b}%
)\delta^{lj}+b^j x^l-\nabla^l(\pmb{x}\times \pmb{a})^j, \\
\\
\displaystyle \frac{\delta\mathscr{J}^j}{\delta a^l}=\epsilon^{jlk}\nabla^k(%
\pmb{x}\cdot \pmb{\pi})+(\pmb{x}\times\pmb{\nabla})^l\pi^j-\epsilon^{jlk}x^k %
\pmb{\nabla}\cdot\pmb{\pi}.%
\end{array}%
\end{eqnarray}
The explicit structure of $\mathscr{P}^0$ can be found in Eq.~(\ref%
{transalationalcharges}). Because of Eq.~(\ref{elliminationao}) we will
ignore the contribution proportional to $a_0.$ Having this in mind, the
respective derivatives turn out to be
\begin{eqnarray}  \label{derivatives4}
\frac{\delta\mathscr{P}^0}{\delta \pi^l}&=&-e^l,\qquad \frac{\delta%
\mathscr{P}^0}{\delta a^l}=(\pmb{\nabla}\times\pmb{h})^l.
\end{eqnarray}
Substituting Eqs.~(\ref{derivatives1}) and  (\ref{derivatives4}) into $%
\mathscr{I}_{1,2}^i$,  one finds
\begin{eqnarray}
\mathscr{I}_1^i&=&\int d^3x \left[-\epsilon^{ijk}e^j\nabla^k(\pmb{x}\cdot%
\pmb{\pi})-e^j(\pmb{x}\times\pmb{\nabla})^j\pi^i\right.  \notag \\
&&+\left.\epsilon^{ijk}x^k e^j\pmb{\nabla}\cdot\pmb{\pi}\right],  \label{a34}
\\
\mathscr{I}_2^i&=&\int d^3x \left[b^ix^j(\pmb{\nabla}\times\pmb{h})^j-(%
\pmb{\nabla}\times\pmb{h})^i(\pmb{x}\cdot\pmb{b})\right.  \notag \\
&&-\left.(\pmb{\nabla}\times\pmb{h})^l\nabla^l(\pmb{x}\times \pmb{a})^i%
\right].  \label{fhgo59333}
\end{eqnarray}
The last term in Eq.~(\ref{fhgo59333}) vanishes identically provided an
integration by parts.

We expand the derivative in the first integrand of $\mathscr{I}_1^i$ and use
the vectorial identity
\begin{eqnarray}  \label{vectorailindettzioa}
&&\pmb{u}\times[(\pmb{x}\times\pmb{\nabla})\times\pmb{o}]=u^l(\pmb{x}\times%
\pmb{\nabla})o^l-\pmb{u}\cdot(\pmb{x}\times\pmb{\nabla})\pmb{o}  \notag \\
&&\qquad=-\epsilon^{ijk}u^jx^k (\pmb{\nabla}\cdot\pmb{o})+%
\epsilon^{ijk}u^jx^m\nabla^ko^m.
\end{eqnarray}
As a consequence,
\begin{eqnarray}  \label{a37}
\mathscr{I}_1^i&=&\int d^3x \left\{-(\pmb{e}\times\pmb{\pi})^i-e^l(\pmb{x}%
\times\pmb{\nabla})^i\pi^l\right\}.
\end{eqnarray}%
Let us turn our attention to $\mathscr{I}_2^i.$ Note that an integration by
parts in the second term of the latter leads to
\begin{eqnarray}
\mathscr{I}_2^i&=&\int d^3x \left[b^ix^j(\pmb{\nabla}\times\pmb{h})^j+(%
\pmb{h}\times\pmb{b})^i+\epsilon^{ijk} h^kx^l\nabla^jb^l\right].  \notag \\
\end{eqnarray}
Adding to this expression a vanishing term $-\int d^3x\epsilon^{ijk} h^j x^k%
\pmb{\nabla}\cdot\pmb{b}$ and using Eq.~(\ref{vectorailindettzioa}),  we find
\begin{eqnarray}  \label{a39}
\mathscr{I}_2^i&=&\int d^3x \left[(\pmb{h}\times\pmb{b})^i-h^l(\pmb{x}\times%
\pmb{\nabla})^ib^l\right].
\end{eqnarray}

Now, we focus ourselves on  $\mathscr{I}_3$. Expanding the sum over $\alpha$
and $\beta$ and taking into account Eq.~(\ref{inversematricsdd}),  it reduces
to
\begin{eqnarray}
\mathscr{I}_3&=&\int d^3yd^3 z\nabla_k^{\pmb{y}}\left\{\mathscr{J}^i,\pi_k(%
\pmb{y})\right\}\frac{1}{\nabla^{\pmb{y}}\varepsilon\nabla^{\pmb{y}}}%
\delta^{(3)}(\pmb{y}-\pmb{z})\nabla_m^{\pmb{z}}\varepsilon_{mn}  \notag \\
&\times& \left\{a_n(\pmb{z}),\mathscr{P}^0\right\}-\int d^3yd^3z\nabla_m^{%
\pmb{y}}\varepsilon_{mn}\left\{\mathscr{J}^i,a_n(\pmb{y})\right\}  \notag \\
&\times&\frac{1}{\nabla^{\pmb{y}}\varepsilon\nabla^{\pmb{y}}}\delta^{(3)}(%
\pmb{y}-\pmb{z})\nabla_k^{\pmb{z}}\left\{\pi_k(\pmb{z}),\mathscr{P}%
^0\right\}.
\end{eqnarray}
With the help of Eqs.~(\ref{63}) and   (\ref{eq.1far}),  it is simple to
obtain the following identities: 
\begin{eqnarray}
\nabla_k^{\pmb{y}}\left\{\mathscr{J}^i,\pi_k(\pmb{y})\right\}=0, \quad
\nabla_k^{\pmb{z}}\left\{\pi_k(\pmb{z}),\mathscr{P}^0\right\}=0.
\label{PPmomentum}
\end{eqnarray}
In correspondence,  $\mathscr{I}_3$ vanishes identically as well. We then
substitute Eqs.~(\ref{a37}) and  (\ref{a39}) into Eq.~(\ref{a33}). As a
matter of fact, 
\begin{eqnarray}  \label{a40}
\left\{\mathscr{J}^i,\mathscr{P}^0\right\}_*=\int d ^3x\left\{(\pmb{x}\times%
\pmb{\nabla})^i\Theta^{00}+\pmb{\pi}\times \pmb{e}+\pmb{h}\times\pmb{b}%
\right\},  \notag \\
\end{eqnarray}%
where the identity
\begin{eqnarray}  \label{seondindetiftioid}
&&\left(\pmb{x}\times\pmb{\nabla}\right)\Theta^{00}=-e^l(\pmb{x}\times%
\pmb{\nabla})\pi^l+h^l(\pmb{x}\times\pmb{\nabla})b^l
\end{eqnarray}
has been taken into account. Note that an integration by parts carries out a
change of sign in the integral $\int d^3 x \left(\pmb{x}\times \pmb{\nabla}%
\right)\Theta^{00}=-\int d^3 x \left(\pmb{x}\times \pmb{\nabla}%
\right)\Theta^{00}.$ Therefore, the latter vanishes identically and can be
ignored in Eq.~(\ref{a40}). This operation leaves us with the terms which
appear in Eq.~ (\ref{gpbracketbetwJH}).


\subsection{The case of $\left\{\pmb{\mathscr{K}},\mathscr{P}^0\right\}_*$
\label{EMA2}}


Let us consider the Dirac bracket involved in Eq.~(\ref{boostnewton-law1}).
To determine the latter,  we express the photon boost in Eq.~(\ref%
{Lorentzgenerators2}) in the following form:
\begin{eqnarray}
\begin{array}{c}
\pmb{\mathscr{K}}=\pmb{\mathscr{K}}_{\mathscr{P}}+\pmb{\mathscr{K}}%
_{\Theta^{00}}, \\
\\
\displaystyle \pmb{\mathscr{K}}_{\mathscr{P}}=-x^0\pmb{\mathscr{P}},\quad %
\pmb{\mathscr{K}}_{\Theta^{00}}=\int d^3x (\pmb{x} \Theta^{00})%
\end{array}
\label{a2variation}
\end{eqnarray}
where $\pmb{\mathscr{P}}$ is the spatial translation generator and $%
\Theta^{00}$ is the energy density see Eq.~(\ref{transalationalcharges}). As
before, we neglect the contributions involving the Lagrangian multiplier $%
a_0.$ Considering Eq.~(\ref{a2variation}),  one finds
\begin{eqnarray}  \label{inkj1}
\left\{\mathscr{K}^i,\mathscr{P}^0\right\}_*=\left\{\mathscr{K}^i_{%
\mathscr{P}},\mathscr{P}^0\right\}_*+\left\{\mathscr{K}^i_{\Theta^{00}},%
\mathscr{P}^0\right\}_*
\end{eqnarray}
The first bracket on  the right-hand side can be computed by applying Eq.~(%
\ref{poissbracktraslakdu}). According to this equation, $\left\{\mathscr{P}%
^i,\mathscr{P}^0\right\}_*$ vanishes identically. Thus
\begin{equation}
\left\{\mathscr{K}^i_{\mathscr{P}},\mathscr{P}^0\right\}_*=-x^0\left\{%
\mathscr{P}^i,\mathscr{P}^0\right\}_*=0.
\end{equation}

In order to analyze the remaining terms on the right-hand side of Eq.~(\ref%
{inkj1}),  it is rather convenient to have at our disposal the following set
of derivatives: 
\begin{eqnarray}  \label{derivatives2}
\frac{\mathscr{K}^i_{\Theta^{00}}}{\delta \pi^l}=-x^ie^l,\quad \frac{\delta%
\mathscr{K}_{\Theta^{00}}^i}{\delta a^l}=-\epsilon^{ilm}h^m+x^i(\nabla\times%
\pmb{h})^l.  \notag \\
\end{eqnarray}

With the above expressions in mind,  one can undertake the calculation of the
second bracket in Eq.~(\ref{inkj1}). We write the latter as
\begin{equation}
\left\{\mathscr{K}^i_{\Theta^{00}},\mathscr{P}^0\right\}_*=\left\{\mathscr{K}%
^i_{\Theta^{00}},\mathscr{P}^0\right\}+\mathpzc{W}
\end{equation}
with
\begin{equation}  \label{adicionalKP0}
\mathpzc{W}\equiv-\int d^3yd^3z\left\{ \mathscr{K}_{\Theta^{00}}^i,\varphi_%
\alpha(\pmb{y})\right\}C_{\alpha\beta}^{-1}(\pmb{y},\pmb{z}%
)\left\{\varphi_\beta(\pmb{z}), \mathscr{P}^0 \right\}.
\end{equation}
From Eqs.~(\ref{derivatives2}) and  (\ref{derivatives4}),  it is
straightforward to get
\begin{eqnarray}
\left\{\mathscr{K}^i,\mathscr{P}^0\right\}&=&\int d^3x (\pmb{e}\times\pmb{h}%
)^i.  \label{a33variationfgsa}
\end{eqnarray}
Withthe help of Eq.~(\ref{inversematricsdd}),  we express $\mathpzc{W}$ as
\begin{eqnarray}
\mathpzc{W}&=&\int d^3yd^3 z\nabla_k^{\pmb{y}}\left\{\mathscr{K}%
^i_{\Theta^{00}},\pi_k(\pmb{y})\right\}\frac{1}{\nabla^{\pmb{y}%
}\varepsilon\nabla^{\pmb{y}}}\delta^{(3)}(\pmb{y}-\pmb{z})\nabla_m^{\pmb{z}%
}\varepsilon_{mn}  \notag \\
&\times& \left\{a_n(\pmb{z}),\mathscr{P}^0\right\}-\int d^3yd^3z\nabla_m^{%
\pmb{y}}\varepsilon_{mn}\left\{\mathscr{K}^i_{\Theta^{00}},a_n(\pmb{y}%
)\right\}  \notag \\
&\times&\frac{1}{\nabla^{\pmb{y}}\varepsilon\nabla^{\pmb{y}}}\delta^{(3)}(%
\pmb{y}-\pmb{z})\nabla_k^{\pmb{z}}\left\{\pi_k(\pmb{z}),\mathscr{P}^0\right\}
\end{eqnarray}
Observe that the last integral vanishes identically because $\nabla_k^{%
\pmb{z}}\left\{\pi_k(\pmb{z}),\mathscr{P}^0\right\}=0$ [see Eq.~(\ref%
{PPmomentum})]. The first integral vanishes identically as well because
\begin{eqnarray}
&&\nabla_k^{\pmb{y}}\left\{\mathscr{K}^i_{\Theta^{00}},\pi_k(\pmb{y}%
)\right\}=\nabla_k^{\pmb{y}}\frac{\delta \mathscr{K}^i_{\Theta^{00}}}{\delta
a_k(\pmb{y})}  \notag \\
&&\qquad\qquad=x^i\pmb{\nabla}\cdot(\pmb{\nabla}\times\pmb{h})=0.
\label{identitykpi}
\end{eqnarray}%
We then conclude that $\mathpzc{W}=0$, and the Dirac bracket $\left\{%
\mathscr{K}^i,\mathscr{P}^0\right\}_*$ is just as it appears in Eq.~(\ref%
{boostnewton-law2}).


\section{DERIVING THE ALTERATIONS TO THE  LORENTZ ALGEBRAIC RELATIONS \label%
{MLA}}



\subsection{Dirac bracket between $\pmb{\mathscr{K}}$ and $\pmb{\mathscr{J}}$
\label{appendA}}


In order to determine the Dirac bracket $\left\{\pmb{\mathscr{K}},%
\pmb{\mathscr{J}}\right\}_*$ we consider Eq.~(\ref{a2variation}). This
allows us to express
\begin{eqnarray}  \label{inkj}
\left\{\mathscr{K}^i,\mathscr{J}^j\right\}_*=\left\{\mathscr{K}^i_{%
\mathscr{P}},\mathscr{J}^j\right\}_*+\left\{\mathscr{K}^i_{\Theta^{00}},%
\mathscr{J}^j\right\}_*.
\end{eqnarray}
The first bracket on the right-hand side can be computed by applying Eq.~(%
\ref{preliminary1PBSdf}). Indeed, according to this equation, $\left\{%
\mathscr{P}^i,\mathscr{J}^j\right\}_*=\epsilon^{ijk}\mathscr{P}^k$ provided
that the fields  vanish  at  infinity. Therefore, 
\begin{eqnarray}  \label{jfoopaops,x,q03030}
\left\{\mathscr{K}^i_{\mathscr{P}},\mathscr{J}^j\right\}_*=\epsilon^{ijk}\left[%
-x^0\mathscr{P}^k\right].
\end{eqnarray}

Analogously to Eq.~(\ref{a33}), we split  the second Dirac bracket in Eq.~(%
\ref{inkj}) into three different contributions:
\begin{eqnarray}  \label{inkjkdjgf}
\left\{\mathscr{K}^i_{\Theta^{00}},\mathscr{J}^j\right\}_*=\mathcal{I}_1-%
\mathcal{I}_2+\mathcal{I}_3,
\end{eqnarray}
with
\begin{eqnarray}
\begin{array}{c}
\displaystyle \mathcal{I}_1=\int d^3x \frac{\delta \mathscr{K}%
_{\Theta^{00}}^i}{\delta a_l}\frac{\delta \mathscr{J}^j}{\delta\pi^l}, \ \
\mathcal{I}_2=\int d^3x \frac{\delta \mathscr{K}_{\Theta^{00}}^i}{\delta\pi_l%
}\frac{\delta \mathscr{J}^j}{\delta a^l}, \\
\\
\displaystyle \mathcal{I}_3=-\int d^3yd^3z\left\{\pmb{\mathscr{K}}%
,\varphi_\alpha(\pmb{y})\right\}C_{\alpha\beta}^{-1}(\pmb{y},\pmb{z}%
)\left\{\varphi_\beta(\pmb{z}), \pmb{\mathscr{J}} \right\}.%
\end{array}%
\end{eqnarray}
The explicit insertion of the derivatives involved in $\mathcal{I}_1$ allows us 
to write
\begin{eqnarray}
\mathcal{I}_1&=&\int d^3x\left[\epsilon^{ijm}h^m(\pmb{x}\cdot\pmb{b}%
)-\epsilon^{ilm}h^m b^jx^l-x^i(\pmb{\nabla}\times\pmb{h})^j\right.  \notag \\
&&\quad\left.\times(\pmb{x}\cdot\pmb{b})+x^i(\pmb{\nabla}\times\pmb{h}%
)^lb^jx^l-x^i(\pmb{\nabla}\times\pmb{h})^l\nabla^l(\pmb{x}\times\pmb{a}%
)^j\right.  \notag \\
&&\quad\left.+\epsilon^{ilm}h_m\nabla^l(\pmb{x}\times\pmb{a})^j\right]
\label{larga1}
\end{eqnarray}
Thanks to the vectorial identity given in Eq.~(\ref{vectorailindettzioa}), 
we are able to express the integral of the first four terms in this equation
as
\begin{eqnarray}
&&-\int d^3x\left[x^i\left(h^l(\pmb{x}\times\pmb{\nabla})^jb^l+(\pmb{h}%
\times \pmb{b})^j\right)\right]  \label{identilarga1}
\end{eqnarray}
Note that the fifth and sixth terms in Eq.~(\ref{larga1}) cancel each other, 
since an integration by parts leads to
\begin{eqnarray}
\int d^3 x x^i(\pmb{\nabla}\times\pmb{h})^l\nabla^l(\pmb{x}\times \pmb{a}%
)^j=\int d^3 x \epsilon^{ilm} h^m \nabla_l(\pmb{x}\times \pmb{a})^j.  \notag
\end{eqnarray}
Once this fact is taken into account and Eq.~(\ref{identilarga1}) is
inserted into Eq.~(\ref{larga1}),  its right-hand side acquires the following
structure: 
\begin{eqnarray}
\mathcal{I}_1&=&\int d^3x\left\{-x^i\left[h^l(\pmb{x}\times\pmb{\nabla}%
)^jb^l+(\pmb{h}\times\pmb{b})^j\right]\right\}.  \label{largafinalgthjjl}
\end{eqnarray}
Furthermore, the substitution of Eqs.~(\ref{derivatives1}) and   (\ref%
{derivatives2}) into $\mathcal{I}_2$ yields
\begin{eqnarray}
\mathcal{I}_2&=&\int d^3 x\left\{-x^i\epsilon^{jlk}e^l\nabla^k (\pmb{x}\cdot%
\pmb{\pi})-x^ie^l(\pmb{x}\times\pmb{\nabla})^l\pi^j\right.  \notag \\
&&+\left.x^i\epsilon^{jlk}e^lx^k(\pmb{\nabla}\cdot\pmb{\pi})\right\}.
\end{eqnarray}
Nevertheless, the desirable expression of $\mathcal{I}_2$ is obtained by
expanding the derivative present in the first term and using the identity
given in Eq.~(\ref{vectorailindettzioa}). With these details in mind,  we find
\begin{eqnarray}  \label{muycompacta}
\mathcal{I}_2&=&\int d^3 x\left\{-x^i\left[e^l(\pmb{x}\times\pmb{\nabla}%
)^j\pi^l+(\pmb{e}\times \pmb{\pi})^j\right]\right\}.
\end{eqnarray}

On the other hand, once $\mathcal{I}_3$ is expanded over $\alpha$ and $\beta$, 
one obtains
\begin{eqnarray}
\mathcal{I}_3&=&\int d^3yd^3 z\nabla_k^{\pmb{y}}\left\{\mathscr{K}%
_{\Theta^{00}}^i,\pi_k(\pmb{y})\right\}\frac{1}{\nabla^{\pmb{y}%
}\varepsilon\nabla^{\pmb{y}}}\delta^{(3)}(\pmb{y}-\pmb{z})\nabla_m^{\pmb{z}%
}\varepsilon_{mn}  \notag \\
&\times& \left\{a_n(\pmb{z}),\mathscr{J}^j\right\}-\int d^3yd^3z\nabla_m^{%
\pmb{y}}\varepsilon_{mn}\left\{\mathscr{K}_{\Theta^{00}}^i,a_n(\pmb{y}%
)\right\}  \notag \\
&\times&\frac{1}{\nabla^{\pmb{y}}\varepsilon\nabla^{\pmb{y}}}\delta^{(3)}(%
\pmb{y}-\pmb{z})\nabla_k^{\pmb{z}}\left\{\pi_k(\pmb{z}),\mathscr{J}%
^j\right\},
\end{eqnarray}
where the relevant elements of $C^{-1}_{\alpha\beta}$ [Eq.~(\ref{inversematricsdd}%
)]  were inserted. Thanks to Eq.~(\ref{63}),  both integrals in $\mathcal{I}_3$
vanish identically,  and one ends up with $\mathcal{I}_3=0$. Hence, by
substituting Eqs.~(\ref{largafinalgthjjl}) and  (\ref{muycompacta}) into
Eq.~(\ref{inkjkdjgf}),  we get
\begin{eqnarray}
&&\left\{\mathscr{K}^i,\mathscr{J}^j\right\}_*=\epsilon^{ijk}\mathscr{K}^k
\notag \\
&&\qquad\qquad-\int d^3 x\left\{x^i[(\pmb{\pi}\times\pmb{e})^j+(\pmb{h}\times%
\pmb{b})^j]\right\},   \label{A30}
\end{eqnarray}
where Eq.~(\ref{a2variation}) is  taken into account. We then use the
definition of $\tau^{\mu\nu}$ [Eq.~(\ref{4dtorque})]  to write the second line of
this equation as it stands in Eq.~(\ref{psplitLA1}).


\subsection{Dirac bracket between $\pmb{\mathscr{J}}^i$ and $
\pmb{\mathscr{J}}^j$ \label{appendB}}


We start off by expressing the Dirac bracket between $\pmb{\mathscr{J}}^{i}$
and $\pmb{\mathscr{J}}^{j}$ in terms of two elements
\begin{equation}
\begin{array}{c}
\displaystyle\left\{ \mathscr{J}^{i},\mathscr{J}^{j}\right\} _{\ast
}=I_{1}+I_{2}, \\
\\
\displaystyle I_{1}=\int d^{3}x\left\{ \frac{\delta \mathscr{J}^{i}}{\delta
a^{l}}\frac{\delta \mathscr{J}^{j}}{\delta \pi ^{l}}-\frac{\delta \mathscr{J}%
^{i}}{\delta \pi ^{l}}\frac{\delta \mathscr{J}^{j}}{\delta a^{l}}\right\} ,
\\
\\
\displaystyle I_{2}=-\int d^{3}yd^{3}z\left\{ \pmb{\mathscr{J}},\varphi
_{\alpha }(\pmb{y})\right\} C_{\alpha \beta }^{-1}(\pmb{y},\pmb{z})\left\{
\varphi _{\beta }(\pmb{z}),\pmb{\mathscr{J}}\right\} .%
\end{array}\label{apenb2}
\end{equation}%
An explict substitution of Eq.~(\ref{derivatives1}) into $I_{1}$ allows us  to
write
\begin{eqnarray}
I_{1} &=&\int d^{3}x\left\{ 2\epsilon ^{ijk}\left[ -(\pmb{x}\cdot \pmb{b}%
)\nabla ^{k}(\pmb{x}\cdot \pmb{\pi})+(\pmb{\nabla}\cdot \pmb{\pi})x^{k}(%
\pmb{x}\cdot \pmb{b})\right] \right.  \notag \\
&+&\epsilon ^{ilm}b^{j}x^{l}\nabla ^{m}(\pmb{x}\cdot \pmb{\pi})-\epsilon
^{jlm}b^{i}x^{l}\nabla ^{m}(\pmb{x}\cdot \pmb{\pi})+\epsilon ^{ilm}x^{m}
\notag \\
&\times &(\pmb{\nabla}\cdot \pmb{\pi})\nabla ^{l}(\pmb{x}\times \pmb{a}%
)^{j}-\epsilon ^{jlm}x^{m}(\pmb{\nabla}\cdot \pmb{\pi})\nabla ^{l}(\pmb{x}%
\times \pmb{a})^{i}  \notag \\
&-&(\pmb{x}\cdot \pmb{b})(\pmb{x}\times \pmb{\nabla})^{j}\pi ^{i}-(\pmb{x}%
\cdot \pmb{b})(\pmb{x}\times \pmb{\nabla})^{i}\pi ^{j}-\epsilon ^{ilm}\nabla
^{m}  \notag \\
&\times &(\pmb{x}\cdot \pmb{\pi})\nabla ^{l}(\pmb{x}\times \pmb{a}%
)^{j}+\epsilon ^{jlm}\nabla ^{m}(\pmb{x}\cdot \pmb{\pi})\nabla ^{l}(\pmb{x}%
\times \pmb{a})^{i}  \notag \\
&-&\left. \nabla ^{l}(\pmb{x}\times \pmb{a})^{j}(\pmb{x}\times \pmb{\nabla}%
)^{l}\pi ^{i}+\nabla ^{l}(\pmb{x}\times \pmb{a})^{i}(\pmb{x}\times %
\pmb{\nabla})^{l}\pi ^{j}\right\}  \notag \\
&&
\end{eqnarray}%
where the terms that vanish due to the antisymmetric property of $\epsilon
^{ijk}$ have been omitted. Note that the last four terms of this expression
vanish as well provided an integration by parts. The remaining integrant can
be written as
\begin{eqnarray}
I_{1} &=&\epsilon ^{ijk}\int d^{3}x\left\{ 2\left[ -(\pmb{x}\cdot \pmb{b}%
)\nabla ^{k}(\pmb{x}\cdot \pmb{\pi})+(\pmb{\nabla}\cdot \pmb{\pi})x^{k}(%
\pmb{x}\cdot \pmb{b})\right] \right.  \notag \\
&-&(\pmb{x}\cdot \pmb{\pi})\left[ (\pmb{x}\times \pmb{\nabla})\times \pmb{b}%
\right] ^{k}-(\pmb{\nabla}\cdot \pmb{\pi})\left[ (\pmb{x}\times \pmb{\nabla}%
)\times (\pmb{x}\times \pmb{a})\right] ^{k}  \notag \\
&+&\left. (\pmb{x}\cdot \pmb{b})\left[ (\pmb{x}\times \pmb{\nabla})\times %
\pmb{\pi}\right] ^{k}\right\} .
\end{eqnarray}%
We then use the identity $(\pmb{x}\times \pmb{\nabla})\times \pmb{A}=-\pmb{x}%
\pmb{\nabla}\cdot \pmb{A}+x^{l}\pmb{\nabla}A^{l}.$ As a consequence, 
\begin{eqnarray}
I_{1} &=&\epsilon ^{ijk}\int d^{3}x\left\{ -2(\pmb{x}\cdot \pmb{b})\nabla
^{k}(\pmb{x}\cdot \pmb{\pi})+(\pmb{\nabla}\cdot \pmb{\pi})x^{k}\right.
\notag \\
&\times &(\pmb{x}\cdot \pmb{b})-(\pmb{x}\cdot \pmb{\pi})x^{l}\nabla
^{k}b^{l}-(\pmb{\nabla}\cdot \pmb{\pi})x^{k}(\pmb{x}\cdot \pmb{b})  \notag \\
&-&\left. (\pmb{\nabla}\cdot \pmb{\pi})(\pmb{x}\cdot \pmb{b})x^{l}\nabla
^{k}\left( \pmb{x}\times \pmb{a}\right) +(\pmb{x}\cdot \pmb{b})x^{l}\nabla
^{k}\pi ^{l}\right\} .  \notag \\
&&
\end{eqnarray}%
Expanding the derivative of the first integrand and integrating by parts the
third and fifth terms,  we end up with
\begin{eqnarray}
I_{1} &=&\epsilon ^{ijk}\int d^{3}x\left\{ -(\pmb{x}\cdot \pmb{b})\pi ^{k}+(%
\pmb{x}\cdot \pmb{\pi})b^{k}\right.  \notag  \label{APendkfhfoo} \\
&&+\left. (\pmb{\nabla}\cdot \pmb{\pi})(\pmb{x}\times \pmb{a})^{k}\right\}, 
\end{eqnarray}%
where Eq.~(\ref{apenb1}) was inserted in order to obtain the second line.

Now, it is rather clear that $I_2$ reduces to
\begin{eqnarray}
I_2&=&\int d^3yd^3 z\nabla_k^{\pmb{y}}\left\{\mathscr{J}^i,\pi_k(\pmb{y}%
)\right\}\frac{1}{\nabla^{\pmb{y}}\varepsilon\nabla^{\pmb{y}}}\delta^{(3)}(%
\pmb{y}-\pmb{z})\nabla_m^{\pmb{z}}\varepsilon_{mn}  \notag \\
&\times& \left\{a_n(\pmb{z}),\mathscr{J}^j\right\}-\int d^3yd^3z\nabla_m^{%
\pmb{y}}\varepsilon_{mn}\left\{\mathscr{J}^i,a_n(\pmb{y})\right\}  \notag \\
&\times&\frac{1}{\nabla^{\pmb{y}}\varepsilon\nabla^{\pmb{y}}}\delta^{(3)}(%
\pmb{y}-\pmb{z})\nabla_k^{\pmb{z}}\left\{\pi_k(\pmb{z}),\mathscr{J}%
^j\right\},
\end{eqnarray}
where Eq.~(\ref{inversematricsdd}) is used. Since $\nabla_k^{\pmb{y}}\left\{%
\mathscr{J}^i,\pi_k(\pmb{y})\right\}$ vanishes identically [see Eq.~(\ref{63}%
)],  $I_2$ does not contribute to the Dirac bracket between $\mathscr{J}^i$
and $\mathscr{J}^j.$ Equipped with this result and substituting Eq.~(\ref%
{APendkfhfoo}) into Eq.~(\ref{apenb2}),  we end up with the bracket written in
Eq.~(\ref{psplitLA1}).


\subsection{Dirac bracket between $\pmb{\mathscr{K}}^i$ and $
\pmb{\mathscr{K}}^j$ \label{appendC}}


Let us conclude the derivation of the modified Lorentz algebra by obtaining
the Dirac bracket $\left\{\pmb{\mathscr{K}}^i,\pmb{\mathscr{K}}^j\right\}_*.$
A straightforward substitution of Eq.~(\ref{a2variation}) leads us to write
this as
\begin{eqnarray}
\left\{\mathscr{K}^i,\mathscr{K}^j\right\}_*&=&\left\{\mathscr{K}_{%
\mathscr{P}}^i,\mathscr{K}_{\mathscr{P}}^j\right\}_*+\left\{\mathscr{K}_{%
\mathscr{P}}^i,\mathscr{K}_{\Theta^{00}}^j\right\}_*  \notag \\
&+&\left\{\mathscr{K}_{\Theta^{00}}^i,\mathscr{K}_{\mathscr{P}%
}^j\right\}_*+\left\{\mathscr{K}_{\Theta^{00}}^i,\mathscr{K}%
_{\Theta^{00}}^j\right\}_*.  \label{KKPB}
\end{eqnarray}

The first three brackets are easy to compute. For instance,
\begin{eqnarray}
\left\{\mathscr{K}_{\mathscr{P}}^i,\mathscr{K}_{\mathscr{P}%
}^j\right\}_*=x^{02}\left\{\mathscr{P}^i, \mathscr{P}^j\right\}_*=0
\label{dificil6}
\end{eqnarray}%
where Eq.~(\ref{standardtrasnlationcomuiur}) is  used. Also, by considering
Eq.~(\ref{trucogenial}) and using  integration by parts,  we find that
\begin{eqnarray}
\left\{\mathscr{K}_{\mathscr{P}}^i,\mathscr{K}_{\Theta^{00}}^j\right\}_*=x^0%
\mathscr{P}^0\eta^{ij}.
\end{eqnarray}
As a consequence, 
\begin{eqnarray}
\left\{\mathscr{K}_{\mathscr{P}}^i,\mathscr{K}_{\Theta^{00}}^j\right\}_*+%
\left\{\mathscr{K}_{\Theta^{00}}^i,\mathscr{K}_{\mathscr{P}}^j\right\}_*=0.
\label{dificil5}
\end{eqnarray}

All that remains is  to compute the last bracket on  the right-hand side of Eq.~(\ref%
{KKPB});  i.e.,  the Dirac bracket between $\mathscr{K}_{\Theta^{00}}^i$ and $%
\mathscr{K}_{\Theta^{00}}^j.$ To compute this,  we write
\begin{eqnarray}
\begin{array}{c}
\displaystyle \left\{\mathscr{K}_{\Theta^{00}}^i,\mathscr{K}%
_{\Theta^{00}}^j\right\}_*=\Sigma_1+\Sigma_2, \\
\\
\displaystyle\Sigma_1= \int d^3x\left\{\frac{\delta \mathscr{K}%
_{\Theta^{00}}^i}{\delta a^l}\frac{\delta \mathscr{K}_{\Theta^{00}}^j}{%
\delta\pi^l}-\frac{\delta \mathscr{K}_{\Theta^{00}}^i}{\delta\pi^l}\frac{%
\delta \mathscr{K}_{\Theta^{00}}^j}{\delta a^l}\right\}, \\
\\
\displaystyle \Sigma_2=-\int d^3yd^3z\left\{\pmb{\mathscr{K}}%
_{\Theta^{00}},\varphi_\alpha(\pmb{y})\right\}C_{\alpha\beta}^{-1}(\pmb{y},%
\pmb{z})\left\{\varphi_\beta(\pmb{z}), \pmb{\mathscr{K}}_{\Theta^{00}}
\right\}.%
\end{array}
\notag
\end{eqnarray}%
Inserting Eq.~(\ref{derivatives2}) into $\Sigma_1$,  one finds
\begin{eqnarray}  \label{dificil1}
\Sigma_1&=&\epsilon^{ijk}\int d^3x \left\{-[\pmb{x}\times(\pmb{e}\times%
\pmb{h})]^k\right\}.
\end{eqnarray}
This bracket can be written in a more appropriate form by using the relation
\begin{eqnarray}
(\pmb{e}\times\pmb{h})^i+(\pmb{\pi}\times\pmb{b})^i-a^i \pmb{\nabla}\cdot %
\pmb{\pi}=\tau^{i0},
\end{eqnarray}%
where a term proportional to $a_0$ has  been ignored. Indeed, the
substitution of the expression above into Eq.~(\ref{dificil1}) allows us  to get
\begin{eqnarray}  \label{KKsdasPBv}
\Sigma_1&=&-\epsilon^{ijk}\mathscr{J}^k-\int
d^3x\left(x^i\tau^{j0}-x^j\tau^{i0}\right).  \notag \\
\end{eqnarray}%
Thanks to Eq.~(\ref{inversematricsdd}), $\Sigma_2$ reduces to
\begin{eqnarray}
\Sigma_2&=&\int d^3yd^3 z\nabla_k^{\pmb{y}}\left\{\mathscr{K}^i,\pi_k(\pmb{y}%
)\right\}\frac{1}{\nabla^{\pmb{y}}\varepsilon\nabla^{\pmb{y}}}\delta^{(3)}(%
\pmb{y}-\pmb{z})\nabla_m^{\pmb{z}}\varepsilon_{mn}  \notag \\
&\times& \left\{a_n(\pmb{z}),\mathscr{K}^j\right\}-\int d^3yd^3z\nabla_m^{%
\pmb{y}}\varepsilon_{mn}\left\{\mathscr{K}^i,a_n(\pmb{y})\right\}  \notag \\
&\times&\frac{1}{\nabla^{\pmb{y}}\varepsilon\nabla^{\pmb{y}}}\delta^{(3)}(%
\pmb{y}-\pmb{z})\nabla_k^{\pmb{z}}\left\{\pi_k(\pmb{z}),\mathscr{K}%
^j\right\}.
\end{eqnarray}%
But, since $\nabla_k^{\pmb{y}}\left\{\mathscr{K}^i,\pi_k(\pmb{y})\right\}$
vanishes identically [see Eq.~(\ref{64})],  we can assert that $\Sigma_2$ does
not contribute to the Dirac bracket of $\mathscr{K}^i$ and $\mathscr{K}^j.$
Due to this fact, 
\begin{eqnarray}  \label{KKsdasasdaPBv}
\left\{\mathscr{K}^i,\mathscr{K}^j\right\}_*=-\epsilon^{ijk}\mathscr{J}%
^k-\int d^3x\left(x^i\tau^{j0}-x^j\tau^{i0}\right).  \notag \\
\end{eqnarray}


\section{THE MOMENTUM AND PHASE VELOCITY  OF THE EIGENWAVES  \label{mpvsew}}


We find it convenient to determine a connection between the translational
generator associated with each degree of freedom and its respective phase
velocity. In order to do this,  we express the electric field of each
eigenmode as it is given in Eq.~(\ref{fieldsaveragedfff}). Likewise, the
magnetic field of each eigenwave turns out to be
\begin{eqnarray}  \label{fieldsaveragedfff2}
\pmb{b}^{(\lambda)}(\pmb{x},x^0)=\mathpzc{E}_0^{(\lambda)}\frac{\pmb{b}%
^{(\lambda)}(\pmb{k})}{\vert\pmb{b}^{(\lambda)}(\pmb{k})\vert}\cos[%
\omega_\lambda x^0-\pmb{k}\cdot\pmb{x}].
\end{eqnarray}%
As in Sec. \ref{torquemagnetico}, $\mathpzc{E}_0^{(\lambda)}$ and $%
\omega_\lambda(\pmb{k})$ are the amplitude and  frequency of mode $\lambda$, respectively.  
Besides, whatever the nature of the external field, the unit vectors $\sim%
\pmb{e}^{(\lambda)} (\pmb{k})/\vert\pmb{e}^{(\lambda)}(\pmb{k})\vert$ in Eq.~(%
\ref{fieldsaveragedfff}) and $\sim\pmb{b}^{(\lambda)}(\pmb{k})/\vert\pmb{b}%
^{(\lambda)}(\pmb{k})\vert$ in Eq.~(\ref{fieldsaveragedfff2}) must be understood
as the respective electric and magnetic polarizations. \footnote{%
When the external background is a magneticlike field tensor,  i.e., $\mathfrak{%
F}>0$ and $\mathfrak{G}=0$, the behavior of $\pmb{e}^{(\lambda)}(\pmb{k})$
and $\pmb{b}^{(\lambda)}(\pmb{k})$ can be found below Eq.~(\ref%
{electricpromag}).}  We remark that the plane wave decomposition for the
induction vectors $\pmb{d}$ and $\pmb{h}$ follows from Eqs.~(\ref%
{fieldsaveragedfff}),   (\ref{fieldsaveragedfff2}),  and  (\ref{eq15}).
With these details in mind, the Maxwell equations (\ref{gausslawdiel}) and (%
\ref{maxwell1}) and (\ref{maxwell2}) read
\begin{eqnarray}  \label{momentumME}
\begin{array}{c}
\pmb{k}\cdot\pmb{d}^{(\lambda)}(\pmb{x},x^0)=0, \quad \pmb{k}\times \pmb{e}%
^{(\lambda)}(\pmb{x},x^0)=\omega_\lambda \pmb{b}^{(\lambda)}(\pmb{x},x^0),\\  \\
\pmb{k}\cdot \pmb{b}^{(\lambda)}(\pmb{x},x^0)=0, \quad \pmb{k}\times\pmb{h}%
^{(\lambda)}(\pmb{x},x^0)=-\omega_\lambda \pmb{d}^{(\lambda)}(\pmb{x},x^0).
\end{array}\nonumber\\
\end{eqnarray}
Thanks to the Faraday equation, the momentum associated with each
propagation mode
\begin{equation}  \label{C44444}
\pmb{\mathscr{P}}^{(\lambda)}=\int d^3x\ \left(\pmb{d}^{(\lambda)}\times%
\pmb{b}^{(\lambda)}\right)
\end{equation}%
can be written as
\begin{eqnarray}  \label{avemomentufoeldss}
\pmb{\mathscr{P}}^{(\lambda)}&=&\int d^3 x \frac{1}{\omega_\lambda(\pmb{k})}
\left[\pmb{d}^{(\lambda)}\times\left(\pmb{k}\times\pmb{e}%
^{(\lambda)}\right)\right]  \notag \\
&=&\int d^3x \frac{\pmb{d}^{(\lambda)}\cdot \pmb{e}%
^{(\lambda)}}{ u_\lambda(\pmb{k})}\pmb{n}
\end{eqnarray}
where $u_\lambda=\omega_\lambda/\vert\pmb{k}\vert$ is the phase velocity and
$\pmb{n}=\pmb{k}/\vert\pmb{k}\vert$ denotes the wave vector.

Now, the Ampere law allows us  to obtain the following relation $\pmb{h}%
^{(\lambda)}=u_\lambda \left(\pmb{n}\times \pmb{d}^{(\lambda)}\right)+\left(%
\pmb{n}\cdot \pmb{h}^{(\lambda)}\right)\pmb{n}. $ Multiplying the latter by $%
\pmb{b}$,  we end up with
\begin{eqnarray}
\pmb{h}^{(\lambda)}\cdot\pmb{b}^{(\lambda)}&=&u_\lambda \left(\pmb{n}\times %
\pmb{d}^{(\lambda)}\right)\cdot\pmb{b}^{(\lambda)}=u_\lambda \pmb{n}%
\cdot\left(\pmb{d}^{(\lambda)}\times\pmb{b}^{(\lambda)}\right)  \notag \\
&=&\pmb{d}^{(\lambda)}\cdot \pmb{e}^{(\lambda)}.
\label{relacionimporatnteteee}
\end{eqnarray}
Its substitution into the energy [Eq.~(\ref{transalationalchargesiniciales})] 
yields $\mathscr{P}^{0(\lambda)}=\int d^3x \left[\pmb{d}%
^{(\lambda)}\cdot\pmb{e}^{(\lambda)}\right].\label{energyaveraged} $ We use
this identity to express Eq.~(\ref{avemomentufoeldss}) in the following form: 
\begin{equation}
\pmb{\mathscr{P}}^{(\lambda)}= \frac{\mathscr{P}^{0(\lambda)}}{ u_\lambda(%
\pmb{k})}\pmb{n}.  \label{pfinalpv}
\end{equation}%
Thus, the translation generator associated with each $\Pi_{\mu\nu}$ 
eigenmode turns out to be parallel to the wave vector. Observe, in addition,
that Eq.~(\ref{pfinalpv}) allows to write the phase velocity as $\pmb{u}_\lambda=%
\mathscr{P}^{0(\lambda)}/\vert \pmb{\mathscr{P}}^{(\lambda)} \vert \pmb{n}.$


\section{THE POYNTING VECTOR AND GROUP VELOCITY OF THE EIGENWAVES\label%
{pvgvsew}}


Let us turn our attention to the Poynting vector given in Eq.~(\ref%
{poytinvectordensisty}). In order to simplify our exposition,  we will confine
ourselves to the case in which the external background is a magneticlike
field $(\mathfrak{F}>0,\ \mathfrak{G}=0).$ The results, however, are easily
extensible to the case of an electriclike vector $(\mathfrak{F}<0, \
\mathfrak{G}=0)$. To establish a comparison with the previously discussed
translation generator, it is rather convenient to work with the spatial
integral of Eq.~(\ref{poytinvectordensisty}):
\begin{equation}
\mathpzc{P}^i=\int d^3 x\ \Theta^{i0}(\pmb{x},x^0).
\end{equation}
It also advantageous to express the Poynting vector in terms of $\pmb{\pi}$
and $\pmb{b}.$ This is carried out by inserting the relations $\pmb{e}=-%
\pmb{\pi}/\varepsilon_\perp+ \frac{\mathfrak{L}_{\mathfrak{G}\mathfrak{G}}}{%
\varepsilon_\perp\varepsilon_\parallel}(\pmb{\pi}\cdot\pmb{B})\pmb{B}$ and $%
\pmb{h}=\varepsilon_\perp \pmb{b}-\mathfrak{L}_{\mathfrak{F}\mathfrak{F}%
}\left(\pmb{b}\cdot\pmb{B}\right)\pmb{B}$ into Eq.~(\ref%
{poytinvectordensisty}). Considering these details we obtain
\begin{eqnarray}  \label{dasdcasfasfvfx}
\pmb{\mathpzc{P}}=\pmb{\mathscr{P}}+\pmb{\mathscr{N}}\times\pmb{B}
\end{eqnarray}
where  Eq.~(\ref{transalationalchargesiniciales2}) has been used,  and
\begin{eqnarray}
\pmb{\mathscr{N}}=\int d^3 x\left\{\frac{\mathfrak{L}_{\mathfrak{F}\mathfrak{%
F}}}{\varepsilon_\perp}\left(\pmb{b}\cdot\pmb{B}\right)\pmb{\pi}-\frac{%
\mathfrak{L}_{\mathfrak{G}\mathfrak{G}}}{\varepsilon_\parallel}\left(%
\pmb{\pi}\cdot\pmb{B}\right) \pmb{b}\right\}.  \label{torqueextrage}
\end{eqnarray}
We remark that the last term in Eq.~(\ref{dasdcasfasfvfx}) does not point in
the same direction of $\pmb{k}.$ Of course, each mode has a spatial integral
of the Poynting vector given by $\pmb{\mathpzc{P}}^{(\mathrm{\lambda})}=%
\pmb{\mathscr{P}}^{(\lambda)}+\pmb{\mathscr{N}}^{(\lambda)}\times\pmb{B}$, 
with
\begin{eqnarray}  \label{gugugabi}
\pmb{\mathscr{N}}^{(1)}&=&0, \\
\pmb{\mathscr{N}}^{(2)}&=&-\frac{\mathfrak{L}_{\mathfrak{G}\mathfrak{G}}}{%
\varepsilon_\parallel}\int d^3x \left(\pmb{\pi}^{(2)}\cdot\pmb{B}\right)%
\pmb{b}^{(2)},  \\
\pmb{\mathscr{N}}^{(3)}&=&\frac{\mathfrak{L}_{\mathfrak{F}\mathfrak{F}}}{%
\varepsilon_\perp} \int d^3x \left(\pmb{b}^{(3)}\cdot\pmb{B}\right)\pmb{\pi}%
^{(3)}. \label{gugugabi1}
\end{eqnarray}
To derive these expressions,  we have considered Eq.~(\ref{torqueextrage}) and
the equations below Eq.~(\ref{electricpromag}). One must note that, for the
second and third propagation modes, the following relations hold:  $(\pmb{\pi}%
^{(2)}\cdot\pmb{B})\pmb{b}^{(2)}=-\pmb{B}\times(\pmb{\pi}^{(2)}\times\pmb{b}%
^{(2)})$ and $\ (\pmb{b}^{(3)}\cdot\pmb{B})\pmb{\pi}^{(3)}=\pmb{B}\times(%
\pmb{\pi}^{(3)}\times\pmb{b}^{(3)})$, respectivelly. This  allows us  to express
Eqs.~(\ref{gugugabi})-(\ref{gugugabi1}) as
\begin{eqnarray}  \label{gugugabi2}
\begin{array}{c}
\displaystyle \pmb{\mathscr{N}}^{(1)}=0, \ \ \pmb{\mathscr{N}}^{(2)}=-\frac{%
\mathfrak{L}_{\mathfrak{G}\mathfrak{G}}}{\varepsilon_\parallel}\pmb{B}\times%
\pmb{\mathscr{P}}^{(2)}, \\
\displaystyle \pmb{\mathscr{N}}^{(3)}=-\frac{\mathfrak{L}_{\mathfrak{F}%
\mathfrak{F}}}{\varepsilon_\perp} \pmb{B}\times\pmb{\mathscr{P}}^{(3)},%
\end{array}%
\end{eqnarray}
with $\pmb{\mathscr{P}}^{(\lambda)}$ given in Eq.~(\ref{pfinalpv}). As a
consequence, the spatial integral of the Poynting vector associated with
each eigenmode reads
\begin{eqnarray}  \label{gugugabi3}
\begin{array}{c}
\displaystyle \pmb{\mathpzc{P}}^{(\mathrm{1})}=\pmb{\mathscr{P}}^{(1)}, \ \ %
\pmb{\mathpzc{P}}^{(\mathrm{2})}=\pmb{\mathscr{P}}^{(2)}-\frac{2\mathfrak{F}%
\mathfrak{L}_{\mathfrak{G}\mathfrak{G}}}{\varepsilon_\parallel}%
\pmb{\mathscr{P}}_\perp^{(2)}, \\
\displaystyle \pmb{\mathpzc{P}}^{(\mathrm{3})}=\pmb{\mathscr{P}}^{(3)}-\frac{%
2\mathfrak{F}\mathfrak{L}_{\mathfrak{F}\mathfrak{F}}}{\varepsilon_\perp}%
\pmb{\mathscr{P}}_\perp^{(3)}%
\end{array}%
\end{eqnarray}
In accordance with  the expression above, we can conclude that as long as  
$k_\perp\neq 0$, the direction of the energy propagation in each
physical mode differs from its respective momentum.

To proceed in our analysis,  we consider the center-of-mass energy associated
with the electromagnetic wave. The latter can be defined by
\begin{eqnarray}  \label{centroid}
\pmb{x}_{cm}=\frac{1}{\mathscr{P}^0}\int d^3 x \pmb{x}\Theta^{00}
\end{eqnarray}
where $\Theta^{00}$ and $\mathscr{P}^0$ are given in Eqs.~(\ref{explicit1})
and  (\ref{transalationalchargesiniciales}), respectively. Note that the
derivative with respect to  time allows us  to define the velocity of energy
transport
\begin{eqnarray}
\pmb{v}_{cm}=\frac{d \pmb{x}_{cm}}{dx^0}=\frac{1}{\mathscr{P}^0}\int d^3 x %
\pmb{x}\frac{d\Theta^{00}}{dx^0}
\end{eqnarray}
where the energy conservation $(d\mathscr{P}^0/dx^0=0)$ is taken into
account. Making use of the continuity equation [Eq.~(\ref%
{conservationequataion})] and  integrating by parts,  one obtains
\begin{eqnarray}
\pmb{v}_{cm}=\frac{\pmb{\mathpzc{P}}}{\mathscr{P}^0}=\frac{1}{u}\pmb{n}+%
\frac{\pmb{\mathscr{N}}\times\pmb{B}}{\mathscr{P}^0}  \label{cmvelociv}
\end{eqnarray}
where Eq.~(\ref{dasdcasfasfvfx}) has been considered as well,  and $u=\omega(%
\pmb{k})/\vert\pmb{k}\vert$ denotes the phase velocity of the
small electromagnetic wave. Obviously, the velocity of energy transport
associated with each eigenwave follows from this expression and Eqs.~(\ref%
{gugugabi2}) and (\ref{gugugabi3}). In this context, 
\begin{eqnarray}  \label{modegvet}
\begin{array}{c}
\displaystyle \pmb{v}_{cm 1}=\frac{1}{u_1}\pmb{n},\qquad \pmb{v}_{cm 2}=%
\frac{1}{u_2}\pmb{n}-\frac{2\mathfrak{F}\mathfrak{L}_{\mathfrak{G}\mathfrak{G%
}}}{\varepsilon_\parallel u_{2\perp}}\pmb{n}_\perp, \\
\\
\displaystyle \pmb{v}_{cm 3}=\frac{1}{u_3}\pmb{n}-\frac{2\mathfrak{F}%
\mathfrak{L}_{\mathfrak{F}\mathfrak{F}}}{\varepsilon_\perp u_{3\perp}}\pmb{n}%
_\perp%
\end{array}%
\end{eqnarray}
with $u_{\lambda\perp}=\mathscr{P}^{0(\lambda)}/\vert\pmb{\mathscr{P}}%
_\perp^{(\lambda)}\vert.$

Now, we consider the dispersion equation [Eq.~(\ref{dispequat})] with the infrared
approximation of the vacuum polarization tensor given in Eq.~(\ref{poleig1}).
The corresponding solutions are given by
\begin{eqnarray}  \label{dispersionlaw}
\begin{array}{c}
\omega_1=\vert\pmb{k}\vert, \\
\omega_{2}=\sqrt{\pmb{k}_\parallel^2+\pmb{k}_\perp^2\frac{\mu_\perp^{-1}}{%
\varepsilon_\parallel}},\quad \omega_{3}=\sqrt{\pmb{k}_\parallel^2+\pmb{k}%
_\perp^2\frac{\mu_\parallel^{-1}}{\varepsilon_\perp}}.%
\end{array}%
\end{eqnarray}%
With these expressions in mind, it is a straightforward calculation to show
that the group velocity $\pmb{\mathpzc{v}}_\lambda=\partial\omega_\lambda/%
\partial \pmb{k}$ of each eigenwave  coincides with the respective velocity
of energy transport [Eq.~(\ref{modegvet})].


\begin{thebibliography}{99}
\expandafter\ifx\csname natexlab\endcsname\relax

\fi
\expandafter\ifx\csname bibnamefont\endcsname\relax

\fi
\expandafter\ifx\csname bibfnamefont\endcsname\relax

\fi
\expandafter\ifx\csname citenamefont\endcsname\relax

\fi
\expandafter\ifx\csname url\endcsname\relax

\fi
\expandafter\ifx\csname urlprefix\endcsname\relax

\fi \providecommand{\bibinfo}[2]{#2} \providecommand{\eprint}[2][]{\url{#2}}


\bibitem{batalin} \bibinfo{author}{\bibfnamefont{I.~A.}~%
\bibnamefont{Batalin}} and
\bibinfo{author}{\bibfnamefont{A.~E.}
\bibnamefont{Shabad}}. \bibinfo{journal}{Zh. Eksp. Teor. Fiz.} \textbf{%
\bibinfo{volume}{60}}, \bibinfo{pages}{894} (\bibinfo{year}{1971}).
[\bibinfo{journal}{Sov. Phys. JETP} \textbf{\bibinfo{volume}{33}}, %
\bibinfo{pages}{483} (\bibinfo{year}{1971})].

\bibitem{shabadrc} \bibinfo{author}{\bibfnamefont{A.~E.}~%
\bibnamefont{Shabad}}.
\bibinfo{journal}{Lettere al Nuovo Cimento} \pmb{\bibinfo{volume}{2}}, %
\bibinfo{pages}{457} (\bibinfo{year}{1972}); \bibinfo{author}{%
\bibfnamefont{A.~E.}~\bibnamefont{Shabad}}.
\bibinfo{journal}{Ann. Phys.} \pmb{\bibinfo{volume}{90}}, %
\bibinfo{pages}{166} (\bibinfo{year}{1975}).

\bibitem{shabadpc} \bibinfo{author}{\bibfnamefont{A.~E.}~%
\bibnamefont{Shabad}} and
\bibinfo{author}{\bibfnamefont{V.~V.}
\bibnamefont{Usov}}.
\bibinfo{journal}{Nature (London)}, \pmb{\bibinfo{volume}{295}}, %
\bibinfo{pages}{215}, (\bibinfo{year}{1982}); \bibinfo{author}{%
\bibfnamefont{A.~E.}~\bibnamefont{Shabad}} and \bibinfo{author}{%
\bibfnamefont{V.~V.} \bibnamefont{Usov}}.
\bibinfo{journal}{Astrophys. Space Sci.}, \pmb{\bibinfo{volume}{117}}, %
\bibinfo{pages}{309}, (\bibinfo{year}{1985}); \pmb{\bibinfo{volume}{128}}, %
\bibinfo{pages}{377}, (\bibinfo{year}{1986}); \bibinfo{author}{%
\bibfnamefont{H.}~\bibnamefont{Herold}}, \bibinfo{author}{\bibfnamefont{H.}~%
\bibnamefont{Ruder}} and \bibinfo{author}{\bibfnamefont{G.}~%
\bibnamefont{Wunner}}.
\bibinfo{journal}{Phys. Rev. Lett.}, \pmb{\bibinfo{volume}{54}}, %
\bibinfo{pages}{1452}, (\bibinfo{year}{1985}); \bibinfo{author}{%
\bibfnamefont{V.~V.}~\bibnamefont{Usov}} and \bibinfo{author}{%
\bibfnamefont{D.~B.}~\bibnamefont{Melrose}}.
\bibinfo{journal}{Aust. J.
Phys.}, \pmb{\bibinfo{volume}{48}}, \bibinfo{pages}{571}, (%
\bibinfo{year}{1995});

\bibitem{shabad5} \bibinfo{author}{\bibfnamefont{A.~E.} \bibnamefont{Shabad}}
and \bibinfo{author}{\bibfnamefont{V.~V.} \bibnamefont{Usov}}. %
\bibinfo{journal}{Phys. Rev. Lett.} \pmb{\bibinfo{volume}{98}}, %
\bibinfo{pages}{180403} (\bibinfo{year}{2007}). [arXiv:astro-ph/0607499]. %
\bibinfo{author}{\bibfnamefont{A.~E.}~\bibnamefont{Shabad}} and %
\bibinfo{author}{\bibfnamefont{V.~V.} \bibnamefont{Usov}}.
\bibinfo{journal}{Phys. Rev. D} \pmb{\bibinfo{volume}{77}}, %
\bibinfo{pages}{025001} (\bibinfo{year}{2008}). \bibinfo{author}{%
\bibfnamefont{A.~E.} \bibnamefont{Shabad}} and \bibinfo{author}{%
\bibfnamefont{V.~V.} \bibnamefont{Usov}}.
\emph{%
\bibinfo{title}{``String-Like electrostatic Interaction from QED with
Infinite Magnetic Field.''}} in: ``Particle Physics on the Eve of LHC"
(Proc. of the 13th Lomonosov Conference on Elementary Particle Physics,
Moscow, August 2007), \bibinfo{editor}{World Scientific, Singapore}, %
\bibinfo{pages}{392} (\bibinfo{year}{2009}). arXiv:0801.0115 [hep-th].

\bibinfo{author}{\bibfnamefont{N.} \bibnamefont{Sadooghi}} and
\bibinfo{author}{\bibfnamefont{A.}~\bibnamefont{Sodeiri}
\bibnamefont{Jalili}}.
\bibinfo{journal}{Phys, Rev. D.} \pmb{\bibinfo{volume}{76}}, %
\bibinfo{pages}{065013} (\bibinfo{year}{2007}). [arXiv:0705.4384 [hep-th]]; %
\bibinfo{author}{\bibfnamefont{B.}~\bibnamefont{Machet}} and %
\bibinfo{author}{\bibfnamefont{M.}~\bibnamefont{I.} \bibnamefont{ Vysotsky}}.%
\bibinfo{journal}{Phys, Rev. D.} \pmb{\bibinfo{volume}{83}}, %
\bibinfo{pages}{025022} (\bibinfo{year}{2011}).

\bibitem{PRD2010} \bibinfo{author}{\bibfnamefont{A.~E.}~\bibnamefont{Shabad}}
and \bibinfo{author}{\bibfnamefont{V.~V.}~\bibnamefont{Usov}}.
\bibinfo{journal}{Phys. Rev. D} \pmb{\bibinfo{volume}{81}}, %
\bibinfo{pages}{125008} (\bibinfo{year}{2010});

\bibitem{adler} \bibinfo{author}{\bibfnamefont{S.}~\bibnamefont{L.}~%
\bibnamefont{Adler}},
\bibinfo{author}{\bibfnamefont{J.~N.}
\bibnamefont{Bahcall}},
\bibinfo{author}{\bibfnamefont{C.~G.}
\bibnamefont{Callan}} and
\bibinfo{author}{\bibfnamefont{M.~N.}
\bibnamefont{Rosenbluth}}.
\bibinfo{journal}{Phys. Rev. Lett.} \pmb{\bibinfo{volume}{25}}, %
\bibinfo{pages}{1061} (\bibinfo{year}{1970}); \bibinfo{author}{%
\bibfnamefont{S.}~\bibnamefont{L.}~\bibnamefont{Adler}}.
\bibinfo{journal}{ Ann. Phys.} \pmb{\bibinfo{volume}{67}}, %
\bibinfo{pages}{599} (\bibinfo{year}{1971}); \bibinfo{author}{%
\bibfnamefont{S.}~\bibnamefont{L.}~\bibnamefont{Adler}} and %
\bibinfo{author}{\bibfnamefont{C.}~\bibnamefont{Schubert}}.
\bibinfo{journal}{Phys. Rev. Lett.} \pmb{\bibinfo{volume}{77}}, %
\bibinfo{pages}{1695} (\bibinfo{year}{1996}). [arXiv:hep-th/9605035].

\bibitem{gitshab} \bibinfo{author}{\bibfnamefont{D.}~\bibnamefont{M.}~%
\bibnamefont{Gitman}} and \bibinfo{author}{\bibfnamefont{A.~E.}~%
\bibnamefont{Shabad}}, to be published, arXiv:1209.6289.

\bibitem{stern} \bibinfo{author}{\bibfnamefont{A.}~\bibnamefont{Stern}}. %
\bibinfo{journal}{Phys. Rev. Lett.} \pmb{\bibinfo{volume}{100}}, %
\bibinfo{pages}{061601} (\bibinfo{year}{2008}); For the magnetic moment
carried by a static charge in noncommutative electrodynamics and its
significance see \bibinfo{author}{\bibfnamefont{T.~C.}~\bibnamefont{Adorno}}%
, \bibinfo{author}{\bibfnamefont{D.}~\bibnamefont{M.}~\bibnamefont{Gitman}}, %
\bibinfo{author}{\bibfnamefont{A.~E.}~\bibnamefont{Shabad}} and %
\bibinfo{author}{\bibfnamefont{D.~V.}~\bibnamefont{Vassilevich}}. %
\bibinfo{journal}{Phys. Rev. D.} \pmb{\bibinfo{volume}{84}}, %
\bibinfo{pages}{065031} (\bibinfo{year}{2011}); \textbf{\textit{ibid}}
\textbf{84}, 065003 (2011).  \bibinfo{author}{\bibfnamefont{T.~C.}~\bibnamefont{Adorno}}%
, \bibinfo{author}{\bibfnamefont{D.}~\bibnamefont{M.}~\bibnamefont{Gitman}} and  %
\bibinfo{author}{\bibfnamefont{A.~E.}~\bibnamefont{Shabad}}. %
\bibinfo{journal}{Phys. Rev. D.} \pmb{\bibinfo{volume}{86}}, %
\bibinfo{pages}{027702} (\bibinfo{year}{2012}). 
\bibinfo{author}{\bibfnamefont{F.}~\bibnamefont{Riad}} and  
\bibinfo{author}{\bibfnamefont{M.}~\bibnamefont{M.}~\bibnamefont{Sheikh-Jabbari}}.
\bibinfo{journal}{JHEP} \pmb{\bibinfo{volume}{008}}, %
\bibinfo{pages}{045} (\bibinfo{year}{2000}).  
\bibinfo{author}{\bibfnamefont{A.}~\bibnamefont{Mazumdar}} and
\bibinfo{author}{\bibfnamefont{M.}~\bibnamefont{M.}~\bibnamefont{Sheikh-Jabbari}}.
\bibinfo{journal}{Phys. Rev. Lett.} \pmb{\bibinfo{volume}{87}}, %
\bibinfo{pages}{011301} (\bibinfo{year}{2001}). 


\bibitem{KosteleckyI} \bibinfo{author}{\bibfnamefont{V.~A.}~%
\bibnamefont{Colladay}} and \bibinfo{author}{\bibfnamefont{V.~A.}~%
\bibnamefont{Kostelecky}}.
\bibinfo{journal}{Phys.\ Rev.\ D } \pmb{\bibinfo{volume}{58}}, %
\bibinfo{pages}{116002} (\bibinfo{year}{1998}).

\bibitem{Kostelecky:2007zz} \bibinfo{author}{\bibfnamefont{V.~A.}~%
\bibnamefont{Kostelecky}} and \bibinfo{author}{\bibfnamefont{M.}~%
\bibnamefont{Mewes}}.
\bibinfo{journal}{Phys. Rev. D.} \pmb{\bibinfo{volume}{66}}, %
\bibinfo{pages}{056005} (\bibinfo{year}{2002}); [arXiv:hep-ph/0205211].

\bibitem{Kostelecky} \bibinfo{author}{\bibfnamefont{V.~A.}~%
\bibnamefont{Kostelecky}} and \bibinfo{author}{\bibfnamefont{M.}~%
\bibnamefont{Mewes}}.
\bibinfo{journal}{Phys. Rev. Lett.} \pmb{\bibinfo{volume}{99}}, %
\bibinfo{pages}{011601} (\bibinfo{year}{2007}); [arXiv:astro-ph/0702379]. %
\bibinfo{author}{\bibfnamefont{V.~A.}~\bibnamefont{Kostelecky}} and %
\bibinfo{author}{\bibfnamefont{M.}~\bibnamefont{Mewes}}.
\bibinfo{journal}{Phys. Rev. Lett.} \pmb{\bibinfo{volume}{87}}, %
\bibinfo{pages}{251304} (\bibinfo{year}{2001}); [arXiv:hep-ph/0111026]. %
\bibinfo{author}{\bibfnamefont{A.~G.~M.}~\bibnamefont{Pickering}}.
\bibinfo{journal}{Phys. Rev. Lett.} \pmb{\bibinfo{volume}{91}}, %
\bibinfo{pages}{031801} (\bibinfo{year}{2003}). [arXiv:hep-ph/0212382].

\bibitem{carroll} \bibinfo{author}{\bibfnamefont{S.~M.}~%
\bibnamefont{Carroll}}, \bibinfo{author}{\bibfnamefont{J.~A.}~%
\bibnamefont{Harvey}}, \bibinfo{author}{\bibfnamefont{V.~A.}~%
\bibnamefont{Kostelecky}}, \bibinfo{author}{\bibfnamefont{C.~D.}~%
\bibnamefont{Lane}} and \bibinfo{author}{\bibfnamefont{T.}~%
\bibnamefont{Okamoto}}.
\bibinfo{journal}{Phys. Rev. lett.} \pmb{\bibinfo{volume}{87}}, %
\bibinfo{pages}{141601} (\bibinfo{year}{2001}). [arXiv:hep-th/0105082];
\bibinfo{author}{\bibfnamefont{Z.}~\bibnamefont{Guralnik}}, %
\bibinfo{author}{\bibfnamefont{R.}~\bibnamefont{Jackiw}}, %
\bibinfo{author}{\bibfnamefont{S.~Y.}~\bibnamefont{Pi}} and %
\bibinfo{author}{\bibfnamefont{A.~P.}~\bibnamefont{Polychronakos}}.
\bibinfo{journal}{Phys.  lett. B} \pmb{\bibinfo{volume}{517}}, %
\bibinfo{pages}{450} (\bibinfo{year}{2001}). [arXiv:hep-th/0106044].
\bibinfo{author}{\bibfnamefont{A.}~\bibnamefont{Anisimov}}, %
\bibinfo{author}{\bibfnamefont{T.}~\bibnamefont{Banks}}, \bibinfo{author}{%
\bibfnamefont{M.}~\bibnamefont{Dine}} and \bibinfo{author}{%
\bibfnamefont{M.}~\bibnamefont{Graesser}}.
\bibinfo{journal}{Phys. Rev.  D } \pmb{\bibinfo{volume}{65}}, %
\bibinfo{pages}{085032} (\bibinfo{year}{2002}). [arXiv:hep-ph/0106356].

\bibitem{Gies:2007ua} \bibinfo{author}{\bibfnamefont{R.~D.}~%
\bibnamefont{Peccei}} and \bibinfo{author}{\bibfnamefont{H.~R.}~%
\bibnamefont{Quinn}}.
\bibinfo{journal}{Phys.  Rev.  Lett.} \pmb{\bibinfo{volume}{38}}, %
\bibinfo{pages}{1440} (\bibinfo{year}{1977}); \bibinfo{author}{%
\bibfnamefont{P.}~\bibnamefont{Sikivie}}.
\bibinfo{journal}{Phys.  Rev.  Lett.} \pmb{\bibinfo{volume}{51}}, %
\bibinfo{pages}{1415} (\bibinfo{year}{1983}); \bibinfo{author}{%
\bibfnamefont{P.}~\bibnamefont{Sikivie}}.
\bibinfo{journal}{Phys.  Rev.  D.} \pmb{\bibinfo{volume}{32}}, %
\bibinfo{pages}{2988} (\bibinfo{year}{1985}); \bibinfo{author}{%
\bibfnamefont{H.}~\bibnamefont{Gies}}.
\bibinfo{journal}{J.  Phys. A} \pmb{\bibinfo{volume}{41}}, %
\bibinfo{pages}{164039} (\bibinfo{year}{2008}). [arXiv:0711.1337 [hep-ph]]; %
\bibinfo{author}{\bibfnamefont{M.}~\bibnamefont{Ahlers}}, %
\bibinfo{author}{\bibfnamefont{H.}~\bibnamefont{Gies}}, \bibinfo{author}{%
\bibfnamefont{J.}~\bibnamefont{Jaeckel}}, \bibinfo{author}{%
\bibfnamefont{J.}~\bibnamefont{Redondo}}, and \bibinfo{author}{%
\bibfnamefont{A.}~\bibnamefont{Ringwald}}.
\bibinfo{journal}{Phys. Rev. D.} \pmb{\bibinfo{volume}{77}}, %
\bibinfo{pages}{ 095001} (\bibinfo{year}{2008}). [arXiv:0711.4991 [hep-ph]]; %
\bibinfo{author}{\bibfnamefont{S.~L.}~\bibnamefont{Adler}}, %
\bibinfo{author}{\bibfnamefont{J.}~\bibnamefont{Gamboa}}, %
\bibinfo{author}{\bibfnamefont{F.}~\bibnamefont{Mendez}} and %
\bibinfo{author}{\bibfnamefont{J.}~\bibnamefont{Lopez-Sarrion}}.
\bibinfo{journal}{Ann. Phys.} \pmb{\bibinfo{volume}{323}}, %
\bibinfo{pages}{2851} (\bibinfo{year}{2008}); [arXiv:0801.4739 [hep-ph]].

\bibitem{bastianelli1} 
\bibinfo{author}{\bibfnamefont{M.~E.} \bibnamefont{Gertsenshtein}}.
\bibinfo{journal}{Sov. Phys.  JETP} \pmb{\bibinfo{volume}{14}}, %
\bibinfo{pages}{84} (\bibinfo{year}{1962}). 
\bibinfo{author}{\bibfnamefont{G.} \bibnamefont{Raffelt}} and %
\bibinfo{author}{\bibfnamefont{L.}~\bibnamefont{Stodolsky}}.
\bibinfo{journal}{ Phys. Rev.  D} \pmb{\bibinfo{volume}{37}}, %
\bibinfo{pages}{1237} (\bibinfo{year}{1988}). \bibinfo{author}{%
\bibfnamefont{F.} \bibnamefont{Bastianelli}} and \bibinfo{author}{%
\bibfnamefont{C.}~\bibnamefont{Schubert}}.
\bibinfo{journal}{JHEP} \pmb{\bibinfo{volume}{02}}, \bibinfo{pages}{069} (%
\bibinfo{year}{2005}); [arXiv:gr-qc/0412095]. \bibinfo{author}{%
\bibfnamefont{F.} \bibnamefont{Bastianelli}}, \bibinfo{author}{%
\bibfnamefont{U.} \bibnamefont{Nucamendi}}, \bibinfo{author}{%
\bibfnamefont{C.}~\bibnamefont{Schubert}} and \bibinfo{author}{%
\bibfnamefont{V.~M.}~\bibnamefont{Villanueva}}.
\bibinfo{journal}{JHEP} \pmb{\bibinfo{volume}{11}}, \bibinfo{pages}{099} (%
\bibinfo{year}{2007}); [arXiv:0710.5572 [gr-qc]]. \bibinfo{author}{%
\bibfnamefont{F.} \bibnamefont{Bastianelli}}, \bibinfo{author}{%
\bibfnamefont{U.} \bibnamefont{Nucamendi}}, \bibinfo{author}{%
\bibfnamefont{C.}~\bibnamefont{Schubert}} and \bibinfo{author}{%
\bibfnamefont{V.~M.}~\bibnamefont{Villanueva}}.
\bibinfo{journal}{J. Phys.
A: Math. Theor.}
\bibinfo{pages}{164048} (\bibinfo{year}{2008}); [arXiv:0809.0652 [hep-th]].
\bibinfo{author}{\bibfnamefont{C.} \bibnamefont{Biggio}}, %
\bibinfo{author}{\bibfnamefont{E.} \bibnamefont{Masso}} and %
\bibinfo{author}{\bibfnamefont{J.}~\bibnamefont{Redondo}}. %
\bibinfo{journal}{Phys. Rev.  D.},
\bibinfo{pages}{015012} (\bibinfo{year}{2009}); [arXiv:hep-ph/0604062].

\bibitem{ELI} European Light Infrastructure (ELI): http://\newline
www.extreme-light-infrastructure.eu

\bibitem{hiper} See: http://www.hiperlaser.org/index.asp

\bibitem{Manchester} \bibinfo{author}{\bibfnamefont{R.}~\bibfnamefont{M.}~%
\bibnamefont{Manchester}},
\bibinfo{author}{\bibfnamefont{G.~B.}
\bibnamefont{Hobbs}}, \bibinfo{author}{\bibfnamefont{A.} \bibnamefont{Teoh}}
and \bibinfo{author}{\bibfnamefont{M.} \bibnamefont{Hobbs}}.
\bibinfo{journal}{Astron. J.} \textbf{\bibinfo{volume}{129}}, %
\bibinfo{pages}{1993} (\bibinfo{year}{2005}). 
\bibinfo{author}{\bibfnamefont{C.}~\bibnamefont{Kouveliotou} \emph{et al}}.
\bibinfo{journal}{Nature} \textbf{\bibinfo{volume}{393}}, %
\bibinfo{pages}{235} (\bibinfo{year}{1998}). 
\bibinfo{author}{\bibfnamefont{J.}~\bibfnamefont{S.}~\bibnamefont{Bloom}}, %
\bibinfo{author}{\bibfnamefont{S.~R.} \bibnamefont{Kulkarni}}, %
\bibinfo{author}{\bibfnamefont{F.~A.} \bibnamefont{Harrison}}, %
\bibinfo{author}{\bibfnamefont{T.} \bibnamefont{Prince}} \bibinfo{author}{%
\bibfnamefont{E.~S.} \bibnamefont{Phinney}} and \bibinfo{author}{%
\bibfnamefont{D.~A.} \bibnamefont{Frail}}.
\bibinfo{journal}{Astrophys. J.} \textbf{\bibinfo{volume}{506}}, %
\bibinfo{pages}{L105} (\bibinfo{year}{1998}).

\bibitem{AFO86} \bibinfo{author}{\bibfnamefont{C.~} \bibnamefont{Alcock}}, %
\bibinfo{author}{\bibfnamefont{E.~} \bibnamefont{Farhi}} and %
\bibinfo{author}{\bibfnamefont{A.~} \bibnamefont{Olinto}}. %
\bibinfo{journal}{Astrophys. J.}, \textbf{\bibinfo{volume}{310}}, %
\bibinfo{pages}{261}, (\bibinfo{year}{1986}); 
\bibinfo{author}{\bibfnamefont{Ch.~} \bibnamefont{Kettner}}, %
\bibinfo{author}{\bibfnamefont{F.~} \bibnamefont{Weber}}, %
\bibinfo{author}{\bibfnamefont{M.~K.~} \bibnamefont{Weigel}} and %
\bibinfo{author}{\bibfnamefont{N.~K.~} \bibnamefont{Glendenning}}. %
\bibinfo{journal}{Phys. Rev. D}, \textbf{\bibinfo{volume}{51}}, %
\bibinfo{pages}{1440}, (\bibinfo{year}{1995});
\bibinfo{author}{\bibfnamefont{V.~V.~} \bibnamefont{Usov}}. %
\bibinfo{journal}{Phys. Rev. D}, \textbf{\bibinfo{volume}{70}}, %
\bibinfo{pages}{067301}, (\bibinfo{year}{2004}); 
\bibinfo{author}{\bibfnamefont{V.~V.~} \bibnamefont{Usov}}, %
\bibinfo{author}{\bibfnamefont{T.~} \bibnamefont{Harko}} and %
\bibinfo{author}{\bibfnamefont{K.~S.~} \bibnamefont{Cheng}}. %
\bibinfo{journal}{ Astrophys. J.}, \textbf{\bibinfo{volume}{620}}, %
\bibinfo{pages}{915}, (\bibinfo{year}{2005}).

\bibitem{Dirac:1958sq} \bibinfo{author}{\bibfnamefont{P.~A.~M.}~%
\bibnamefont{Dirac}}.
\bibinfo{journal}{Proc. Roy. Soc. Lond. A}, \pmb{\bibinfo{volume}{246}}, %
\bibinfo{pages}{326} (\bibinfo{year}{1958}).

\bibitem{books} \bibinfo{author}{\bibfnamefont{D.~M.}~\bibnamefont{Gitman}}
and \bibinfo{author}{\bibfnamefont{I.~V.}~\bibnamefont{Tyutin}} \emph{%
\bibinfo{Title}{``Quantization of Fields with Constraints.''}} %
\bibinfo{publisher}{Springer}, (\bibinfo{year}{1990}). \bibinfo{author}{%
\bibfnamefont{M.}~\bibnamefont{Henneaux}} and \bibinfo{author}{%
\bibfnamefont{C.}~\bibnamefont{Teitelboim}} \emph{%
\bibinfo{title}{``Quantization of Gauge Systems.''}} %
\bibinfo{publisher}{Princeton Univ. Press}, (\bibinfo{year}{1992}).

\bibitem{Burnel:2008zz} \bibinfo{author}{\bibfnamefont{A.}~%
\bibnamefont{Burnel}} \emph{%
\bibinfo{Title}{``Noncovariant gauges in
canonical formalism.''}} \bibinfo{publisher}{Springer}, (\bibinfo{year}{2008}%
).

\bibitem{Besting:1989nq} \bibinfo{author}{\bibfnamefont{P.}~%
\bibnamefont{Besting}} and \bibinfo{author}{\bibfnamefont{D.}~%
\bibnamefont{Schutte}}.
\bibinfo{journal}{Phys. Rev. D} \pmb{\bibinfo{volume}{42}}, %
\bibinfo{pages}{594}, (\bibinfo{year}{1990}); 
\bibinfo{author}{\bibfnamefont{M.~G.}~\bibnamefont{Rocha}}, %
\bibinfo{author}{\bibfnamefont{F.~J.}~\bibnamefont{Llanes-Estrada}}, %
\bibinfo{author}{\bibfnamefont{D.}~\bibnamefont{Schuette}} and %
\bibinfo{author}{\bibfnamefont{S.}~\bibnamefont{Villalba-Ch\'avez}}.
\bibinfo{journal}{Eur. Phys. Jour. A} \pmb{\bibinfo{volume}{44}}, \ bibinfo{%
pages}{411}, (\bibinfo{year}{2010}). arXiv:0910.1448 [hep-ph].

\bibitem{selhugo} S. Villalba Ch$\acute{\mathrm{a}}$vez and H. P$\acute{%
\mathrm{e}}$rez-Rojas, arXiv:0604059, 0609008 [hep-th].

\bibitem{hugoelsel} H. P$\acute{\mathrm{e}}$rez-Rojas and E.R. Querts,
Int.J.Mod.Phys. {A} \textbf{21}, 3761 (2006), D \textbf{16}, 165 (2007);
Phys.Rev. D \textbf{79}, 093002 (2009); arXiv:0808.2558[hep-ph];
arXiv:1002.3269[hep-ph]; arXiv:1007.4841[hep-ph].

\bibitem{Chavez:2009ia} \bibinfo{author}{\bibfnamefont{S.}~%
\bibnamefont{Villalba}~\bibnamefont{Ch\'avez}}.
\bibinfo{journal}{Phys. Rev. D} \pmb{\bibinfo{volume}{81}}, %
\bibinfo{pages}{105019}, (\bibinfo{year}{2010}). arXiv:0910.5149 [hep-th].


\bibitem{Weinberg:1995mt} \bibinfo{author}{\bibfnamefont{S.}~%
\bibnamefont{Weinberg}}. \emph{%
\bibinfo{journal}{``The Quantum theory of
fields.''}} \bibinfo{editor}{Cambridge, UK: Univ. Pr.}, (\bibinfo{year}{2001}%
), 609 p.

\bibitem{shabadpolarization} \bibinfo{author}{\bibfnamefont{A.}~%
\bibnamefont{E.}~\bibnamefont{Shabad}}. \emph{%
\bibinfo{title}{``Polarization
of the vacuum  and quantum  relativistic gas in an external field.''}} %
\bibinfo{publisher}{Nova Science Publishers, New York,} (\bibinfo{year}{1991}%
). [\bibinfo{journal}{Trudy Fiz. Inst. im. P.N. Lebedeva} \textbf{%
\bibinfo{tom}{192}}, \bibinfo{page}{5} (\bibinfo{year}{1988})].

\bibitem{PRD2011} \bibinfo{author}{\bibfnamefont{A.~E.}~\bibnamefont{Shabad}}
and \bibinfo{author}{\bibfnamefont{V.~V.}~\bibnamefont{Usov}}.
\bibinfo{journal}{Phys. Rev. D} \pmb{\bibinfo{volume}{83}}, %
\bibinfo{pages}{105006} (\bibinfo{year}{2011}); [arXiv:1101.2343 [hep-th]].
\bibinfo{author}{\bibfnamefont{A.~E.}~\bibnamefont{Shabad}} and %
\bibinfo{author}{\bibfnamefont{V.~V.}~\bibnamefont{Usov}} in \emph{%
\bibinfo{title}{``Convexity of effective Lagrangian in
nonlinear electrodynamics as derived  from causality.''}} arXiv:0911.0640
[hep-th].

\bibitem{footnote} The corresponding equations relating to the most general
case $\mathfrak{F}\neq0,~\mathfrak{G}\neq0$ are written in \cite{batalin},
\cite{shabadpolarization}, \cite{PRD2010}

\bibitem{dipiazza} \bibinfo{author}{\bibfnamefont{A.~Di}~%
\bibnamefont{Piazza}} and \bibinfo{author}{\bibfnamefont{G.}~%
\bibnamefont{Calucci}}.
\bibinfo{journal}{Phys. Rev. D} \pmb{\bibinfo{volume}{66}}, %
\bibinfo{pages}{123006} (\bibinfo{year}{2002}).


\bibitem{Bacry} \bibinfo{author}{\bibfnamefont{H.}~\bibnamefont{Bacry}}, %
\bibinfo{author}{\bibfnamefont{P.}~\bibnamefont{Combe}} and %
\bibinfo{author}{\bibfnamefont{J.}~\bibnamefont{L.}~\bibnamefont{Richard}}.
\bibinfo{journal}{Nuovo Cim.} \pmb{\bibinfo{volume}{A67}}, %
\bibinfo{pages}{267} (\bibinfo{year}{1970}).

\bibitem{shabad4} \bibinfo{author}{\bibfnamefont{A.~E.} \bibnamefont{Shabad}}.%
\bibinfo{journal}{Sov. Phys. JETP} \pmb{\bibinfo{volume}{98}}, %
\bibinfo{pages}{186} (\bibinfo{year}{2004}).

\bibitem{iwo} \bibinfo{author}{\bibfnamefont{I.}~\bibfnamefont{Bialynicki}~%
\bibnamefont{Birula}} and \bibinfo{author}{\bibfnamefont{Z.}~%
\bibfnamefont{Bialynicki}~\bibnamefont{Birula}}. \emph{%
\bibinfo{Title}{``Quantum  Electrodynamics.''}}
\bibinfo{editor}{Pergamon,
Oxford}, (\bibinfo{year}{1975}); 
\bibinfo{author}{\bibfnamefont{J.}~\bibfnamefont{D.}~\bibnamefont{Jackson}},
\emph{\bibinfo{Title}{``Classical Electrodynamics.''}}
\bibinfo{editor}{New
York: John Wiley}, (\bibinfo{year}{1999}).

\bibitem{euler} \bibinfo{author}{\bibfnamefont{W.}~\bibnamefont{Heisenberg}}
and \bibinfo{author}{\bibfnamefont{H.} \bibnamefont{Euler}}.
\bibinfo{journal}{Z. Phys.} \pmb{\bibinfo{volume}{98}}, \bibinfo{pages}{714}
(\bibinfo{year}{1936}).

\bibitem{Schwinger} \bibinfo{author}{\bibfnamefont{J.}~%
\bibnamefont{Schwinger}}.
\bibinfo{journal}{Phys. Rev.} \pmb{\bibinfo{volume}{82}}, %
\bibinfo{pages}{664}, (\bibinfo{year}{1951})

\bibitem{ritus} 
\bibinfo{author}{\bibfnamefont{W.}~\bibnamefont{Dittrich}} and %
\bibinfo{author}{\bibfnamefont{M.} \bibnamefont{Reuter}}. \emph{%
\bibinfo{title}{``Effective Lagrangians in Quantum Electrodynamics.''}} %
\bibinfo{book}{Springer}, (\bibinfo{year}{1985}); \bibinfo{author}{%
\bibfnamefont{V. I.}~\bibnamefont{Ritus}}.
\bibinfo{journal}{Sov. Phys. JETP} \pmb{\bibinfo{volume}{42}}, %
\bibinfo{pages}{774} (\bibinfo{year}{1976}); 
\bibinfo{author}{\bibfnamefont{V. I.}~\bibnamefont{Ritus}}.
\bibinfo{journal}{Proceedings of Workshop on Frontier Tests of Quantum
Electrodynamics and Physics of the Vacuum, Sandansky, Bulgaria.}(%
\bibinfo{year}{1998}). hep-th/9812124,

\bibitem{VillalbaChavez:2010bp} \bibinfo{author}{\bibfnamefont{S.}~%
\bibnamefont{Villalba}~\bibnamefont{Ch\'avez}}.
\bibinfo{journal}{Phys. Lett.  B } \pmb{\bibinfo{volume}{692}}, %
\bibinfo{pages}{317}, (\bibinfo{year}{2010}): [arXiv:1008.0547 [hep-th]].


\bibitem{Kostelecky:2009} \bibinfo{author}{\bibfnamefont{V.~A.}~%
\bibnamefont{Kostelecky}} and \bibinfo{author}{\bibfnamefont{M.}~%
\bibnamefont{Mewes}}.
\bibinfo{journal}{Phys. Rev. D.} \pmb{\bibinfo{volume}{80}}, %
\bibinfo{pages}{015020} (\bibinfo{year}{2009}).


\bibitem{landau} \bibinfo{author}{\bibfnamefont{V.~B.}~%
\bibfnamefont{Berestetsky}}, \bibinfo{author}{\bibfnamefont{E.~M.}~%
\bibfnamefont{Lifshits}} and \bibinfo{author}{\bibfnamefont{L.~P.}~%
\bibnamefont{Pitayevsky}}. \emph{%
\bibinfo{Title}{``Quantum
Electrodynamics.''}} \bibinfo{editor}{Pergamon Press  Oxford, New York}, (%
\bibinfo{year}{1982}).


\bibitem{Klinkhamer:2008ky} \bibinfo{author}{\bibfnamefont{F.~R.}~%
\bibnamefont{Klinkhamer}} and \bibinfo{author}{\bibfnamefont{M.}~%
\bibnamefont{Schreck}}.
\bibinfo{journal}{Phys.\ Rev.\ D} \pmb{\bibinfo{volume}{78}}, %
\bibinfo{pages}{085026} (\bibinfo{year}{2008}); [arXiv:0809.3217 [hep-ph]].

\bibitem{Jerzy} Jerzy Pleba\'{n}ski, 
\emph{Lectures on Nonlinear
Electrodynamics} (Nordita, Copenhagen, 1970).

\bibitem{Borninfeld} \bibinfo{author}{\bibfnamefont{M.}~\bibnamefont{Born}}
and \bibinfo{author}{\bibfnamefont{L.}~\bibnamefont{Infeld}}. %
\bibinfo{journal}{Proc. Roy. Soc. A} \pmb{\bibinfo{volume}{144}}, %
\bibinfo{pages}{425} (\bibinfo{year}{1934}).

\bibitem{gitman} R. Fresneda, D.M. Gitman and A.E. Shabad, to be
published

\bibitem{Kostelecky:2008} \bibinfo{author}{\bibfnamefont{V.~A.}~%
\bibnamefont{Kostelecky}} and \bibinfo{author}{\bibfnamefont{N.}~%
\bibnamefont{Russel}}.
\bibinfo{journal}{Rev.\  Mod.\  Phys.\ } \pmb{\bibinfo{volume}{83}}, %
\bibinfo{pages}{11} (\bibinfo{year}{2011}); arXiv: 0801.0287.

\bibitem{Cameron:1993mr} \bibinfo{author}{\bibfnamefont{R.}~%
\bibnamefont{Cameron}  {\it et al.}}
\bibinfo{journal}{Phys.  Rev.  D } \pmb{\bibinfo{volume}{47}}, %
\bibinfo{pages}{3707 } (\bibinfo{year}{1993});

\bibitem{Zavattini:2007ee} \bibinfo{author}{\bibfnamefont{E.}~%
\bibnamefont{Zavattini}  {\it et al.}  [PVLAS Collaboration]}.
\bibinfo{journal}{Phys.  Rev.  D } \pmb{\bibinfo{volume}{77}}, %
\bibinfo{pages}{032006} (\bibinfo{year}{2008}); [arXiv:0706.3419 [hep-ex]].

\end{thebibliography}
\end{document}